\documentclass[reprint,amsmath,amssymb,prl]{revtex4-1}
\usepackage[colorlinks,citecolor=blue]{hyperref}
\usepackage{times}
\usepackage{units}
\usepackage{ulem}
\usepackage{amsmath}
\usepackage{amssymb}
\usepackage{amsfonts}
\usepackage{epstopdf}
\usepackage{microtype}
\usepackage{bm}
\usepackage{xspace}
\usepackage{minitoc}
\usepackage[english]{babel}

\usepackage{graphicx} 			%include math packages
\graphicspath{{Figures/}}

\usepackage{bm}				%include bold math package
\usepackage{color}			%include colored text support
\usepackage{pstricks}			%include postscript drawings support
\usepackage{float}			%include support for 'floating' objects (tables, figures, etc.)

\newcommand{\avg}[1]{ \langle #1 \rangle }
\newcommand{\ket}[1]{ | #1 \rangle }
\newcommand{\bra}[1]{ \langle #1 | }

\newcommand{\alphass}{ \bar{\alpha} }
\newcommand{\alphasss}{ | \bar{\alpha} |^2 }
\newcommand{\betass}{ \bar{\beta} }
\newcommand{\betasss}{ | \bar{\beta} |^2 }
\newcommand{\DeltaLC}{ \Delta_{db}^{\rm LC} }
\newcommand{\DeltaMP}{ \Delta_{db}^{\rm MP} }
\newcommand{\DeltaCR}{ \Delta_{db}^{\rm crit} }
\newcommand{\skc}[1]{{\color{black}#1}}

\begin{document}

\title{Frequency Combs in a Lumped-Element Josephson-Junction Circuit}
\author{Saeed Khan}
\author{Hakan E. T\"ureci}
%\author{\textsf{author3}}
\affiliation{Department of Electrical Engineering, Princeton University, Princeton, New Jersey 08544, USA}
\date{\today}

\begin{abstract}
We investigate the dynamics of a microwave-driven Josephson junction capacitively coupled to a lumped element LC oscillator. In the regime of driving where the Josephson junction can be approximated as a Kerr oscillator, this minimal nonlinear system has been previously shown to exhibit a bistability in phase and amplitude. In the present study, we characterize the full phase diagram and show that besides a parameter regime exhibiting bistability, there is also a regime of self-oscillations characterized by a frequency comb in its spectrum. We discuss the mechanism of comb generation which appears to be different from those studied in microcavity frequency combs and mode-locked lasers. We then address the fate of the comb-like spectrum in the regime of strong quantum fluctuations, reached when nonlinearity becomes the dominant scale with respect to dissipation. We find that the nonlinearity responsible for the emergence of the frequency combs also leads to its dephasing, leading to broadening and ultimate disappearance of sharp spectral peaks. Our study explores the fundamental question of the impact of quantum fluctuations for quantum systems which do not possess a stable fixed point in the classical limit. 
\end{abstract}

\maketitle

In superconducting quantum circuits, the Josephson junction (JJ) is a lossless nonlinear element that provides critical functionality for various quantum information processing tasks~\cite{clarke_superconducting_2008, bertet_circuit_2011, devoret_superconducting_2013}, from gate operations to readout and amplification\skc{, made possible by controlling JJ dynamics via its embedding circuit and effective drive. For instance, under strong coupling and weak excitation (relative to the intrinsic nonlinearity), JJ-based artificial atoms have enabled Cavity QED implementations~\cite{blais_cavity_2004, wallraff_strong_2004, majer_coupling_2007, dicarlo_demonstration_2009, schuster_ac_2005, gambetta_qubit-photon_2006} that have been extensively discussed using open Jaynes-Cummings or Rabi models in single~\cite{bishop_response_2010} and multi-mode regimes~\cite{sundaresan_beyond_2015}. However, applications employing JJs under strong excitation conditions, for \skc{readout} \cite{reed_high-fidelity_2010, vijay_invited_2009, mallet_single-shot_2009} and quantum-limited amplification \cite{yurke_observation_1989, castellanos-beltran_widely_2007, abdo_directional_2013, eichler_quantum-limited_2014, macklin_nearquantum-limited_2015}, require an understanding of dynamical instabilities that sensitively depend on the model of nonlinearity employed~\cite{fink_observation_2017, fitzpatrick_observation_2017}.}

%That it can play such a multi-faceted role relies on how the action and dynamics of the JJ depend on its electromagnetic environment - the embedding circuit and effective drive (for the SQUID configuration, also flux modulation) parameters. For instance, implementations of Cavity QED using JJ-based artificial atoms \cite{blais_cavity_2004, wallraff_strong_2004} for gate operations~\cite{majer_coupling_2007, dicarlo_demonstration_2009} and readout~\cite{schuster_ac_2005, gambetta_qubit-photon_2006}, rely on strong-coupling conditions and weak JJ excitation (relative to the scale set by its intrinsic nonlinearity). The resulting phenomenology has been extensively discussed using the open Jaynes-Cummings or Rabi model in the single-mode~\cite{bishop_response_2010} and multi-mode regimes~\cite{sundaresan_beyond_2015}. However, a number of applications employing JJs for \skc{readout} \cite{reed_high-fidelity_2010, vijay_invited_2009, mallet_single-shot_2009} and quantum-limited amplification \cite{yurke_observation_1989, castellanos-beltran_widely_2007, abdo_directional_2013, eichler_quantum-limited_2014, macklin_nearquantum-limited_2015} require strong excitation conditions. Such conditions require an understanding of various dynamical instabilities that sensitively depend on the model of nonlinearity employed~\cite{bishop_response_2010, fink_observation_2017, fitzpatrick_observation_2017}.

Here we investigate the dynamics of a shunted JJ when capacitively coupled to a \skc{microwave-driven} linear resonator. The dynamics of such a system under \skc{strong driving} have been theoretically \cite{bishop_response_2010, boissonneault_improved_2010} and experimentally \cite{reed_high-fidelity_2010} studied in the context of a high-power read-out scheme, and are found to exhibit a bistability between two states with distinct phase and amplitude. In the adiabatic regime where mode coupling is weaker than losses, the coupled system maps to a single coherently-driven Kerr oscillator with renormalized parameters, exhibiting precisely this bistability~\cite{dykman_theory_1979, siddiqi_RF-driven_2004, siddiqi_direct_2005, vijay_invited_2009}. However, we find that in the strong-coupling regime \skc{the nonlinear mode acquires a retarded self-interaction mediated by the linear mode, which changes} the classical phase diagram dramatically: for certain drive and detuning ranges the system may have \textit{no} stable fixed points, a phase not exhibited by \skc{the} single coherently-driven Kerr oscillator~\cite{dykman_fluctuating_2012}. \skc{In this dynamical regime}, nonzero frequency instabilities emerge as limit cycles, yielding discrete, equally spaced comb-like spectra in the frequency domain.

%~\cite{dykman_theory_1979, siddiqi_RF-driven_2004, siddiqi_direct_2005, vijay_invited_2009}.
%The above-quoted result needs to be put into perspective with respect to other schemes for generating  comb-like spectra in coherently-driven microresonators~\cite{delhaye_optical_2007, delhaye_full_2008, levy_cmos-compatible_2010, herr_temporal_2014, erickson_frequency_2014} and incoherently-pumped mode-locked lasers~\cite{haus_mode-locking_2000, faist_quantum_2016}. Comb formation is often understood as an instability towards symmetric sideband growth via four-wave mixing, in an underlying resonator geometry supporting multiple spatial modes~\cite{chembo_modal_2010, mansuripur_single-mode_2016}. In many of these cases, the nonlinearity is a distributed nonlinearity (while exceptions have been discussed as well~\cite{rayanov_frequency_2015}). Our results indicate that the minimal manifestation of Kerr-mediated comb formation is embodied in a Kerr-oscillator coupled to a linear oscillator.

Such comb formation in coherently-driven microresonators~\cite{delhaye_optical_2007, delhaye_full_2008, levy_cmos-compatible_2010, herr_temporal_2014, erickson_frequency_2014} and incoherently-pumped mode-locked lasers~\cite{haus_mode-locking_2000, faist_quantum_2016} is often understood as an instability towards symmetric sideband growth via four-wave mixing, in an underlying resonator geometry supporting multiple spatial modes~\cite{chembo_modal_2010, mansuripur_single-mode_2016} \skc{and a distributed nonlinearity} (while exceptions have been discussed as well~\cite{rayanov_frequency_2015, ganesan_frequency_2017, ganesan_phononic_2017}). Our results indicate that the minimal manifestation of Kerr-mediated comb formation is embodied in a Kerr-oscillator coupled to a linear oscillator.

\skc{While limit cycles~\cite{d.sc_lxxxviii._1926, strogatz_nonlinear_1994} and their modification under classical noise~\cite{louca_stable_2015, rayanov_frequency_2015} have been extensively studied in classical systems, they are far less explored in the deep quantum regime~\cite{lorch_laser_2014, navarrete-benlloch_general_2017} accessible to the lumped element JJ circuit discussed here.} Using Master equation and phase-space simulations, we investigate the fate of comb-like spectra as the nonlinearity is tuned from weak to strong (equivalently, high to low mode occupation at the instability threshold), so that the system moves from an expected semiclassical regime towards a well-defined quantum regime where a single photon can in principle trigger the comb instability. We find that while the nonlinearity, together with strong enough coupling to the linear mode, is necessary for limit cycles to emerge, this very nonlinearity introduces quantum noise that dephases the limit cycle; for weak noise, the dephasing time typically scales inversely with the strength of the nonlinearity.

\begin{figure}[t]
\includegraphics[scale=0.19]{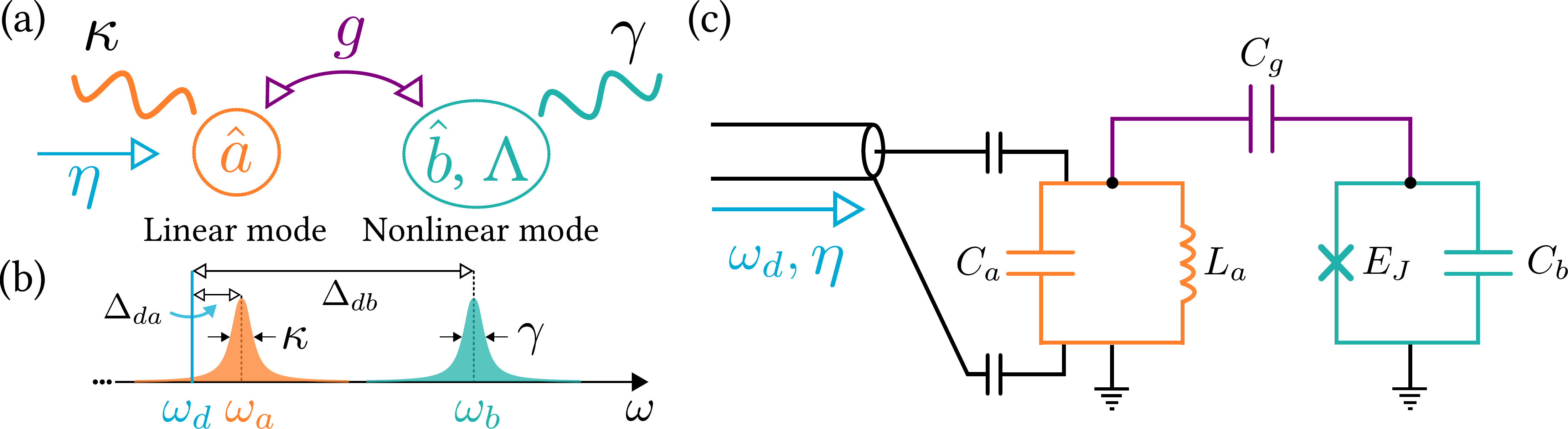}
\caption{(a) Schematic representation of \skc{the two-mode system}. (b) Mode and drive frequencies. (c) Lumped element circuit QED implementation of (a). }
\label{fig:schematic}
\end{figure}

\textit{Model} - The model we study is realizable in lumped element setups [Fig.~\ref{fig:schematic}~(c)], as well as JJ-embedded transmission-line resonators~\cite{bourassa_josephson-junction-embedded_2012}. We assume that the nonlinearity of the junction can be approximated by its lowest order Kerr nonlinearity. The resulting model described by the Hamiltonian $\hat{\mathcal{H}} = \hat{\mathcal{H}}_a + \hat{\mathcal{H}}_b + \hat{\mathcal{H}}_g + \hat{\mathcal{H}}_d$ (See \skc{SM}~\cite{khan_supplementary_2017}) then is generic, consisting of a driven linear oscillator $a$ (frequency $\omega_a$) coupled to a nonlinear oscillator $b$ (frequency $\omega_b$) [Fig.~\ref{fig:schematic}~(a)]. $\hat{\mathcal{H}}_d = \eta(\hat{a}+\hat{a}^{\dagger})$ describes the drive (strength $\eta$) in the frame rotating at the drive frequency $\omega_d$. The corresponding drive frame Hamiltonians of the modes are $\hat{\mathcal{H}}_a = -\Delta_{da}\hat{a}^{\dagger}\hat{a}$ and $\hat{\mathcal{H}}_b =-\Delta_{db} \hat{b}^{\dagger}\hat{b} - \frac{\Lambda}{2}\hat{b}^{\dagger}\hat{b}^{\dagger}\hat{b}\hat{b}$ respectively, with $\Lambda > 0$ being the strength of the nonlinearity, and the frequency detunings defined as $\Delta_{da/db} = \omega_d - \omega_{a/b}$. The two oscillators are coupled linearly, with Hamiltonian $\hat{\mathcal{H}}_g =  g( \hat{a}^{\dagger} \hat{b} + \hat{b}^{\dagger} \hat{a})$. The system dynamics including damping for both the linear (rate $\kappa$) and nonlinear mode (rate $\gamma$) are then governed by the Master equation $\dot{\hat{\rho}} = -i\left[\mathcal{H},\hat{\rho} \right] + \kappa \mathcal{D}[\hat{a}]\hat{\rho} + \gamma \mathcal{D}[\hat{b}]\hat{\rho}$, where $\mathcal{D}[\hat{o}]$ is the standard dissipative superoperator $\mathcal{D}[\hat{o}]\rho = \hat{o}\rho \hat{o}^{\dagger} - \frac{1}{2} \left\{\hat{o}^{\dagger}\hat{o},\rho \right\}$. 

We begin with the classical dynamics of the two-mode system, obtaining operator equations of motion and making the replacement \skc{$( \avg{\hat{a}},\avg{\hat{b}} ) \to ( \alpha,\beta )$}. This yields:
\begin{equation}
\dot{\alpha} = i\Delta_{da}\alpha -\frac{\kappa}{2}\alpha -ig\beta - i\eta;~\dot{\beta} = i\Delta_{db}\beta - \frac{\gamma}{2}\beta  + i\Lambda|\beta|^2\beta -ig\alpha 
\label{classicalEOMs}
\end{equation}
Note that Eqs.~(\ref{classicalEOMs}) are invariant if $\Lambda \to \Lambda/c$ \textit{and} $(\alpha,\beta,\eta) \to \sqrt{c}(\alpha,\beta,\eta)$, for $c \in \mathbb{R}^+$~\cite{khan_supplementary_2017}. Physically, scaling $\Lambda \to \Lambda/c$ and $\eta\to\sqrt{c}\eta$ yields the same \textit{classical} dynamics, except with mode amplitudes scaled by $\sqrt{c}$. This simple $\Lambda$-dependence is not true of the \textit{quantum} dynamics, as we shall see later. Next, the linearity of mode $\hat{a}$ and the coupling allows it to be integrated out exactly, leading to an effective dynamical equation for the nonlinear mode:
\begin{align}
\dot{\beta} &= i\Delta_{db}\beta - \frac{\gamma}{2}\beta  + i\Lambda|\beta|^2\beta +g\chi_a\eta \left[e^{\left(i\Delta_{da}-\frac{\kappa}{2}\right)t} - 1\right]  \nonumber \\
	    &~~~~-g^2\!\! \int_0^t d\tau~F(\tau) \beta(t-\tau),
\label{betaNonMarkov}
\end{align}
where the linear \skc{mode} susceptibility $\chi_a^{-1} = -i\Delta_{da} + \kappa/2$. \skc{The first line is} the classical equation of motion for a coherently-driven Kerr oscillator; the drive term is $\propto g \chi_a$ since the linear mode is driven. More interesting is the term in the second line, which describes the delayed self-interaction of the nonlinear mode - mediated by the linear mode - with a memory kernel $F(\tau) = e^{ \left(i\Delta_{da}-\frac{\kappa}{2}\right)\tau}$. 

\skc{When $F(\tau)$} decays rapidly relative to the timescale of system dynamics ($\kappa \gg g$), we may set $\beta(t-\tau)\approx \beta(t)$ within a Markov approximation; this is also equivalent to adiabatically eliminating the linear mode ($\dot{\alpha} \approx 0$). We then obtain an effective Markov regime equation for the (long-time) dynamics of the nonlinear mode:
\begin{equation}
\dot{\beta} = i\widetilde{\Delta}_{db} \beta - \frac{\widetilde{\gamma}}{2} \beta + i\Lambda|\beta|^2\beta - \widetilde{\eta}
\label{betaMarkov}
\end{equation}
This is the classical dynamical equation for a renormalized Kerr oscillator, with modified detuning $\widetilde{\Delta}_{db} = \omega_d - \left( \omega_b + g^2 \Delta_{da} |\chi_a|^2\right)$, damping $\widetilde{\gamma} = \gamma + g^2\kappa|\chi_a|^2$ and drive $\widetilde{\eta} = g\chi_a \eta$. Therefore, when the linear mode can be adiabatically eliminated, the two-mode system behaves like an effective Kerr oscillator~\cite{drummond_quantum_1980}. 

From here, the classical fixed points $\left(\alphass,\betass\right)$ of the two-mode system are found by setting $\dot{\alpha} = \dot{\beta} = 0$, or equivalently setting $\dot{\beta}=0$ in the Markov regime equation, Eq.~(\ref{betaMarkov}). The equation relating the fixed points $\betasss$ to the drive strength $|\eta|^2$ is found to be the standard cubic polynomial for a Kerr oscillator, with the modified parameters defined earlier~\cite{khan_supplementary_2017}. The relationship is single-valued for \skc{$\Delta_{db} > \DeltaMP$ but becomes multivalued for $\Delta_{db} < \DeltaMP$, defining a region of multiple fixed points; here the critical detuning} $\Delta_{db}^{\rm MP} = -\frac{\sqrt{3}}{2} \left(\gamma + g^2\kappa|\chi_a|^2 \right)  + g^2\Delta_{da}|\chi_a|^2$~\cite{khan_supplementary_2017}. Dropping terms $\propto g^2$ arising from the linear mode yields the standard result for a single driven Kerr oscillator.

\textit{Stability analysis} - To \skc{treat} the memory term \skc{exactly we perform a Laplace domain linear stability analysis} around the above fixed points; details can be found in~\cite{khan_supplementary_2017}. Instability is determined by the dominant pole (pole with largest real part) of the linearized dynamical matrix. For a resonantly driven linear mode, $\Delta_{da} = 0$, an analysis of the real and imaginary parts of the poles separately allows the phase diagram to be mapped out entirely analytically; we focus on this case from here on \skc{(for non-zero $\Delta_{da}$, see SM~\cite{khan_supplementary_2017})}. Two distinct parameter regimes are obtained, determined by the relative strength of $g$ and $\kappa$.

\begin{figure}
\includegraphics[scale=0.155]{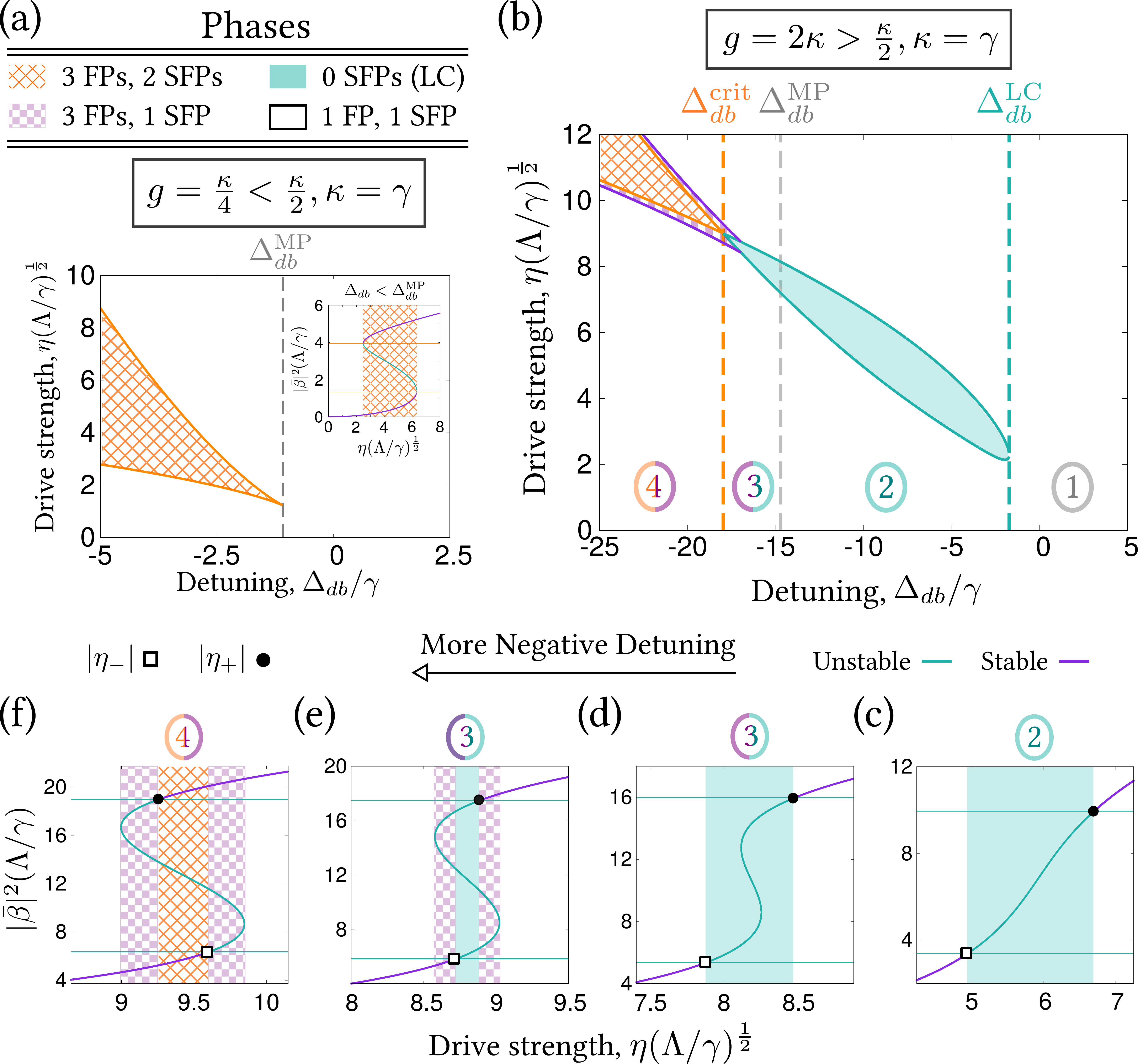}
\caption{Phase diagram in $\eta$-$\Delta_{db}$ space. The possible phases are \skc{listed} in the top-left table. (a) Phase diagram for $g < \frac{\kappa}{2}$ (here, $g = \frac{\kappa}{4}$). (b) Phase diagram for $g > \frac{\kappa}{2}$ (here, $g = 2\kappa$). (c) through (f) Plots of $\betasss$ against $\eta$ showing the change in the S-curve as $\Delta_{db}$ is swept across the four dynamical regions in (b). Green (purple) segments of the S-curve depict unstable (stable) $\betasss$ values.}
\label{fig:betaPhase}
\end{figure}

For $g < \kappa/2$, the typical phase diagram in $\eta$-$\Delta_{db}$ space is shown in Fig.~\ref{fig:betaPhase}~(a). For $\Delta_{db} > \DeltaMP$, the system has only one fixed point (FP), as discussed earlier; the stability analysis indicates that this FP is always stable. For $\Delta_{db} < \DeltaMP$, the orange hatched region emerges where one of the system's FPs is unstable, and the unstable pole $s$ has $\text{Im}~s = 0$. In this region, the typical curve relating $\betasss$ to $\eta$ (S-curve) is shown in the inset, with green (purple) segments showing unstable (stable) FPs. The unstable FPs coincide \textit{exactly} with the region of 3 total FPs; dynamically, instabilities from the unstable branch settle into one of the two stable fixed points at the same drive strength. This is precisely the stability diagram of the \skc{modified Kerr} oscillator defined by Eq.~(\ref{betaMarkov}).

Much more interesting is the case $g > \kappa/2$, for which the phase diagram is shown in Fig.~\ref{fig:betaPhase}~(b). We first consider $\Delta_{db} > \DeltaMP$, where the classical equations admit only 1 FP. We find that for $\Delta_{db}$ above a minimum critical detuning $\DeltaLC = -\frac{\sqrt{3}}{2}\left(\gamma+\kappa \right)$, the single fixed point that exists is never unstable (region~1). For $\DeltaMP < \Delta_{db} < \DeltaLC$ (region 2), this is no longer the case. A typical $\betasss$-$\eta$ plot in region~2, Fig.~\ref{fig:betaPhase}~(c), shows emergent unstable $\betasss$ values in green, where the system has only one, \textit{unstable} FP, hinting at the emergence of limit cycle solutions; this regime is not possible for the single \skc{coherently-driven} Kerr oscillator. The \textit{minimum} and \textit{maximum} \textit{unstable} $\betasss$ values occur at drive strengths $\eta_-$ (open square) and $\eta_+$ (filled circle) respectively. At these drives, the dominant pole reaches the threshold of instability, now with nonzero $\text{Im}~s = \pm\sqrt{g^2-\kappa^2/4} \equiv \pm\Omega$. 

Note that for $\eta > \eta_+$ and $\eta < \eta_-$, \skc{the system always has at least one stable fixed point} (lying on one of the upward pointing purple segments). Also, in region~2, the unstable green segments lie entirely in the drive range $\eta_-<\eta<\eta_+$. However, with more negative detuning, the latter does not remain so. For $\Delta_{db} < \DeltaMP$ (regions 3, 4), the S-curve can be multivalued, as seen in Fig.~\ref{fig:betaPhase}~(d), and eventually further deforms to Fig.~\ref{fig:betaPhase}~(e), where unstable $\betasss$ are \textit{not} all contained in the range $\eta_-<\eta<\eta_+$. Here, for $\eta > \eta_+$ and $\eta<\eta_-$ the system now admits 3 FPs, of which \textit{only one} is stable (checkered purple). On the other hand, \textit{within} the range $\eta_- < \eta < \eta_+$ (shaded green as before), \textit{all three fixed points are unstable}. This is clearly seen in Fig.~\ref{fig:betaPhase}~(e): only green segments of the S-curve lie in the green shaded region.

%With increasing detuning, the S-curve deforms, and for $\Delta_{db} < \DeltaMP$ the system admits multiple fixed points, where a single drive value can correspond to three different $\betasss$ values, as clearly seen in Fig.~\ref{fig:betaPhase}~(e). The S-curve can turn over enough such that not all unstable $\betasss$ values are entirely contained within the drive range $\eta_-<\eta<\eta_+$. It can further be shown that for $\eta > \eta_+$ and $\eta < \eta_-$, there is always atleast one stable fixed point of the system (lying on one of the upward pointing purple segments). Therefore, once the S-curve has turned over sufficiently, a new region emerges where the system admits 3 FPs, of which \textit{only one} is stable (checkered purple). On the other hand, \textit{within} the range $\eta_- < \eta < \eta_+$ (shaded green as before), \textit{all three fixed points are unstable}. This is clearly seen in Fig.~\ref{fig:betaPhase}~(d): only green segments of the S-curve lie in the green shaded region.

Furthermore, the range $\eta_- < \eta < \eta_+$ is detuning dependent; as $\Delta_{db}$ becomes more negative, this region shrinks, and vanishes when $\eta_- = \eta_+$. We mark this as the terminal boundary of region 3, which occurs at a critical detuning $\DeltaCR =  - \sqrt{ (\DeltaMP)^2 + \frac{1}{2} \left[ (\DeltaMP)^2-(\DeltaLC)^2\right]  }$ \skc{[dashed orange line in Fig.~\ref{fig:betaPhase}~(b)].} 

\begin{figure}
\includegraphics[scale=0.38]{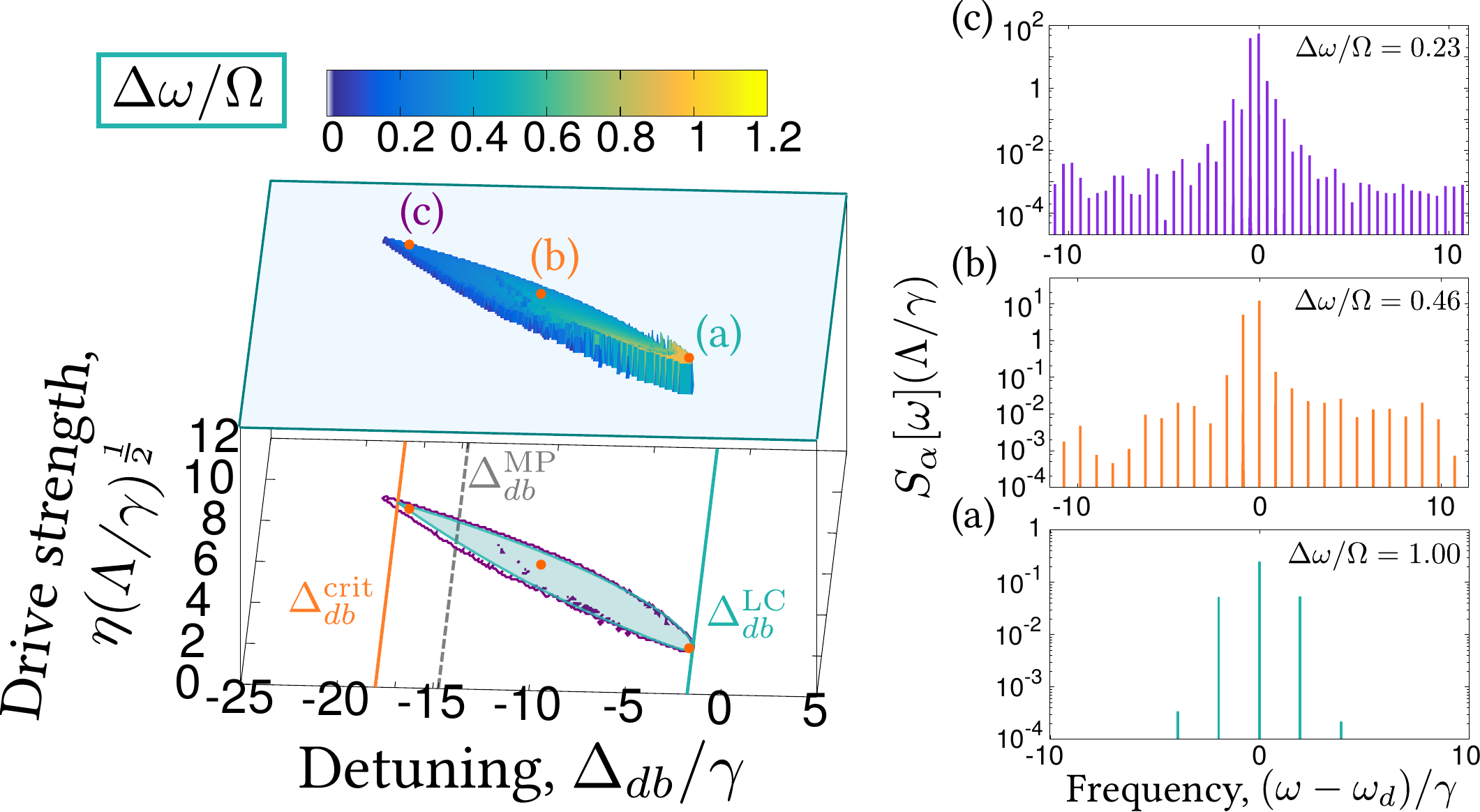}
\caption{Numerically simulated phase diagram in $\eta$-$\Delta_{db}$ space for $g = 2\kappa > \frac{\kappa}{2}$. Surface plot shows spectral spacing $\Delta\omega$ obtained from $S_{\alpha}[\omega]$; \skc{the `island' of multifrequency solutions overlaps perfectly with the analytically-predicted unstable region below (shaded green)}. Plots (a) through (c) show $S_{\alpha}[\omega]$ at the correspondingly labelled points on the phase diagram.  }
\label{fig:numPhase}
\end{figure}

\begin{figure*}
\includegraphics[scale=0.2]{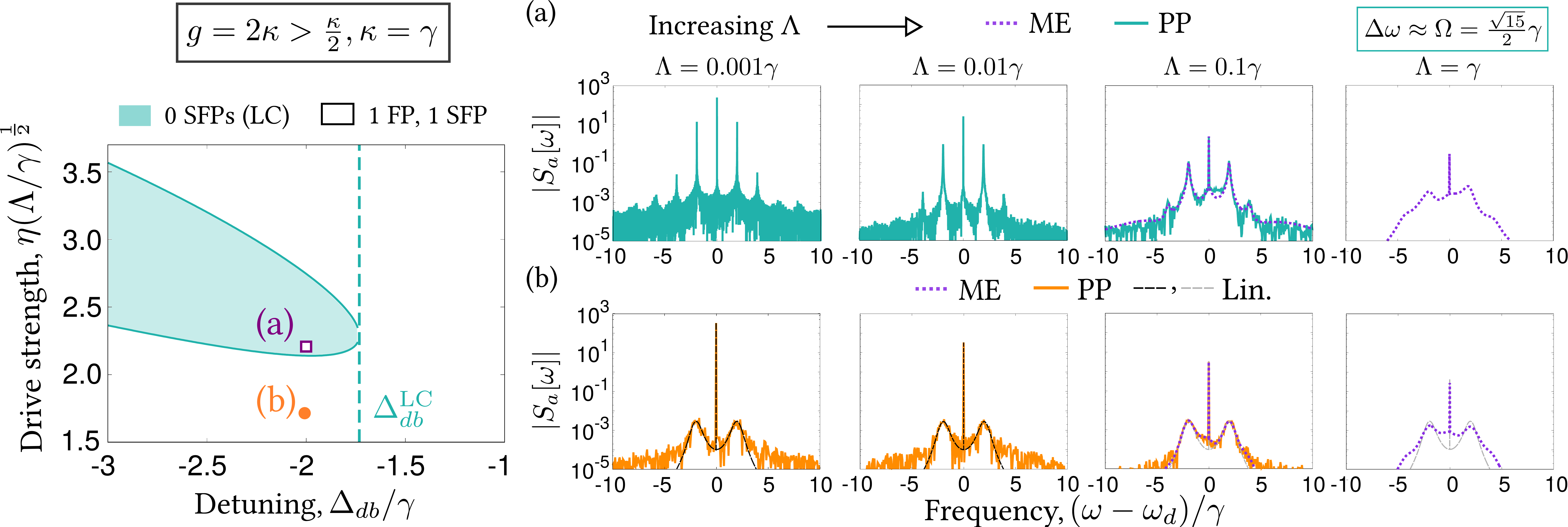}
\caption{$|S_a[\omega]|$, logscale, (a) within, and (b) outside the \skc{limit cycle region}, as a function of increasing nonlinearity $\Lambda$ from left to right. Master equation (ME), SDE simulations (PP), and stable regime linearized spectra (Lin.) are shown.}
\label{fig:corrSpec}
\end{figure*}

For $\Delta_{db} < \DeltaCR$, region~4 begins, where $\eta_- > \eta_+$. The S-curve typically looks like Fig.~\ref{fig:betaPhase}~(f). Since at least one stable fixed point always exists for $\eta > \eta_+$ and $\eta < \eta_-$, and since $\eta_- > \eta_+$ in region~4, we easily conclude that at least one \textit{stable} fixed point (SFP) now exists for \textit{all} driving strengths. The green shaded region with 0 SFPs thus gives way to the orange hatched region with 3 FPs, 2 SFPs. We note that unlike $\DeltaLC$, which is a strict minimal detuning for instability, $\DeltaCR$ is \textit{not} a strict \textit{maximal} detuning. Beyond $\DeltaCR$, limit cycle solutions can ostensibly still emerge, since unstable fixed points with nonzero $\text{Im}~s$ still exist. However, if excursions from these unstable fixed points are large enough, the system can always find a stable fixed point to settle into in this region.

Finally, as $g \to \kappa/2$, $\DeltaCR, \DeltaMP, \DeltaLC$ all become equal; both green and purple regions shrink and eventually vanish, such that for $g < \kappa/2$, only the orange hatched region persists, and we recover the phase diagram in Fig.~\ref{fig:betaPhase}~(a).

% Numerics

Dynamics in the unstable region can be studied numerically by simulating Eqs.~(\ref{classicalEOMs}). We calculate the steady state power spectrum of the linear mode, $S_{\alpha}[\omega] = |\mathcal{F}\left\{ \alpha(t) \right\} |^2$, where $\mathcal{F}\left\{\alpha(\tau) \right\} = \int_{-\infty}^{\infty} d\tau~e^{-i\omega \tau} \alpha(\tau)$ is the Fourier transform. This quantity is of particular relevance for circuit QED realizations of our model, where $S_{\alpha}[\omega]$ is the power spectrum of the resonator mode, which can be directly monitored in experiments \cite{da_silva_schemes_2010}. For each spectrum $S_{\alpha}[\omega]$, the frequency spacing $\Delta\omega$ is plotted in $\eta$-$\Delta_{db}$ space in Fig.~\ref{fig:numPhase}, scaled by $\Omega$. We find multifrequency limit cycles in a region that has excellent agreement with the (green shaded)  analytic region of 0 SFPs. The spacing $\Delta\omega$ is close to $\Omega$ for $\Delta_{db} \sim \DeltaLC$, but decreases as $\Delta_{db}$ becomes more negative; this reduction is observed for general parameters in this system (see additional phase diagrams included in~\cite{khan_supplementary_2017}).

% Quantum regime

\textit{Quantum regime} - To study the modification of limit cycle dynamics in the quantum regime, we employ both Master equation simulations and a stochastic approach based on the positive-$P$ representation of the density matrix $\rho$~\cite{carmichael_statistical_2002}. The latter allows access to normal-ordered operator averages and correlation functions via a set of stochastic differential equations (SDEs) for the independent complex variables $\vec{\alpha} \equiv (\alpha,\alpha^{\dagger},\beta,\beta^{\dagger})$:
\begin{equation}
d\vec{\alpha} = \vec{A}~dt+ \sqrt{\Lambda}~\bm{B}~\vec{dW}
\label{sdes}
\end{equation}
The drift vector $\vec{A}$ describes deterministic classical evolution, equivalent to Eqs.~(\ref{classicalEOMs}). Then, by construction, any quantum effects must appear as stochastic `noise' terms, involving the vector of independent Wiener increments $\vec{dW}$. The scale and nature of this noise is set by the matrix $\bm{B}$. In the absence of thermal noise (which we neglect), $\bm{B} = \sqrt{i}~{\rm diag} \left(0, 0, \beta, i\beta^{\dagger} \right)$ is purely quantum in origin, and depends on $\beta, \beta^{\dagger}$ (as opposed to being constant for thermal noise). Eqs.~(\ref{sdes}) are thus driven by \textit{multiplicative} noise. Crucially, the classical and quantum contributions depend differently on the nonlinearity. Scaling $\Lambda \to \Lambda/c$ \textit{and} $(\vec{\alpha},\eta) \to \sqrt{c}(\vec{\alpha},\eta)$ in Eqs.~(\ref{sdes}) leaves the classical drift term unchanged as discussed earlier, but scales the stochastic term by a $\frac{1}{\sqrt{c}}$ factor~\cite{khan_supplementary_2017}. Hence decreasing $\Lambda$ ($c>1$; equivalently, increasing mode occupation) suppresses the relative impact of quantum dynamics, moving the system closer to effectively classical evolution.

By varying $\Lambda$ and $\eta$ according to the above scaling, we can stay on \textit{fixed} positions on the \textit{classical} phase diagram (a) just within the limit cycle region, and (b) just outside [See Fig.~\ref{fig:corrSpec}], while modifying the stochastic dynamics. For regimes of weak nonlinearity relative to the damping rates, $\Lambda \sim [0.001, 0.1]\gamma, \kappa$, we find that Eqs.~(\ref{sdes}) may be simulated controllably; here, modal occupations of $O([10^2,10^3])$ make Master equation and even Monte Carlo simulations unfeasible. For stronger nonlinearities $\Lambda \gtrsim \gamma$, $P$ representation SDE simulations run into notorious difficulties~\cite{gilchrist_positive_1997, gardiner_quantum_2004}; however, weaker excitation numbers then render density matrix simulations tractable again~\cite{johansson_qutip_2013}. Combining the two methods yields a complete picture of dynamics as the nonlinearity is increased to the quantum regime. The precise scaling of stochastic terms with nonlinearity does depend on the nonlinear model employed~\cite{kamal_signal--pump_2009, navarrete-benlloch_general_2017, casteels_quantum_2017}; \skc{more generally, our} analysis may be regarded as a study of dynamics under transition from high to low modal occupations. We analyze again the linear mode power spectrum, using the Wiener-Khinchin theorem: $S_{a}[\omega] = \int_{-\infty}^{\infty} d\tau~e^{-i\omega \tau} \avg{\hat{a}^{\dagger}(\tau)\hat{a}(0)}$. For SDE simulations, the required correlation function is determined via averaging, $\avg{\hat{a}^{\dagger}(\tau)\hat{a}(0)} = \lim_{t\to\infty} \frac{1}{N_s}\sum_{i=1}^{N_s} \alpha_i^{\dagger}(t+\tau)\alpha_i(t)$, summing over at least $N_s = 10^5$ trajectories for each calculation.

Within the limit cycle region [Fig.~\ref{fig:corrSpec}~(a)], for the weakest nonlinearity $\Lambda = 0.001\gamma$, the spectrum appears close to the classical result [Fig.~\ref{fig:numPhase}~(c)]. However, as the nonlinearity becomes stronger, the peaks in the discrete spectrum broaden. A weak-noise phase dynamics analysis~\cite{khan_supplementary_2017} indicates a phase diffusion time $\propto 1/\Lambda$ (equivalently, comb peak linewidths $\propto\Lambda$), with a proportionality coefficient of order one determined by local properties of the limit cycle attractor. Outside the unstable region [Fig.~\ref{fig:corrSpec}~(b)], the classical FP with mode occupations $(\alphasss,\betasss)$ is stable. Fluctuations around this FP yield a quantum spectrum that we determine analytically~\cite{khan_supplementary_2017} using a linearized analysis~\cite{chaturvedi_stochastic_1977, drummond_quantum_1980, carmichael_statistical_2002}; the result agrees well with $S_a[\omega]$ for weak nonlinearities, but deviates as $\Lambda$ increases and the fluctuations are no longer small relative to $(\alphasss,\betasss)$. For intermediate $\Lambda = 0.1\gamma$, we are able to compare SDE and Master equation simulations in both regions, finding very good agreement.

\textit{Conclusion} - The driven, \textit{strongly-coupled} nonlinear system of a Kerr oscillator and a linear mode admits a phase with no classical SFPs. Here, classical dynamics feature limit cycles with sharp peaks in the mode spectra; however, the \textit{quantum} dynamics introduce dephasing due to the very nonlinearity that gives birth to the limit cycles, broadening and ultimately washing out these spectral peaks as $\Lambda$ increases, even if all other noise sources are absent. Our study is relevant for on-chip microwave domain frequency comb generation using quantum circuits with weak nonlinearities (realized in recent circuit QED experiments~\cite{eichler_controlling_2014, eichler_quantum-limited_2014, zhou_high-gain_2014}), and for further understanding stable operating regimes of JJ-based nonlinear multimode circuit QED systems.

\begin{acknowledgments}

We would like to thank Howard J. Carmichael and Mike Metcalfe for useful discussions. This work was supported by the US Department of Energy, Office of Basic Energy Sciences, Division of Materials Sciences and Engineering under Award No. DE-SC0016011.

\end{acknowledgments}

\newpage

\onecolumngrid

\begin{center}
{\large \textbf{Supplementary Material for ``Frequency Combs in a Lumped-Element Josephson-Junction Circuit''} } \\

\vspace{\baselineskip}

Saeed Khan and Hakan E. T\"ureci \\
{\small
\textit{Department of Electrical Engineering, Princeton University, Princeton, New Jersey 08544, USA} \\
(Dated: April 11, 2018) }
\end{center}

\vspace{\baselineskip}

\twocolumngrid

\appendix

\setcounter{equation}{0}

\setcounter{page}{1}
\thispagestyle{empty}

\tableofcontents

\section{Circuit QED Implementation}
\label{app:cQED}

The two-mode system considered in this paper can be studied in modern circuit QED experiments, with a particularly simple implementation using lumped element circuits~[See Fig.~1~(c) in the main text]. The driven linear mode can be realized with a linear LC circuit coupled to a transmission line, while the nonlinear mode can be realized by shunting a Josephson junction with a capacitor. The two circuits are then capacitively coupled to engineer a quadrature-quadrature coupling~\cite{girvin_circuit_2014}.

To derive the effective system Hamiltonian, we begin with the classical Lagrangian $\mathbb{L}$ for the (undriven) circuit above:
\begin{align}
\mathbb{L} &= \frac{1}{2} C_a {\dot{\Phi}}_a^2 - \frac{{\Phi}_a^2}{2L_a} +  \frac{1}{2} C_b {\dot{\Phi}}_b^2  + E_J \cos \left(2\pi\frac{{\Phi}_b}{\Phi_0} \right) \nonumber \\
  &+ \frac{1}{2}C_g \left[ {\dot{\Phi}}_a-{\dot{\Phi}}_b \right]^2
\end{align}
Here, ${\Phi}_{a,b}$ are the node flux variables for the linear and nonlinear circuit respectively. The first line then contains contributions from the linear and nonlinear mode respectively, while the second line describes the capacitive coupling. It is useful to rewrite the capacitive sector of the Lagrangian in matrix form:
\begin{align}
\mathbb{L} &= \frac{1}{2} \bm{{\dot{\Phi}}}~\mathbf{C}~\bm{{\dot{\Phi}}}^T - \frac{{\Phi}_a^2}{2L_a} + E_J \cos \left(2\pi\frac{{\Phi}_b}{\Phi_0} \right) 
\end{align}
where $\bm{{\dot{\Phi}}} = \{{\dot{\Phi}}_a,{\dot{\Phi}}_b\}$, and $\mathbf{C}$ is the capacitance matrix given by:
\begin{align}
\mathbf{C} = 
\begin{pmatrix}
C_a + C_g & -C_g \\
-C_g & C_b + C_g 
\end{pmatrix}
\end{align}
To obtain the Hamiltonian, we first find the canonical momenta, the charge operators $\{\hat{Q}_a,\hat{Q}_b\}$, defined in the usual way:
\begin{align}
{Q}_{a,b} = \frac{d \mathbb{L}}{d{\dot{\Phi}}_{a,b}} = \sum_{j=a,b} C_{(a,b)j} {\dot{\Phi}}_j \implies \mathbf{{Q}} = \mathbf{C} \bm{{\dot{\Phi}}}
\end{align}
where $\mathbf{{Q}} = \{{Q}_a,{Q}_b\}$. We may now promote the charge and flux variables to operators, ${Q}_i \to {\hat{Q}_i}$, ${\Phi}_i \to {\hat{\Phi}_i}$, satisfying the usual canonical commutation relations $[\hat{\Phi}_i,\hat{Q}_j] = i\delta_{ij}$. Then, the Hamiltonian $\mathbb{H}$ is obtained in the standard way via Legendre transformation of the Lagrangian:
\begin{align}
\mathbb{H} = \sum_{j=a,b} \hat{Q}_j \hat{\dot{\Phi}}_j - \mathbb{L}
\end{align}
which takes the form:
\begin{align}
\mathbb{H} = \frac{1}{2}\mathbf{\hat{Q}} \mathbf{C^{-1}}\mathbf{\hat{Q}}^T + \frac{\hat{\Phi}_a^2}{2L_a} - E_J\cos\left(2\pi\frac{\hat{\Phi}_b}{\Phi_0} \right)
\end{align}
The inverse of the capacitance matrix $\mathbf{C^{-1}}$ is non-diagonal, and leads to the effective coupling term we desire. In particular,
\begin{align}
\mathbf{C^{-1}} = \frac{1}{ C_aC_b+ C_g(C_a+C_b) }
\begin{pmatrix}
C_b+C_g & C_g \\
C_g & C_a + C_g
\end{pmatrix}
\end{align}
Then, the Hamiltonian can be written as:
\begin{align}
\mathbb{H} = \frac{\hat{Q}_a^2}{2\widetilde{C}_a} + \frac{\hat{\Phi}_a^2}{2L_a} + \frac{\hat{Q}_b^2}{2\widetilde{C}_b} - E_J \cos \left( 2\pi\frac{\hat{\Phi}_b}{\Phi_0} \right)  + \mathcal{G} \hat{Q}_a\hat{Q}_b
\end{align}
where we have defined the effective capacitances $\widetilde{C}_a$, $\widetilde{C}_b$:
\begin{align}
\frac{1}{\widetilde{C}_{a/b}} &= \frac{C_{b/a}+C_g}{ C_aC_b+ C_g(C_a+C_b) }
\end{align}
and a coupling strength $\mathcal{G}$:
\begin{align}
\mathcal{G} = \frac{C_g}{ C_aC_b+ C_g(C_a+C_b) }
\end{align}
which is proportional to the coupling capacitance $C_g$, as expected. Taking $C_g \to 0$ returns us to the uncoupled two-mode system.

Next, we expand the Josephson potential to second order to extract the nonlinear term explicitly. This yields:
\begin{align}
\mathbb{H} = \frac{\hat{Q}_a^2}{2\widetilde{C}_a} + \frac{\hat{\Phi}_a^2}{2L_a} + \frac{\hat{Q}_b^2}{2\widetilde{C}_b} + \frac{\hat{\Phi}_b^2}{2\widetilde{L}_b} - \frac{\widetilde{\Lambda}}{24}\hat{\Phi}_b^4 + \mathcal{G} \hat{Q}_a\hat{Q}_b
\end{align}
where we have ignored constant terms. The effective linear inductance and strength of the nonlinear term are respectively given by:
\begin{align}
\frac{1}{\widetilde{L}_b} = E_J\left(\frac{2\pi}{\Phi_0}\right)^2,~\widetilde{\Lambda} = E_J \left(\frac{ 2\pi }{\Phi_0}\right)^4
\end{align}
Lastly, to put the Hamiltonian in a more familiar form, we define the usual creation (and annihilation) operators for each mode as:
\begin{align}
\hat{a} &= \frac{1}{\sqrt{2\hbar L_a\omega_a}}\hat{\Phi}_a + i \frac{1}{\sqrt{2\hbar \widetilde{C}_a\omega_a}}\hat{Q}_a \nonumber \\
\hat{b} &= \frac{1}{\sqrt{2\hbar \widetilde{L}_b\omega_b}}\hat{\Phi}_b + i \frac{1}{\sqrt{2\hbar \widetilde{C}_b\omega_b}}\hat{Q}_b
\end{align}
where $[\hat{a},\hat{a}^{\dagger}] = [\hat{b},\hat{b}^{\dagger}] = 1$. In this basis, and after defining the natural frequencies of each mode as usual:
\begin{align}
\omega_a^2 = \frac{1}{L_a\widetilde{C}_a}~,~\omega_b^2 = \frac{1}{\widetilde{L}_b\widetilde{C}_b}
\end{align}
the Hamiltonian $\mathbb{H}$ transforms to (setting $\hbar = 1$):
\begin{align}
\mathbb{H} &= \omega_a \hat{a}^{\dagger}\hat{a} + \omega_b \hat{b}^{\dagger}\hat{b} - \frac{\Lambda}{2} \hat{b}^{\dagger}\hat{b}^{\dagger}\hat{b} \hat{b} - g(\hat{a}-\hat{a}^{\dagger})(\hat{b}-\hat{b}^{\dagger}) \nonumber \\
&+O[\Lambda (\hat{b}^{\dagger})^n\hat{b}^m, n \neq m ]
\end{align}
In the above, we have again dropped constant terms and the modification ($\propto \Lambda$) of $\omega_b$. We have also pre-emptively extracted out the contribution of the nonlinear term that would survive in a rotating wave approximation (RWA) performed in the frame rotating at some driving frequency $\omega_d$; the terms on the second line will be quickly oscillating in this frame and may be dropped. The coupling term is also written in its full, quadrature-quadrature form; performing the RWA will yield the form used in this paper, $\hat{\mathcal{H}}_g$. The nonlinearity and coupling in terms of circuit parameters respectively take the form:
\begin{align}
\Lambda = \frac{1}{8}\widetilde{\Lambda}\widetilde{L}_b^2\omega_b^2~,~g = \frac{1}{2}\mathcal{G}\sqrt{\widetilde{C}_a\widetilde{C}_b\omega_a\omega_b} 
\end{align}
Using the explicit form of $\widetilde{\Lambda}$ and $\mathcal{G}$ respectively, these constants simplify to:
\begin{align}
\Lambda = \frac{1}{8\widetilde{C}_b}\left(\frac{2\pi}{\Phi_0}\right)^2,~g = \frac{C_g\sqrt{ \omega_a\omega_b} }{\sqrt{(C_a+C_g)(C_b+C_g)}} 
\end{align}
The simple circuit schematic shown here may be modified for additional experimental control; for example, the use of a SQUID in place of the Josephson junction allows \textit{in situ} tuning of the Josephson energy $E_J$ using an external magnetic flux. Shunting the Josephson junction or SQUID with an additional linear inductance allows dilution of the nonlinearity, which would be necessary to explore the limit cycle dynamics studied in this paper. Similarly, employing a chain of coupled SQUIDs also allows access to weaker nonlinearities~\cite{eichler_controlling_2014}.

\section{Linearized equations in Laplace domain}
\label{app:laplace}

To analyze the stability of the fixed point solutions to Eqs.~(1) in the main text, we perform a linear stability analysis. The variables $\alpha$ and $\beta$ are expanded about the steady state values $(\alphass, \betass)$:
\begin{align}
\alpha(t) &= \alphass + \delta\alpha(t)  \nonumber \\
\beta(t)  &= \betass  + \delta\beta(t) 
\label{linExp}
\end{align}
where $\delta\alpha(t)$, $\delta\beta(t)$ are small fluctuations around the fixed point. Our interest, then, is in determining whether these fluctuations grow or decay with time. The dynamics of the fluctuations are described by the linearized versions of Eqs.~(1), which take the form:
\begin{align}
\delta\dot{\alpha} &= i\Delta_{da}\delta\alpha - \frac{\kappa}{2}\delta\alpha -ig\delta\beta \nonumber \\
\delta\dot{\beta} &= i\Delta_{db}\delta\beta - \frac{\gamma}{2}\delta\beta + i2\Lambda \betasss\delta\beta + i\Lambda\betass^2\delta\beta^*-ig\delta\alpha
\label{linEqs}
\end{align}
While a stability analysis of the above 4-by-4 system ($\delta\alpha, \delta\beta$ are complex) can be performed directly, we proceed via an alternative, and eventually more intuitive analysis. As before, we integrate out the linear mode, thereby simplifying to a single equation of motion for fluctuations of the nonlinear mode:
\begin{align}
\delta\dot{\beta} &= i\Delta_{db}\delta\beta - \frac{\gamma}{2}\delta\beta + i2\Lambda\betasss\delta\beta + i\Lambda\betass^2\delta\beta^* \nonumber \\
&~~~-g^2\!\! \int_0^t d\tau~F(\tau)\delta\beta(t-\tau) 
\label{dBMem}
\end{align}
Here we have ignored the initial value term $\delta\alpha(0)$; the resulting equation is simply the linearized version of Eq.~(2) in the main text. For the analysis of stability, the appearance of the convolution term hints at an efficient solution in the Laplace domain. Defining the Laplace transform as
\begin{align}
\mathcal{L}\left\{G(t) \right\} \equiv G[s] =  \int_0^{\infty} dt~e^{-st} G(t),
\end{align}
we introduce the Laplace-transformed fluctuation terms: $\mathcal{L}\{\delta\beta(t)\} = \delta\beta[s], \mathcal{L}\{\delta\beta^*(t)\} = \delta\beta^*[s]$), and the Laplace transform of the memory kernel in Eq.~(\ref{dBMem}):
\begin{align}
F[s] = \mathcal{L}\left\{ F(\tau)  \right\} = \frac{ \left(s+\frac{\kappa}{2} \right) + i\Delta_{da} }{ \left( s + \frac{\kappa}{2}\right)^2 + \Delta_{da}^2 }
\label{Fs}
\end{align}
We have written the above transform in a non-standard but equivalent form which proves convenient later. With these definitions, Eq.~(\ref{dBMem}) and its conjugate in the Laplace domain may be conveniently written in the following matrix form:
\begin{align}
\mathcal{M}[s]
\begin{pmatrix}
\delta\beta[s] \\
\delta\beta^*[s]
\end{pmatrix} = 
\begin{pmatrix}
\delta\beta(0) \\
\delta\beta^*(0)
\end{pmatrix} 
\end{align} 
The vector on the right hand side contains the initial values of the fluctuating terms. The dynamical matrix in the Laplace domain, $\mathcal{M}$, takes the form:
\begin{widetext}
\begin{align}
\mathcal{M} = 
\begin{bmatrix}
s - i\Delta_{db} + \frac{\gamma}{2} - i2\Lambda\betasss + g^2F[s] & -i\Lambda\betass^2 \\
i\Lambda(\betass^*)^2 & s + i\Delta_{db} + \frac{\gamma}{2} + i2\Lambda\betasss + g^2F^*[s]  
\end{bmatrix}
\end{align}
\end{widetext}
Here, we define $F^*[s] = \mathcal{L}\left\{\mathcal{F}^*(\tau) \right\} \neq \left(F[s] \right)^*$. The poles of the transfer function that determine unstable solutions are given by the zeros of the determinant of $\mathcal{M}[s]$; these satisfy the equation:
\begin{widetext}
\begin{align}
\left[ \left( s - i\Delta_{db} + \frac{\gamma}{2} - i2\Lambda\betasss + g^2F[s] \right)\left(s + i\Delta_{db} + \frac{\gamma}{2} + i2\Lambda\betasss + g^2F^*[s] \right) - \Lambda^2|\betass|^4 \right] = 0
\label{poles}
\end{align} 
\end{widetext}
The above is a quartic polynomial in $s$, but a redefinition allows us to simplify it greatly; using the form of $F[s]$ in Eq.~(\ref{Fs}), we may define effective detuning and damping terms that are now dependent on $s$:
\begin{align}
\widetilde{\Delta}_{db}[s] &= \Delta_{db} - G[s]\Delta_{da}, \nonumber \\
\widetilde{\gamma}[s] &= \gamma + 2G[s] \left( s + \frac{\kappa}{2} \right), \nonumber \\
G[s] &= \frac{g^2}{ \left( s + \frac{\kappa}{2} \right)^2 + \Delta_{da}^2 }
\label{sParams}
\end{align} 
With this redefinition, Eq.~(\ref{poles}) becomes exactly equivalent to the equation determining stability for a single Kerr oscillator (this must be so, since all the $g$-dependence that comes from the linear mode has been absorbed into the effective detuning and damping terms). Then, the solutions for $s$ are formally given by:
\begin{align}
s = -\frac{\widetilde{\gamma}[s]}{2} \pm \sqrt{ \Lambda^2|\betass|^4 - \left( \widetilde{\Delta}_{db}[s] + 2\Lambda\betasss \right)^2 }
\label{sSol}
\end{align}
Note that this equation is still non-trivial to solve, since it is an implicit equation for $s$. However, a complete analysis can be carried out in the regime where the drive is resonant with the linear mode, $\Delta_{da} = 0$. Then, $\widetilde{\Delta}_{db}[s] = \Delta_{db}$, namely the $s$-dependence of the detuning term drops out.

\section{Region of instability in $\betasss$-$\Delta_{db}$ space}
\label{app:linStab}

The philosophy of the linear stability analysis we perform is standard: we are only interested in the pole $s$ with largest real part, and whether this real part can become positive - indicating an unstable fixed point - as system parameters are changed. If such an instability is possible, we wish to map out the region in parameter space where it occurs. 

For convenience, we define the square root term in Eq.~(\ref{sSol}), for the condition $\Delta_{da} = 0$, as:
\begin{align}
D(\Delta_{db},\betasss) = \sqrt{ \Lambda^2|\betass|^4 - \left( \Delta_{db} + 2\Lambda\betasss \right)^2 }
\label{D}
\end{align}
and indicate the dependence on $(\Delta_{db}, \betasss)$ only where appropriate. We now proceed to analyzing instabilities in this regime. In doing so, we will first consider $\betasss$ to be a free (positive-definite) parameter, and map out phase diagrams in $\betasss$-$\Delta_{db}$ space. In the next Appendix section, we include details of relating $\betasss$ back to the driving strength. Note that once $\Delta_{da} = 0$, the argument of the square root term is purely real, and hence the square root term can only yield a purely real \textit{or} a purely imaginary result. Clearly, for both small enough and large enough $\betasss$, the argument of the square root is negative, yielding a purely imaginary contribution. In this case, writing $s = \text{Re}~s + i~\text{Im}~s$, where $\text{Re}~s,\text{Im}~s\in \mathbb{R}$, and comparing the real parts of Eq.~(\ref{sSol}) yields:
\begin{align}
{\rm Re}~s \stackrel{!}{=} -\frac{1}{2} \left( \gamma + \frac{ 2g^2 \left(\text{Re}~s+\frac{\kappa}{2} \right) }{\left({\rm Im}~s \right)^2 + \left( \text{Re}~s + \frac{\kappa}{2} \right)^2 }\right) 
\end{align}
Note that to have $\text{Re}~s > 0$, we require the left hand side to be positive definite; however, this requirement forces the right hand side to be negative definite, thereby making our demand inconsistent. Hence $\text{Re}~s$ is always negative when the square root term is purely imaginary, indicating an absence of instability of the classical fixed points in this case.

For an intermediate range of $\betasss$ values, the square root term becomes positive and its contribution is \textit{purely real}. Here, the equations for the real and imaginary parts of Eq.~(\ref{sSol}) yield:
\begin{align}
{\rm Re}~s = -\frac{1}{2} \left(\gamma + \frac{2g^2 \left( \text{Re}~s + \frac{\kappa}{2}\right) }{\left({\rm Im~}s\right)^2 + \left(\text{Re}~s + \frac{\kappa}{2}\right)^2 } \right) \pm D  
\label{ReS}
\end{align}
and
\begin{align}
&{\rm Im}~s = \frac{g^2{\rm Im}~s }{ \left({\rm Im}~s\right)^2 + \left(\text{Re}~s + \frac{\kappa}{2} \right)^2 }
\end{align}
respectively. The second equation can immediately be solved for $\text{Im}~s$,
\begin{align}
{\rm Im}~s = 0, \pm\sqrt{ g^2 - \left(\text{Re}~s + \kappa/2\right)^2 } 
\label{ims}
\end{align}
These are the only possible values of $\text{Im}~s$ for a pole when $D$ is purely real. As discussed in the main text, this defines two separate two parameter regimes. If $g > \kappa/2$, it is possible to have \textit{unstable} poles ($\text{Re}~s > 0$), which have nonzero imaginary parts. However, if $g < \kappa/2$, the only value of $\text{Im}~s$ possible for an unstable pole is $\text{Im}~s = 0$.

\subsubsection{$g < \kappa/2$ case}

We begin by considering what will turn out to be the simpler regime, where $g < \kappa/2$. As discussed, in this case only poles with $\text{Im}~s = 0$ can become unstable. With this restriction, Eq.~(\ref{ReS}) for $\text{Re}~s$ can be written in the form:
\begin{align}
\frac{ \left[\text{Re}~s + \frac{\kappa}{2}\left(1 - \frac{4g^2}{\kappa^2} \right) \right] \text{Re}~s }{\text{Re}~s + \frac{\kappa}{2} } = -\frac{1}{2}\left(\gamma+ \frac{4g^2}{\kappa} \right) \pm D  
\label{zeroCond}
\end{align}
Note that if $g < \kappa/2$, then $4g^2/\kappa < 1$, and hence requiring $\text{Re}~s \geq 0$ necessarily forces the left and right hand sides of the above equation to be greater than or equal to zero as well. This is only possible if we take the positive sign above. Furthermore, this allows us to define a necessary condition for instability of a pole:
\begin{align}
D(\Delta_{db},\betasss) \geq \frac{1}{2}\left( \gamma + \frac{4g^2}{\kappa} \right)
\label{regionMPDef}
\end{align}
This condition is in fact a quadratic inequality which defines the range of $\betasss$ values, bounded by $|\bar{\beta}_{\pm}^{\rm MP}|^2$, within which a pole can become unstable,
\begin{align}
|\bar{\beta}_{-}^{\rm MP}|^2 < \betasss < |\bar{\beta}_{+}^{\rm MP}|^2~;~\text{Re}~s > 0~\&~\text{Im}~s = 0
\label{MPCond}
\end{align}
The values $|\bar{\beta}_{\pm}^{\rm MP}|^2$ are defined via:
\begin{align}
\Lambda |\bar{\beta}_{\pm}^{\rm MP}|^2 = -\frac{2}{3}\Delta_{db} \pm \frac{1}{3} \sqrt{\Delta_{db}^2 - \frac{3}{4}\left(\gamma + \frac{4g^2}{\kappa} \right)^2 }
\label{betaMP}
\end{align}
We emphasize that within the region defined by Eq.~(\ref{MPCond}) \textit{not} all poles will have positive real parts; rather this defines the only region where a pole \textit{could} become unstable. Outside this region, all poles must necessarily have negative real parts. To see that one pole does in fact become unstable, we can simply solve Eq.~(\ref{zeroCond}), which is a quadratic in $\text{Re}~s$. This yields:
\begin{align}
2\text{Re}~s = &\left( D - \frac{\gamma+\kappa}{2} \right) \nonumber \\
	       &\pm \sqrt{ \left( D - \frac{\gamma+\kappa}{2} \right)^2 + 2\kappa \left[ D - \frac{1}{2}\left(\gamma + \frac{4g^2}{\kappa} \right)   \right] }
\end{align}
Within the region given by Eq.~(\ref{regionMPDef}), the term in square brackets is positive definite; \skc{thus} the discriminant of the quadratic is positive, guaranteeing two distinct real roots. The solution obtained by choosing the plus sign above yields the unstable pole with positive real part.

\begin{figure}
\includegraphics[scale=0.25]{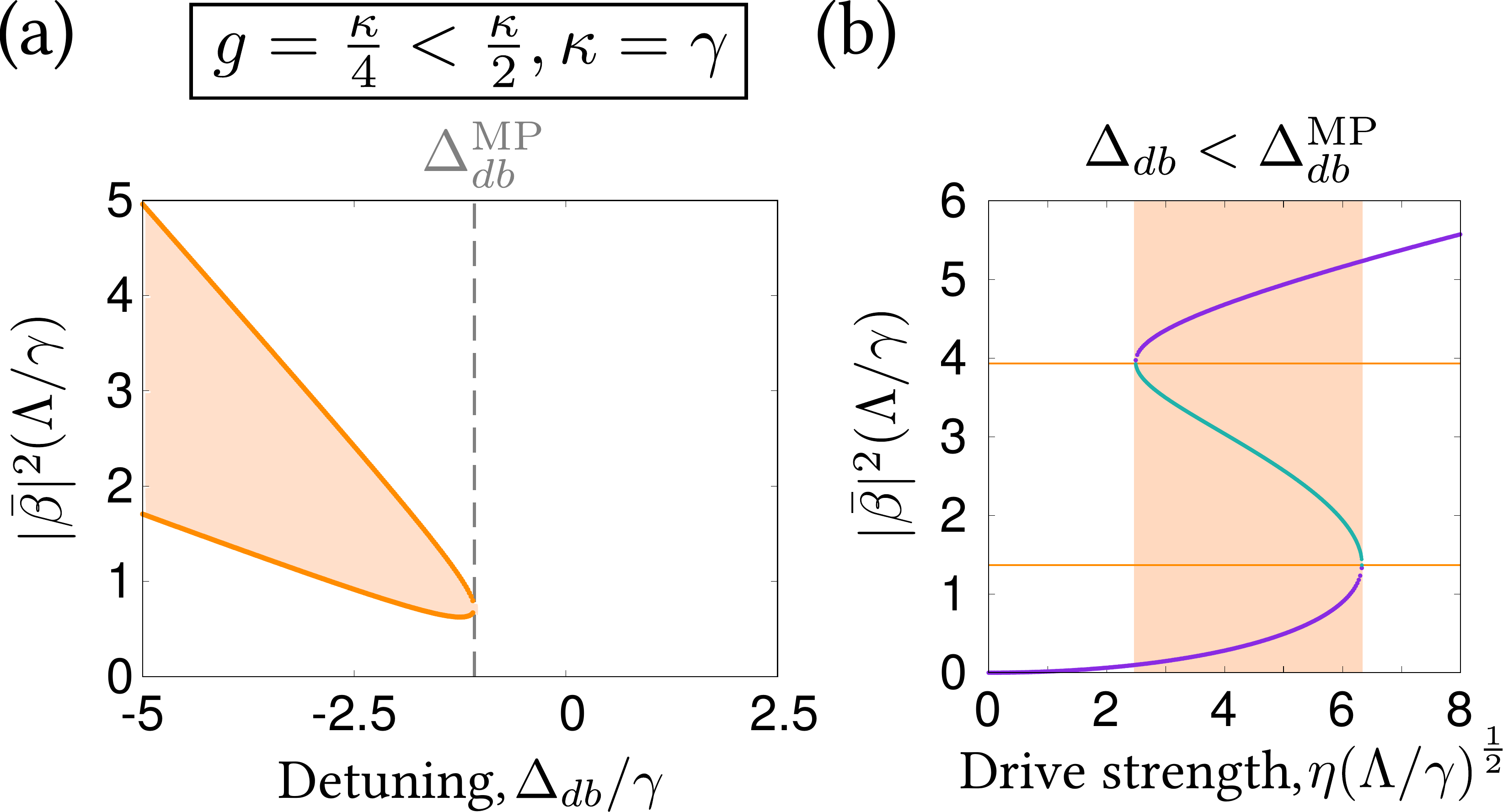}
\caption{(a) Region of instability of the two mode system in $\betasss$-$\Delta_{db}$ space for $g < \frac{\kappa}{2}$ (here, $g =\frac{\kappa}{4}$) and $\Delta_{da} = 0$. In the orange shaded region, three fixed points exist, two of which are stable, while in the blank region only a single, stable fixed point exists. (b) Typical plot of $\betasss$ against $\eta$ (S-curve); here $\Delta_{db} < \DeltaMP$, so that multiple fixed points exist for driving strengths in the orange shaded region. The downward branch, in green, consists of exactly those $\betasss$ values that lie within the shaded region in (a) at the given detuning.}
\label{fig:betaNoLC}
\end{figure}

The quadratic inequality in Eq.~(\ref{MPCond}) has another important consequence; the instability region it defines can only be reached below a critical detuning,
\begin{align}
\Delta_{db} < -\frac{\sqrt{3}}{2} \left(\gamma + \frac{4g^2}{\kappa} \right) = \DeltaMP~~~~~[\Delta_{da} = 0]
\end{align}
which is exactly the critical detuning (with $\Delta_{da} = 0$) for the existence of multiple fixed points of the two-mode system, as described in the main text. In fact, we have essentially recovered the results one would find from a linear stability analysis of the single Kerr oscillator, with modified parameters: instabilities are zero-frequency instabilities (unstable poles have vanishing imaginary part), and only exist concurrently with regions where multiple fixed points are possible, namely for $\Delta_{db} < \DeltaMP$. Furthermore, the unstable fixed point values $\betasss$ satisfying Eq.~(\ref{MPCond}) are precisely those that lie on the downward branch of the multivalued S-curve, again exactly like the single Kerr oscillator. In a typical plot of $\betasss$ against $\eta$ in Fig.~\ref{fig:betaNoLC}~(b), this branch is indicated in green. Instabilities from this branch drive the system to settle into one of the two other fixed points, lying on the upward branches of the S-curve, which are stable [purple segments in Fig.~\ref{fig:betaNoLC}~(b)]. This is the origin of bistability in the standard Kerr oscillator, and it appears that for $g < \kappa/2$, the same physics manifests in the two-mode system.

\subsubsection{$g > \kappa/2$ case}

We consider next the case where $g > \kappa/2$ such that $\text{Im}~s \neq 0$ is \textit{possible} for unstable poles. Poles in this case can either have zero or nonzero imaginary parts. Starting with poles of the latter form, we ask whether these can become unstable, which requires analyzing $\text{Re}~s$ as given by Eq.~(\ref{ReS}), with $\text{Im}~s = \pm \sqrt{g^2 - \left(\text{Re}~s + \kappa/2\right)^2} \neq 0$. We find:
\begin{align}
2{\rm Re}~s = -\frac{1}{2} \left(\gamma + \kappa  \right) + D  
\label{realSNZ}
\end{align}
To allow for instability of these poles, $\text{Re}~s > 0$, we again must choose the positive sign for $D$ above, which provides the condition:
\begin{align}
D\left(\Delta_{db}, \betasss \right) > \frac{1}{2}\left(\gamma + \kappa \right)
\label{LCCond}
\end{align}
This is a quadratic inequality in $\betasss$ very similar to Eq.~(\ref{MPCond}); it is satisfied for values of $\betasss$ in the range,
\begin{align}
&|\bar{\beta}_{-}^{\rm LC}|^2 \leq \betasss \leq |\bar{\beta}_{+}^{\rm LC}|^2
\end{align}
where the values $|\bar{\beta}_{\pm}^{\rm LC}|^2$ are given by:
\begin{align}
\Lambda |\beta_{\pm}^{\rm LC}|^2 = -\frac{2}{3}\Delta_{db} \pm \frac{1}{3} \sqrt{\Delta_{db}^2 - \frac{3}{4}\left(\gamma + \kappa \right)^2 }
\label{betaLC}
\end{align}
Most importantly, the discriminant of the quadratic enforces that the instability is possible \textit{only} below a critical detuning given by:
\begin{align}
\Delta_{db}^{\rm LC} = -\frac{\sqrt{3}}{2} \left(\gamma + \kappa \right)
\end{align}
Hence, poles with $\text{Im}~s \neq 0$ can in fact become unstable, for $\betasss$ satisfying Eq.~(\ref{MPCond}), \textit{provided} $\Delta_{db} < \DeltaLC$. This is now a critical detuning for the threshold of these non-zero frequency instabilities. Furthermore, for $g > \kappa/2$, the critical detuning $\DeltaLC > \DeltaMP$, so that these poles can become unstable even in the regime of a single fixed point. 

However, the above is not the full story, since $\text{Im}~s$ depends on $\text{Re}~s$. As $\text{Re}~s$ increases, it is possible for it to be large enough that $\text{Im}~s \to 0$. In this case, $\text{Re}~s$ is no longer given by Eq.~(\ref{realSNZ}), but rather by Eq.~(\ref{zeroCond}), since the former assumed $\text{Im}~s \neq 0$. Therefore, while the above conclusions regarding the critical detuning and the \textit{boundary} of the instability region are correct, we cannot \textit{a priori} say that for all $\betasss$ satisfying Eq.~(\ref{LCCond}), at least one unstable pole must exist. This is simply because $\text{Re}~s$ is not given by Eq.~(\ref{realSNZ}) for \textit{all} points in this region.

To proceed, we first find the boundary where $\text{Re}~s \to g-\kappa/2$, and thus $\text{Im}~s \to 0$. To do so, we rewrite Eq.~(\ref{zeroCond}) in the equivalent form:
\begin{align}
\frac{\left[\text{Re}~s - (g-\frac{\kappa}{2}) \right]^2 }{\text{Re}~s + \frac{\kappa}{2} } = -\frac{1}{2}\left(\gamma + 4g -\kappa \right) \pm D
\label{reX}
\end{align}
Requiring $\text{Re}~s > g-\kappa/2$ forces the left hand side to be positive-definite; this demands of the right hand side:
\begin{align}
D\left(\Delta_{db}, \betasss \right) > \frac{1}{2}\left(\gamma + 4g -\kappa \right)
\label{DeltaX}
\end{align}
The above therefore once again defines a quadratic inequality for the region where $\text{Re}~s$ can be greater than $g-\kappa/2$. It is satisfied for $\betasss$ values in the range:
\begin{align}
|\bar{\beta}_-^{\rm X}|^2 < \betasss < |\bar{\beta}_+^{\rm X}|^2
\label{XCond}
\end{align}
where:
\begin{align}
\Lambda|\bar{\beta}_{\pm}^{\rm X}|^2 = -\frac{2}{3}\Delta_{db} \pm \sqrt{\Delta_{db}^2-\frac{3}{4}\left(\gamma+4g-\kappa \right)^2 }
\end{align}
The above necessitates $\Delta_{db} < \Delta_{db}^{\rm X}$, where
\begin{align}
\Delta^{\rm X}_{db} = -\frac{\sqrt{3}}{2} \left(\gamma+4g-\kappa \right)
\end{align}
At the boundary of the region defined by Eq.~(\ref{XCond}), two poles have real part equal to $g-\kappa/2$, due to being repeated roots as seen in Eq.~(\ref{reX}). Within this region we can find the real parts of the poles by solving Eq.~(\ref{reX}), which yields:
\begin{align}
\text{Re}~s = &\left(g-\frac{\kappa}{2}\right) + \frac{1}{2}\left[D - \frac{\gamma+4g-\kappa}{2}  \right] \nonumber \\
	      &\pm \frac{1}{2} \sqrt{ \left[D - \frac{\gamma+4g-\kappa}{2}  \right]^2 + 4g\left[D - \frac{\gamma+4g-\kappa}{2}  \right]  }
\end{align}
Again, the terms in square brackets are positive within the region defined by Eq.~(\ref{DeltaX}), which forces the solution obtained by choosing the plus sign to be greater than $g-\kappa/2$. Since this necessarily implies $\text{Re}~s > 0$ for $g>\kappa/2$, we have shown that within this region, there is always at least one pole with positive real part (in fact real part greater than $g-\kappa/2$). 

\begin{figure}
\includegraphics[scale=0.27]{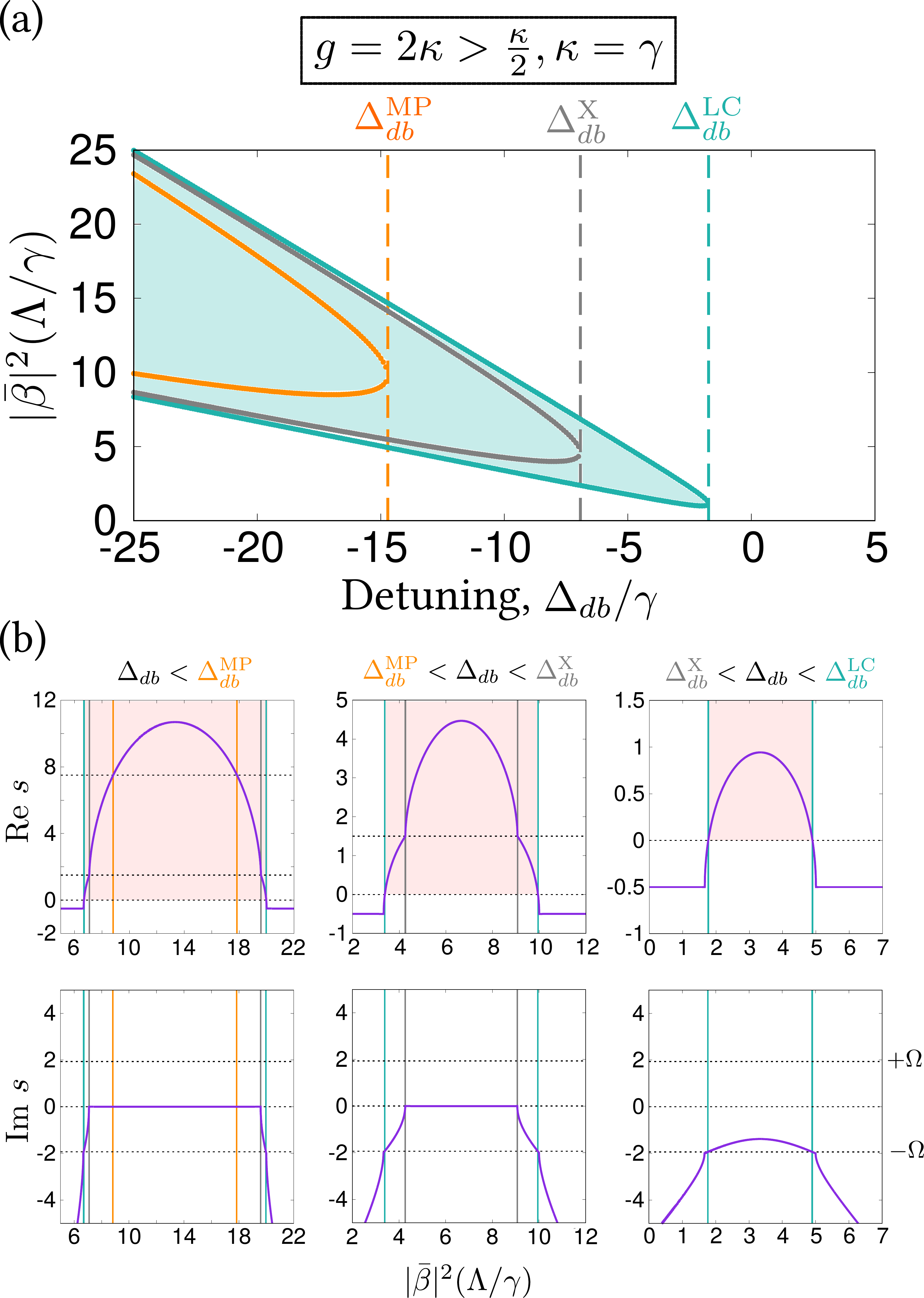}
\caption{(a) Region of instability in $\betasss$-$\Delta_{db}$ space for $g > \kappa/2$ (here, $g=2\kappa$) and $\Delta_{da} = 0$. All unstable $\betasss$ values lie in the green shaded region, enclosed by the green curve. (b) Typical flow of the real part (top panel) and imaginary part (bottom panel) of the dominant pole, as a function of fixed point value $\betasss$ for different detuning ranges indicated above the plots (see text for details). Red shaded region is where $\text{Re}~s > 0$.}
\label{fig:poleFlowLC}
\end{figure}

To summarize our results visually, we plot in Fig.~\ref{fig:poleFlowLC}~(a) the various regions obtained in $\betasss$-$\Delta_{db}$ space. The green shaded region is defined by Eq.~(\ref{LCCond}). \textit{On} the green curve bounding this region, poles with $\text{Im}~s = \pm\sqrt{g^2-\kappa^2/4}$ just reach threshold. The region defined by Eq.~(\ref{XCond}) is enclosed by the grey curve, while the region defined by Eq.~(\ref{MPCond}) is bounded by the orange curve. 

The typical flow of the unstable pole (pole with largest real part) as a function of $\betasss$ is indicated in Fig.~\ref{fig:poleFlowLC}~(b), for detunings in the various regions of Fig.~\ref{fig:poleFlowLC}~(a). For $\Delta_{db} > \DeltaLC$, no pole becomes unstable, and therefore the flow is not plotted. For $\Delta^{\rm X}_{db} < \Delta_{db} < \DeltaLC$, the unstable region lying within the green curve is accessible. The boundary of this region is depicted by the vertical green lines in the top panel of Fig.~\ref{fig:poleFlowLC}~(b). The pole becomes unstable once it crosses this green curve: its real part becomes positive (red shaded region), and remains so within the green boundary. The imaginary part of the pole at the crossing is equal to $\Omega$, and reduces in magnitude within the unstable region. Note that we have not shown the complex conjugate pole with imaginary part $-\Omega$, for clarity. 

For $\DeltaMP < \Delta_{db} < \Delta^{\rm X}_{db}$, the grey curve is now also accessible. Its boundaries are indicated by the vertical grey lines, which always lie within the green lines (since the grey curve is enclosed by the green curve). For $\betasss$ values within the grey curve, the real part of the pole remains positive; however, the imaginary part vanishes at the grey boundary, and remains so for all points within this boundary. This behaviour of the unstable pole is very similar for $\Delta_{db} < \DeltaMP$ as well. 

Crucially, our overall analysis indicates that the pole with largest real part always has a positive real part within the region enclosed by the green curve, and has a negative real part outside this region. Thus, for the case where $g > \kappa/2$, for \textit{all} values of $\betasss$ satisfying Eq.~(\ref{LCCond}) [green shaded region in Fig.~\ref{fig:betaNoLC}~(c)], there is at least one unstable pole. Eq.~(\ref{LCCond}) hence defines the unstable $\betasss$ values employed in the discussion of Fig.~2 of the main text.

\section{Analytic phase boundaries and critical detunings in $\eta$-$\Delta_{db}$ space}
\label{app:stabilityCurves}

As discussed in the main text, the critical drive values corresponding to the maximum and minimum unstable $\betasss$ values are crucial in determining the phase boundaries in $\eta$-$\Delta_{db}$ space. To do so requires solving for the fixed points $(\alphass,\betass)$ of Eqs.~(1) of the main text. The equation which relates $\betasss$ to the drive strength $\eta$ is a cubic polynomial, and hence may be solved analytically; it takes the form:
\begin{align}
\left[ \left(\widetilde{\Delta}_{db} + \Lambda \betasss \right)^2 + \left( \frac{\widetilde{\gamma}}{2} \right)^2 \right] \betasss = |\widetilde{\eta}|^2
\label{cubic}
\end{align}
where the parameters $\widetilde{\Delta}_{db}$, $\widetilde{\gamma}$, and $\widetilde{\eta}$ are all as defined in the main text. To proceed, we now rewrite $\Lambda \betasss = y$ and simplify to find:
\begin{align}
y^3 + 2\widetilde{\Delta}_{db}y^2 + \left( \widetilde{\Delta}_{db}^2 + \frac{\widetilde{\gamma}^2}{4} \right)y - \Lambda |\widetilde{\eta}|^2 = 0
\end{align} 
A standard transformation,
\begin{align}
z = y + \frac{2\widetilde{\Delta}_{db}}{3}
\end{align}
puts the above equation into the `depressed cubic' form (with vanishing quadratic term):
\begin{align}
z^3 + pz + q = 0
\label{depressedCubic}
\end{align}
where we have defined:
\begin{align}
p &= +\left[ \frac{\widetilde{\gamma}^2}{4} - \frac{1}{3}\widetilde{\Delta}_{db}^2 \right] \nonumber \\
q &= -\left[ \frac{2}{27}\widetilde{\Delta}_{db}^3 + \frac{\widetilde{\Delta}_{db}\widetilde{\gamma}^2}{6} + \Lambda |\widetilde{\eta}|^2 \right]
\end{align}
The discriminant for Eq.~(\ref{depressedCubic}) is $\mathcal{S} = -4p^3 - 27q^2$, which separates its solutions into two regions. For $\mathcal{S} > 0$, the cubic equation has three real roots, which indicates the region of multiple fixed points. This requires $p < 0$, which implies $\widetilde{\Delta}_{db} < -\frac{\sqrt{3}}{2} \widetilde{\gamma}$, as mentioned in the main text. In this region, the real roots of the cubic equation may be expressed in the form:
\begin{align}
z_k = 2\sqrt{ -\frac{p}{3} } \cos \left[ \frac{1}{3} \cos^{-1} \left( \frac{3q}{2p}\sqrt{-\frac{3}{p} }\right) - \frac{2\pi k}{3} \right],~p < 0
\label{realRoots}
\end{align}
where $k = 0,1,2$ indexes the three roots. 

If instead $p > 0$, then the discriminant is necessarily negative, and the cubic equation has only one real root. An analytic result for the one real root that always exists may be written in terms of hyperbolic functions:
\begin{align}
z_0 = -2\sqrt{ \frac{p}{3} } \sinh \left[ \frac{1}{3} \sinh^{-1} \left( \frac{3q}{2p}\sqrt{ \frac{3}{p} }  \right) \right],~p > 0
\label{sinhSol}
\end{align}
While the above form strictly holds for $p > 0$, it also extends to the region with multiple real roots, but with complex hyperbolic functions. Undoing our transformations from $z \to y \to \betasss$, the final solution for $\betasss$ in terms of $\eta$ is given by:
\begin{align}
\Lambda \betasss = -\frac{2}{3}\widetilde{\Delta}_{db} - 2\sqrt{ \frac{p}{3} } \sinh \left[ \frac{1}{3} \sinh^{-1} \left( \frac{3q}{2p}\sqrt{ \frac{3}{p} }  \right) \right]
\end{align}

Having solved Eq.~(\ref{cubic}), we will now specialize to the resonantly-driven linear mode case, $\Delta_{da}  =0$, so that $\widetilde{\Delta}_{db} \to \Delta_{db}$. Then, we may impose the conditions on $\betasss$ required by Eq.~(\ref{MPCond}) and Eq.~(\ref{LCCond}). The maximum and minimum values $|\bar{\beta}_{\pm}^{\rm c}|^2$ defined by these conditions may be encapsulated by the single equation:
\begin{align}
\Lambda|\bar{\beta}_{\pm}^c|^2 = -\frac{2}{3}\Delta_{db} \pm \frac{1}{3}\sqrt{ \Delta_{db}^2 - \frac{3}{4}\Gamma^2 }
\end{align}
where for $\Gamma = \gamma + 4g^2/\kappa$ the above reproduces Eq.~(\ref{betaMP}), and for $\Gamma = \gamma + \kappa$ it becomes Eq.~(\ref{betaLC}) instead. Equating these values to our obtained solution, we have:
\begin{align}
- 2\sqrt{ \frac{p}{3} } \sinh \left[ \frac{1}{3} \sinh^{-1} \left( \frac{3q}{2p}\sqrt{ \frac{3}{p} }  \right) \right] = \pm \frac{1}{3}\sqrt{ \Delta_{db}^2 - \frac{3}{4}\Gamma^2 }
\end{align}
Rearranging, we can solve for $q$,
\begin{align}
q = 2 \left( \frac{p}{3} \right)^{3/2} \sinh \left\{ 3\sinh^{-1}\left[  \mp \frac{1}{ 2\sqrt{3p} }\sqrt{ \Delta_{db}^2 - \frac{3}{4}\Gamma^2 }\right] \right\}
\end{align}
Now, using the explicit form of $q$, we find for the drive strengths $|\widetilde{\eta}_{\pm}|$,
\begin{widetext}
\begin{align}
\Lambda|\widetilde{\eta}_{\pm}^c|^2 = \Lambda g^2 |\chi_a|^2 |\eta_{\pm}^c|^2 = - \left[  \frac{2}{27}\Delta_{db}^3 + \frac{1}{6}\Delta_{db}\widetilde{\gamma}^2 + 2 \left( \frac{p}{3} \right)^{3/2} \sinh \left\{ 3\sinh^{-1}\left[  \mp \frac{1}{ 2\sqrt{3p} }\sqrt{ \Delta_{db}^2 - \frac{3}{4}\Gamma^2 }\right] \right\} \right]
\label{etaSol}
\end{align}
\end{widetext}
Before discussing how the above result relates to the phase diagrams in the main text, we note that the discriminant of Eq.~(\ref{depressedCubic}) vanishes ($\mathcal{S} = 0$) precisely at driving strengths $|\eta_{\pm}^{\rm MP}|$ given by Eq.~(\ref{etaSol}) for $\Gamma = \gamma + 4g^2/\kappa$. It is then easily found that within the driving range $|\eta_-^{\rm MP}| < |\eta| < |\eta_+^{\rm MP}|$, the discriminant is negative ($\mathcal{S} < 0$); thus, only within this range does the system exhibit three fixed points. 

\subsubsection{$g<\kappa/2$ case}

We now begin with the simpler $g<\kappa/2$ case, where the instability region is determined by Eq.~(\ref{MPCond}). Letting $\Gamma = \gamma + 4g^2/\kappa$ in Eq.~(\ref{etaSol}) yields the drive strengths $|\eta_{\pm}^{\rm MP}|$ as mentioned earlier; these correspond to the boundary values $|\beta_{\pm}^{\rm MP}|^2$ respectively, as set by Eq.~(\ref{MPCond}). Thus $|\eta_+^{\rm MP}|$ and $|\eta_-^{\rm MP}|$ determine the upper and lower segments respectively of the orange curve on the phase diagram, Fig.~2~(a) of the main text. Note that the region of instability coincides exactly with the region of multiple fixed points, a standard result for the Kerr oscillator, recovered for the two-mode system provided $g<\kappa/2$.

\subsubsection{$g>\kappa/2$ case}

For $g>\kappa/2$, the instability region is determined by Eq.~(\ref{LCCond}). Setting $\Gamma = \gamma+\kappa$, we obtain $|\eta_{\pm}|$ as defined in the main text, namely the drive strengths corresponding to $\betasss$ values satisfying the boundaries of Eq.~(\ref{LCCond}). When $\Delta_{db} = \DeltaLC$, the two drive values are equal, $|\eta_+| = |\eta_-|$, as the limit cycle instability first emerges. Beyond the critical detuning $\DeltaLC$, the expressions for $|\eta_+|$ and $|\eta_-|$ in Eq.~(\ref{etaSol}) define the upper and lower segments respectively of the green curve in Fig.~2~(b) of the main text. As mentioned there, with more negative detuning the two drive strengths again become equal at $\DeltaCR$. To compute this detuning, we simply equate $|\eta_+|$ and $|\eta_-|$ as found via Eq.~(\ref{etaSol}), but using now the real valued roots for $p<0$ given by Eq.~(\ref{realRoots}). We obtain:
\begin{align}
&\frac{2p}{3}\sqrt{-\frac{p}{3} } \cos \left[ 3 \cos^{-1}\!\! \left(+ \frac{1}{2\sqrt{-3p}} \sqrt{(\DeltaCR)^2\! -\! \frac{3}{4}\Gamma^2 }  \right) \right] = \nonumber \\
&\frac{2p}{3}\sqrt{-\frac{p}{3} } \cos \left[ 3 \cos^{-1}\!\! \left(- \frac{1}{2\sqrt{-3p}} \sqrt{(\DeltaCR)^2\! -\! \frac{3}{4}\Gamma^2 }  \right) \right] 
\end{align}
where $\Gamma = \gamma+\kappa$. Simplifying, we have:
\begin{align}
&3\cos^{-1} \left( +\frac{1}{2\sqrt{-3p}} \sqrt{(\DeltaCR)^2 - \frac{3}{4}\Gamma^2 } \right) = \nonumber \\
&3\cos^{-1} \left( -\frac{1}{2\sqrt{-3p}} \sqrt{(\DeltaCR)^2 - \frac{3}{4}\Gamma^2 } \right) + 2 \pi m
\end{align}
where $m = 0, 1, 2$. For $m=0$, the above equation is satisfied at the critical detuning $\DeltaLC$, as we already know. The only remaining distinct solutions, for $m=1,2$, turn out to be identical. Hence choosing $m=1$, and simplifying using the explicit form of $p$ and some trigonometry, we can solve for $\DeltaCR$:
\begin{align}
\DeltaCR &= -\sqrt{(\DeltaMP)^2 + \frac{1}{2}\left[(\DeltaMP)^2- (\DeltaLC)^2\right]}  \nonumber \\
\implies \DeltaCR &= -\sqrt{ \frac{9}{8} \left(\gamma + \frac{4g^2}{\kappa} \right)^2 - \frac{3}{8} \left(\gamma+\kappa\right)^2 } 
\end{align}

Once $\Delta_{db} < \DeltaCR$, we have $|\eta_-| > |\eta_+|$; thus $|\eta_+|$ and $|\eta_-|$ now describe the \textit{lower} and \textit{upper} segments respectively (order reversed) of the orange curve in Fig.~2~(b). The final part of the phase diagram is the purple curve; the drive values $|\eta_+^{\rm MP}|$ and $|\eta_-^{\rm MP}|$ give the upper and lower segments respectively of this curve. Let $\Delta_{db}^{\rm P}$ be the critical detuning where this curve first emerges on the phase diagram; beyond this detuning, the driving range for 3 FPs, $|\eta_-^{\rm MP}| < |\eta| < |\eta_+^{\rm MP}|$ lies within this purple curve. Therefore, beyond $\Delta_{db}^{\rm P}$, the purple curve separates regions of one FP from regions of 3 FPs. This change is accompanied by a change in sign of the discriminant $\mathcal{S}$ of Eq.~(\ref{depressedCubic}), such that the discriminant vanishes for points on the purple curve. Note further that at $\Delta_{db}^{\rm P}$, the purple and green curves intersect; thus at this critical detuning, points lying on the green curve must concurrently cause the discriminant $\mathcal{S}$ to vanish. The latter constraint requires:
\begin{align}
q = \sqrt{- \frac{4}{27}p^3 }~~~~~~~~~[\mathcal{S} = 0]
\end{align}
Using the expression for real roots in this parameter regime, Eq.~(\ref{realRoots}), we find:
\begin{align}
q = \frac{2p}{3}\sqrt{-\frac{p}{3} } \cos \left[ 3 \cos^{-1} \left(\pm \frac{1}{2\sqrt{-3p}} \sqrt{(\Delta_{db}^{\rm P})^2 - \frac{3}{4}\Gamma^2 }  \right) \right]
\end{align}
where $\Gamma = \gamma + \kappa$ for points on the green curve. The above relation between $q$ and $p$ for $\mathcal{S} = 0$, together with the explicit form of $p$, then implies:
\begin{align}
\pm \frac{1}{2} \sqrt{ \frac{(\Delta_{db}^{\rm P})^2 - \frac{3}{4}\Gamma^2 }{(\Delta_{db}^{\rm P})^2 - \frac{3}{4}\widetilde{\gamma}^2} } = \cos\left[ \frac{1}{3}\cos^{-1}\left( 1 \right) \right] 
\end{align}
The only nontrivial solution arises if $\cos^{-1}(1) = 0$. Finally, solving for $\Delta_{db}^{\rm P}$ yields:
\begin{align}
\Delta_{db}^{\rm P} &= -\sqrt{(\DeltaMP)^2 + \frac{1}{3}\left[(\DeltaMP)^2-(\DeltaLC)^2 \right]}  \nonumber \\
\implies \Delta_{db}^{\rm P} &= -\sqrt{ \left( \gamma + \frac{4g^2}{\kappa} \right)^2 - \frac{1}{4} \left(\gamma+\kappa\right)^2} 
\end{align}
The first line above indicates that $\Delta_{db}^{\rm P}$ is always more negative than $\DeltaMP$, since $(\DeltaMP)^2 > (\DeltaLC)^2$. Comparison with $\DeltaCR$ also indicates that $\Delta_{db}^{\rm P} > \DeltaCR$, and therefore $\DeltaCR < \Delta_{db}^{\rm P} < \DeltaMP$, placing it in region 3 of Fig.~2~(b) of the main text. For clarity in that figure, we have omitted explicitly labelling this detuning there.

As an aside, we can also use Eq.~(\ref{etaSol}) to obtain the \textit{critical} drive strength values needed at the minimal detunings for each instability region. For the region with multiple fixed points, the critical drive strength $|\eta_d^{\rm MP}|$ is obtained by setting $\Delta_{db} = -\frac{\sqrt{3}}{2}\Gamma$, with $\Gamma = \gamma+ 4g^2/\kappa$:
\begin{align}
|\widetilde{\eta}_d^{\rm MP}|^2 = \frac{1}{3\sqrt{3}} \frac{\widetilde{\gamma}^3}{\Lambda} = \frac{1}{3\sqrt{3}} \frac{\left( \gamma + 4g^2/\kappa \right)^3}{\Lambda}
\end{align}
which we recognize as the usual result for a single Kerr oscillator with effective damping rate $\gamma + 4g^2/\kappa$. On the other hand, at the threshold of the limit cycle instability, we can find the required drive strength $|\widetilde{\eta}_d^{\rm LC}|$ by setting $\Delta_{db} = -\frac{\sqrt{3}}{2} \Gamma$, with $\Gamma = \gamma + \kappa$ instead. This yields:
\begin{align}
|\widetilde{\eta}_d^{\rm LC}|^2 &= \!\left[ \frac{1}{4} \frac{ (\gamma+\kappa)^2 }{ (\gamma + 4g^2/\kappa)^2 } + \frac{3}{4} \right]\!\! \left( \gamma + \kappa \right) \!\frac{1}{3\sqrt{3}}\frac{1}{\Lambda}\!\left( \gamma + 4g^2/\kappa \right)^2\!  \nonumber \\
|\widetilde{\eta}_d^{\rm LC}|^2 &= \!\left[ \frac{1}{4} \left( \frac{ \gamma+\kappa }{ \gamma + 4g^2/\kappa }\right)^3  + \frac{3}{4} \left( \frac{\gamma+\kappa}{\gamma + 4g^2/\kappa} \right) \right] |\widetilde{\eta}_d^{\rm MP}|^2 
\end{align}
The factor in square brackets is always less than unity for $g > \kappa/2$.

\section{Supplementary phase diagrams}
\label{app:suppPhaseDiagrams}

In this brief section we include additional numerically simulated phase diagrams, and comparisons with stability analysis predictions, for some parameter regimes beyond those considered in the main text. In the phase diagrams included here, the green shaded region indicates the predicted instability region where the system no stable fixed points, while the purple contour plot separates the numerical limit cycle solutions from single frequency solutions. The frequency spacing $\Delta\omega$ in each plot is scaled by $\Omega = \sqrt{g^2 -\kappa^2/4}$, as in the main text. In this section, we consider $\Delta_{da} = 0$.

\begin{figure}[t]
\includegraphics[scale=0.6]{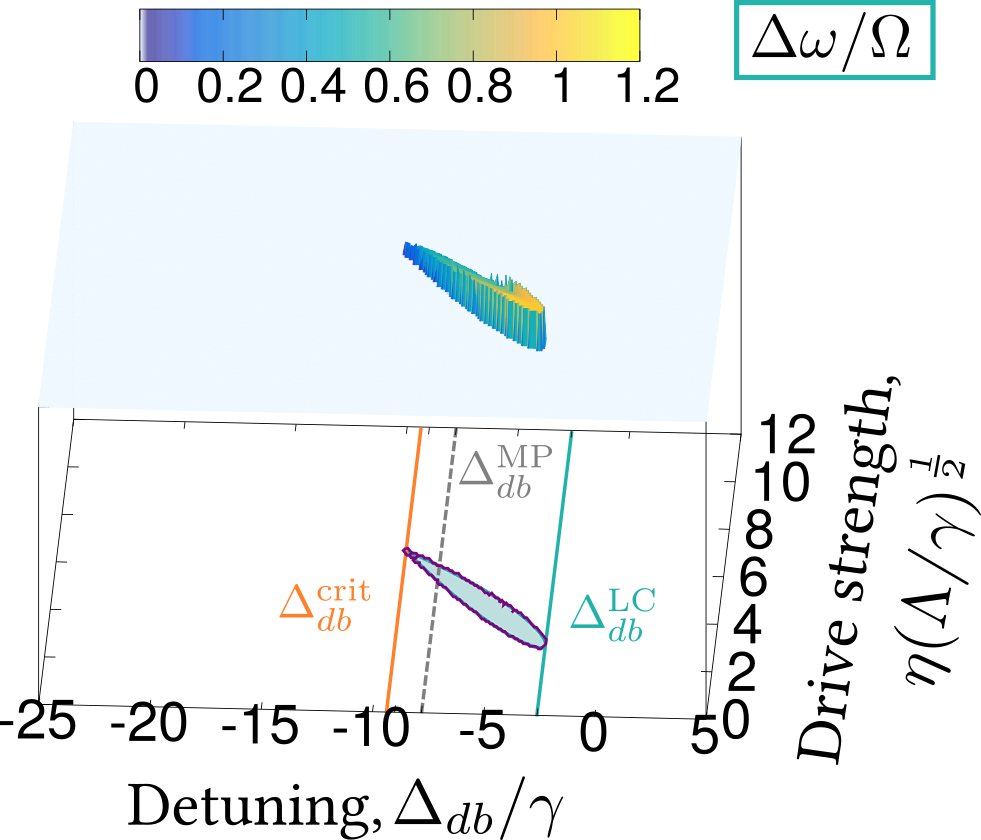}
\caption{Phase diagram in $\eta$-$\Delta_{db}$ space for $g = 2\gamma$, $\kappa = 2\gamma$ so that $g=\kappa$, and the linear mode being resonantly driven ($\Delta_{da} = 0$). Purple curve indicates numerical boundary of instability, and green shaded region is the analytical prediction for the 0 SFP region.}
\label{fig:kappa2}
\end{figure}

\begin{figure}[t]
\includegraphics[scale=0.6]{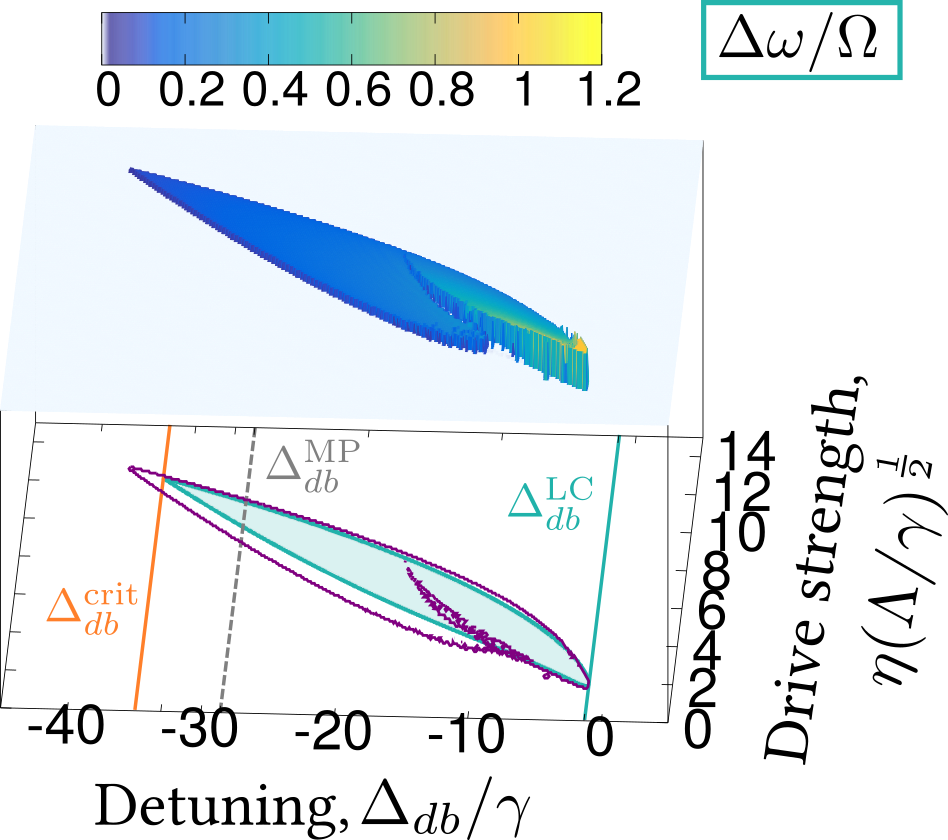}
\caption{Phase diagram in $\eta$-$\Delta_{db}$ space for $g = 2\gamma$, $\kappa = \frac{1}{2}\gamma$ so that $g = 4\kappa$, with the linear mode being resonantly driven ($\Delta_{da} = 0 $). Purple curve indicates numerical boundary of instability, and green shaded region is the analytical prediction for the 0 SFP region.}
\label{fig:kappa0.5}
\end{figure}

Fig.~\ref{fig:kappa2} shows the numerical phase diagram in $\eta$-$\Delta_{db}$ space for $g=\kappa = 2\gamma$, together with the analytical prediction for the unstable region in shaded green. Similarly, in Fig.~\ref{fig:kappa0.5}, we show the phase diagram for stronger coupling $g = 4\kappa$, by taking $g = 2\gamma, \kappa = \frac{1}{2}\gamma$. Overall, very good agreement is observed; note however that in the latter case there is some discrepancy near the instability boundary. Generally, we find that for weaker values of $\kappa$, dynamics at points near the instability boundary are quite sensitive to initial conditions. The results here clearly indicate how the value of $g$ relative to $\frac{\kappa}{2}$ controls the size of the unstable region in $\eta$-$\Delta_{db}$ space.

\section{Instabilities for non-resonantly driven linear mode ($\Delta_{da} \neq 0$)}
\label{app:nzDeltaDA}

So far we have taken the linear mode to be resonantly driven, namely $\Delta_{da} = 0$. In this case, the linear mode susceptibility $\chi_a^{-1} = -i\Delta_{da}+\kappa/2$ becomes purely real, so that Markovian-regime frequency shift of the nonlinear mode due to the linear mode vanishes, namely $\widetilde{\Delta}_{db} = \Delta_{db}$. The same effect simplifies the linear stability analysis, as the $s$-dependence of $\widetilde{\Delta}_{db}[s]$ in Eq.~(\ref{sSol}) drops out. In this section, we consider the case where $\Delta_{da} \neq 0$. The linear oscillator now acquires a complex-valued susceptibility. For the analysis of fixed points of the two-mode system, governed by Eq.~(\ref{cubic}), this is simply accounted for by retaining the $\Delta_{da}$ dependence of $\widetilde{\Delta}_{db}$. However, the linear stability analysis becomes more involved, since the argument of the square root becomes $s$-dependent. 

\textit{At threshold} where $\text{Re}~s=0$, the real and imaginary parts of Eq.~(\ref{sSol}) yield an equation for the (unknown) value of $\text{Im}~s$, together with an equation describing the unstable $\betasss$ values as a function of $\Delta_{db}$. However, the system still admits poles with $\text{Im}~s=0$ at threshold. In this case, Eq.~(\ref{sSol}) simplifies, yielding the boundary of instability in $\betasss$-$\Delta_{db}$ space for such zero frequency instabilities. This region exists provided $\Delta_{db} < \Delta_{db}^{\rm MP}[\Delta_{da}]$, and the resulting boundary values are given by:
\begin{align}
\Lambda|\bar{\beta}_{\pm}^{\rm MP}[\Delta_{da}]|^2 = -\frac{2}{3}\widetilde{\Delta}_{db} \pm \frac{1}{3} \sqrt{\widetilde{\Delta}_{db}^2 - \frac{3}{4}\widetilde{\gamma}^2 }
\end{align}
The square brackets indicate the dependence of $|\bar{\beta}_{\pm}^{\rm MP}|^2$ and $\Delta_{db}^{\rm MP}$ on $\Delta_{da}$. Note that the above reduces exactly to Eq.~(\ref{betaMP}) when $\Delta_{da} \to 0$, as expected. This result is not surprising: unstable poles with $\text{Im}~s=0$ occur for the usual Kerr multistability, where an unstable fixed point coincides with two stable fixed points. Here, the Kerr oscillator appears with parameters modified by the linear mode. Driving the linear mode off-resonance contributes to a slightly different modification of these parameters than in the $\Delta_{da} = 0$ case, namely by inducing a frequency shift and reducing the amplitude of the modification $\propto |\chi_a|$. 

For poles with nonzero imaginary part, Eq.~(\ref{sSol}) must be solved for $\text{Im}~s$. Squaring Eq.~(\ref{sSol}) and setting its imaginary part to zero yields an expression for $(\text{Im}~s)^2$ of a pole at threshold:
\begin{align}
&(\text{Im}~s)^2 = g^2\!\left(\frac{\gamma-\kappa}{2\gamma} \right)-\left[\left(\frac{\kappa}{2} \right)^2\!\!\!-\Delta_{da}^2 \right] \pm \nonumber \\
& \sqrt{g^4\!\left(\frac{\gamma+\kappa}{2\gamma} \right)^2\!\!\!- \kappa^2\Delta_{da}^2-\frac{2g^2\kappa}{\gamma}\Delta_{da}\left[\Delta_{da}+\Delta_{db}+2\Lambda\betasss \right] } 
\label{nzIMS}
\end{align}
For $\Delta_{da} \to 0$, the positive solution above yields precisely the nonzero result found in Eq.~(\ref{ims}) for $\text{Re}~s = 0$. Immediately, a difficulty relative to that case becomes apparent: $\text{Im}~s$ now depends on (the bare) $\Delta_{db}$ and $\betasss$ as well. This renders determining the instability curve in $\betasss$-$\Delta_{db}$ space an analytically more challenging task. The results simplify somewhat in the case where $\gamma=\kappa$, to which we now specialize. Poles with $\text{Im}~s$ defined by Eq.~(\ref{nzIMS}) become unstable for $\betasss$ values satisfying:
\begin{align}
\left[D(\Delta_{db},\betasss)\right]^2 = 2g^2+\frac{\kappa^2}{2}+\Delta_{da}^2-2(\text{Im}~s)^2
\label{quartic}
\end{align}
where $D(\Delta_{db},\betasss)$ is defined as in Eq.~(\ref{D}). In writing the above, we have assumed both that the argument of the square root in Eq.~(\ref{nzIMS}) is always positive, and that $\text{Im}~s \neq 0$. This equation reduces to Eq.~(\ref{LCCond}) with $\gamma=\kappa$ if we take $\Delta_{da} \to 0$ and retain the real solution for $\text{Im}~s$, as it must. Eq.~(\ref{LCCond}) is a simple (bi-)quadratic in $\betasss$ and thus easily solvable; Eq.~(\ref{quartic}), on the other hand, is a true quartic equation in $\betasss$ due to the specific dependence of $\text{Im}~s$ on $\betasss$. While technically accessible, the full solutions of Eq.~(\ref{quartic}) are quite unwieldy, and we do not include them explicitly here.

\begin{figure}[t]
\includegraphics[scale=0.315]{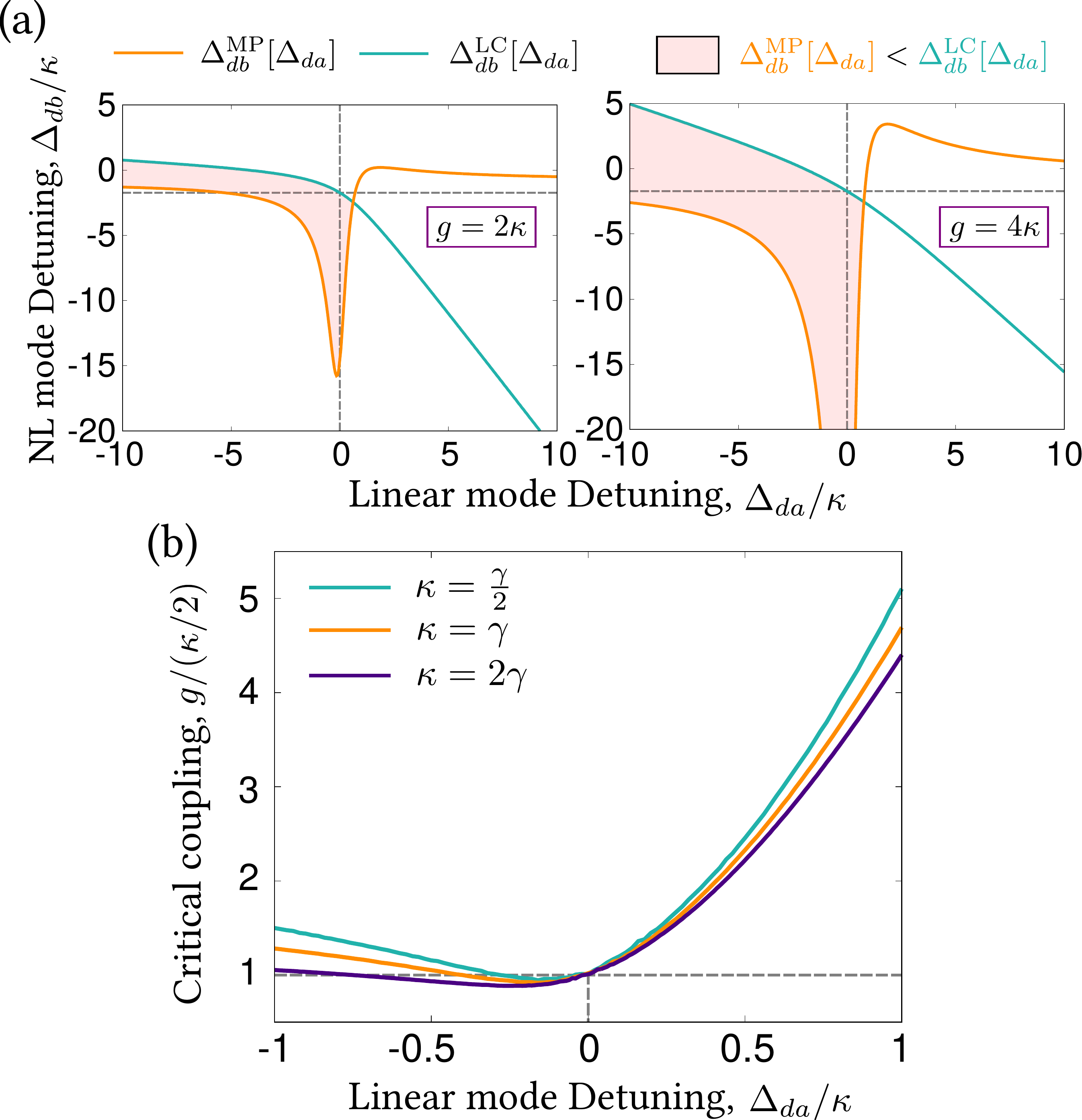}
\caption{(a) Plot of $\DeltaLC[\Delta_{da}]$ (solid green) and $\DeltaMP[\Delta_{da}]$ (solid orange) as functions of $\Delta_{da}$, all scaled by $\kappa$, for $g = 2\kappa$ (left panel) and $g=4\kappa$ (right panel). Here, $\gamma=\kappa$. (b) Plot of critical coupling strength $g$, scaled by $\kappa/2$, above which 0 SFP region emerges, as a function of $\Delta_{da}/\kappa$, for $\kappa = [0.5, 1, 2]\gamma$ (green, orange, and purple curves respectively).}
\label{fig:minG}
\end{figure}

Quite generally, the instability curve in $\betasss$-$\Delta_{db}$ space defined by Eq.~(\ref{quartic}) is like the green curve in Fig.~\ref{fig:poleFlowLC}~(a). In particular, the detuning at which the green curve first appears is where Eq.~(\ref{quartic}) has \textit{repeated} real, positive roots. Using analytic expressions for the roots of Eq.~(\ref{quartic}) (via Ferrari's method, for example), this condition can be checked for varying $\Delta_{db}$ to determine the minimal detuning below which poles with $\text{Im}~s\neq0$ can become unstable. We plot in solid green this minimal detuning $\DeltaLC[\Delta_{da}]$ in Fig.~\ref{fig:minG}~(a), for $g=2\kappa$ in the left panel and $g=4\kappa$ in the right, as a function of $\Delta_{da}$. Interestingly, unlike the $\Delta_{da} = 0$ case the minimal detuning now depends on the coupling strength $g$ as well. When $\Delta_{da} = 0$, both curves pass through $\DeltaLC = -\frac{\sqrt{3}}{2}(\gamma+\kappa)$ (here $\gamma=\kappa$), indicated by the dashed horizontal line. Away from $\Delta_{da}=0$, note the asymmetric behaviour of $\DeltaLC[\Delta_{da}]$: this ultimately stems from the dependence of $\chi_a$ on the sign of $\Delta_{da}$, which also manifests in $\widetilde{\Delta}_{db}[s]$ in Eq.~(\ref{sSol}). The same also leads to an asymmetry in $\DeltaMP[\Delta_{da}]$, which is plotted in solid orange in both panels. In the red shaded region, we have $\DeltaMP[\Delta_{da}] < \DeltaLC[\Delta_{da}]$; within this detuning range, the two-mode system has only one fixed point (since the detuning is above $\DeltaMP[\Delta_{da}]$), and for some region in $\betasss$-$\Delta_{db}$ space this fixed point becomes unstable, leading to a region with zero stable fixed points (0 SFPs). For increased $g$, this red shaded region is larger, indicating how stronger coupling should lead to an enlarged 0 SFP region, just as in the $\Delta_{da} = 0$ case. Note also that $\DeltaMP[\Delta_{da}]$ reaches a minimum for $\Delta_{da} = (\frac{\sqrt{3}}{2}-1 )\kappa$ (irrespective of $\gamma$), and here the width of the red shaded region is maximised.

Interestingly, the opposite case of $\DeltaLC[\Delta_{da}] < \DeltaMP[\Delta_{da}]$ is \textit{not} sufficient to guarantee that the 0 SFP region no longer exists. This is different from the $\Delta_{da} = 0$ case. There, $\DeltaMP > \DeltaLC$ only if $g$ was smaller than its critical value of $\kappa/2$; when this happened, poles with $\text{Im}~s \neq 0$ could no longer be unstable, and their instability curve in $\betasss$-$\Delta_{db}$ space no longer existed. For $\Delta_{da} \neq 0$, this is no longer the case; the instability region exists even when $\DeltaLC[\Delta_{da}] < \DeltaMP[\Delta_{da}]$. To then determine whether an analogous minimal coupling strength $g$ exists for $\Delta_{da} \neq 0$, we must find whether 0 SFP regions exist in the full phase diagram in $\eta$-$\Delta_{db}$ space for varying $g$. 

At this point, we find it easiest to proceed numerically, assessing stability in $\eta$-$\Delta_{db}$ space by finding the eigenvalues of the 4-by-4 system stability matrix given by Eqs.~(\ref{linEqs}). In Fig.~\ref{fig:minG}~(b), we plot the critical coupling strength above which a region with 0 SFPs does emerge, scaled by its value for $\Delta_{da} = 0$, which is indicated by the horizontal dashed line. The results are plotted as a function of $\Delta_{da}$ for varying values of $\kappa/\gamma$ (we do not impose $\gamma=\kappa$ for these numerical results); note that the behaviour for the different values of $\kappa$ appears very similar. We see again an asymmetric dependence on the linear mode detuning $\Delta_{da}$, with a much sharper increase in the critical $g$ for a blue-detuned drive on the linear mode. Curiously, a minimum is observed in the red-detuned case, for $\Delta_{da} \approx (\frac{\sqrt{3}}{2}-1)\kappa$, where $\DeltaMP[\Delta_{da}]$ is most negative.

\begin{figure}
\includegraphics[scale=0.6]{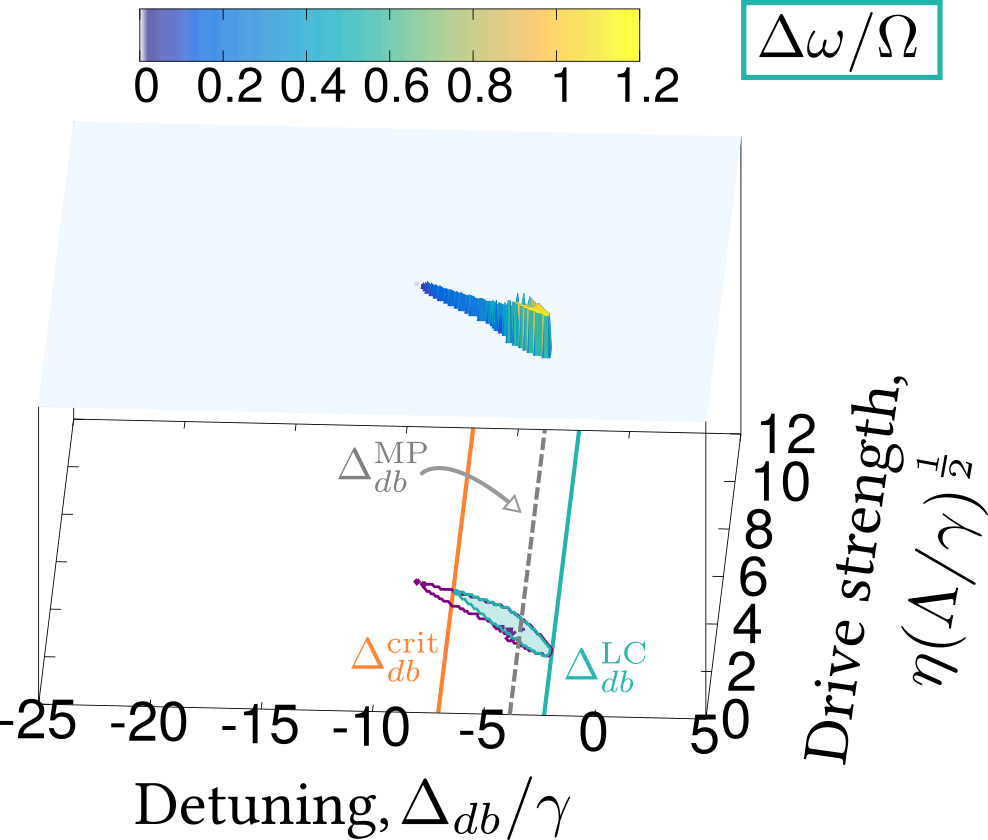}
\caption{Phase diagram in $\eta$-$\Delta_{db}$ space for $g = 2\gamma$, $\kappa = \gamma$, and blue-detuned drive on the linear mode, $\Delta_{da} = +\frac{1}{2}\gamma$. Purple curve indicates numerical boundary of instability, and green shaded region is the theoretical prediction for the 0 SFP region.}
\label{fig:Delta0.5}
\end{figure}

\begin{figure}
\includegraphics[scale=0.6]{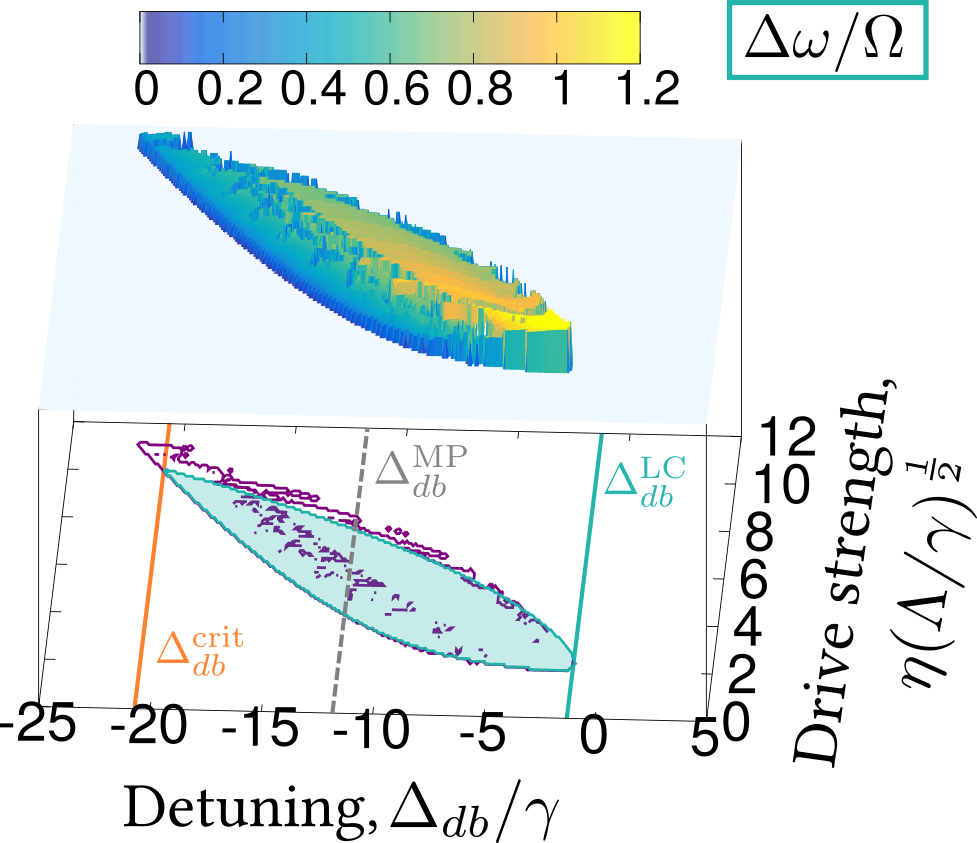}
\caption{Phase diagram in $\eta$-$\Delta_{db}$ space for $g = 2\gamma$, $\kappa = \gamma$, and red-detuned drive on the linear mode, $\Delta_{da} = -\frac{1}{2}\gamma$. Purple curve indicates numerical boundary of instability, and green shaded region is the theoretical prediction for the 0 SFP region.}
\label{fig:Delta-0.5}
\end{figure}

The analysis thus far indicates that above a critical coupling $g$ and below a critical detuning $\DeltaLC[\Delta_{da}]$, \textit{some} region in $\eta$-$\Delta_{db}$ space will have a 0 SFP phase, and that negative $\Delta_{da}$ is preferred for the existence of this phase. However, the size of the unstable region has not been explicitly determined yet. Here we again resort to a numerical determination of the 0 SFP phase in $\eta$-$\Delta_{db}$ space via the linear stability analysis, and compare it with full numerical simulations. In Fig.~\ref{fig:Delta0.5}, we plot the phase diagram for $g = 2\gamma ,\kappa = \gamma$, and $\Delta_{da} = +\frac{1}{2}\gamma$. Here, the green shaded region denotes the 0 SFP phase. For simplicity we still scale the frequency spacing $\Delta\omega$ by $\Omega = \sqrt{g^2-\kappa^2/4}$. A similar plot but now for $\Delta_{da} = -\frac{1}{2}\gamma$ is included in Fig.~\ref{fig:Delta-0.5}. The numerical results follow closely the stability analysis predictions. Clearly the instability region is much smaller for $\Delta_{da} > 0$. For larger $\Delta_{da}$ values (not shown), the instability area generally shrinks, regardless of sign, and requires larger coupling strength $g$ to emerge as well, as indicated in Fig.~\ref{fig:minG}~(b).

\section{Phase diagram in $\eta$-$\Delta_{da}$ space, and dispersive regime}
\label{app:disp}

Thus far we have studied phase diagrams as a function of varying $\Delta_{db}$, with fixed $\Delta_{da}$; experimentally, this amounts to varying $\omega_b$ for fixed $\omega_d$ and $\omega_a$ (in circuit QED, such protocols may be implemented using SQUID-based nonlinear elements). Alternatively one could fix $\omega_a$, $\omega_b$, while making a sweep of the drive frequency $\omega_d$. In this case, both $\Delta_{db}$ and $\Delta_{da}$ vary, and therefore the appearance of the phases discussed in the main text is more complex.

\begin{figure*}
\includegraphics[scale=0.35]{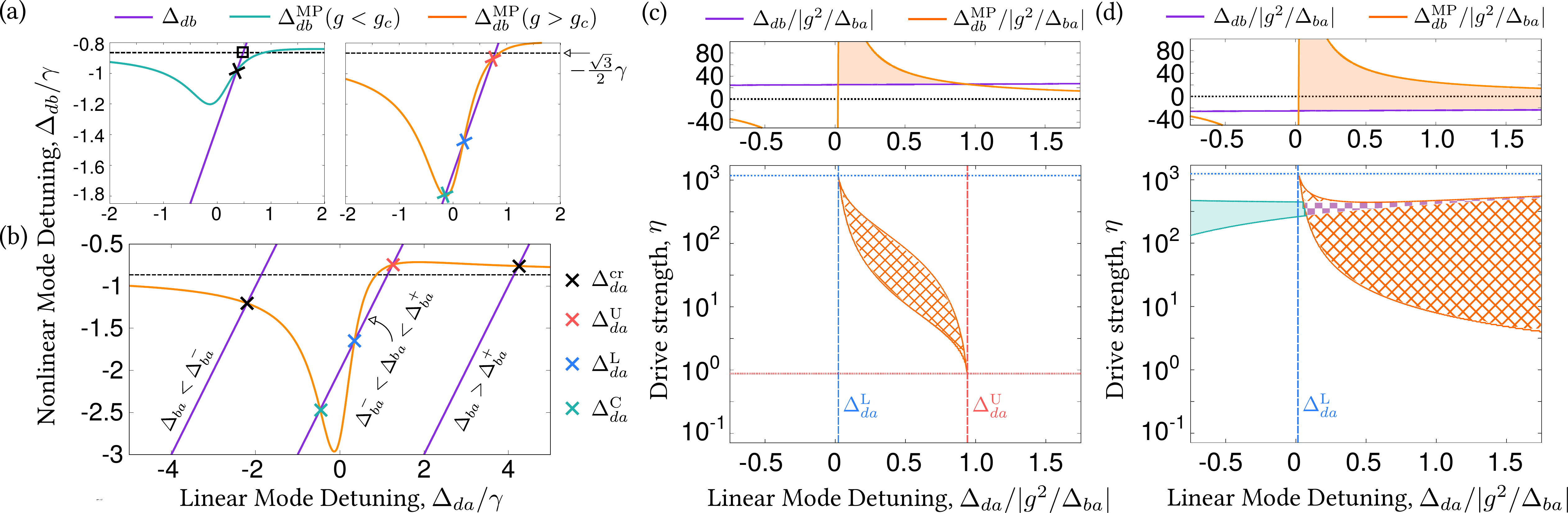}
\caption{(a) Plot of $\Delta_{db}$-line (purple) and $\DeltaMP$-curve (green, orange) as a function of $\Delta_{da}$, for $g < g_c$ in the left panel and $g>g_c$ in the right. (b) Same for $g>g_c$, for three different values of $\Delta_{ba}$. Intersections indicated by crosses are roots of Eq.~(\ref{cubicDeltaDA}); see text for details. (c) Phase diagram in $\eta$-$\Delta_{da}$ space for red-detuned nonlinear mode relative to the linear mode, $\Delta_{ba} = - 1000\kappa$. (d) Same for blue-detuned nonlinear mode, $\Delta_{ba} = + 1000\kappa$. Other parameters are $g = 200\kappa$, $\Lambda = 200\kappa$, and $\gamma = 0.01\kappa$, with $\kappa/(2\pi) = 1~$MHz; these are chosen for correspondence with Ref.~\cite{bishop_response_2010}. Phases are as defined in Fig.~2 of the main text: orange hatched region has 3 FPs, 2 SFPs, purple checkered region has 3 FPs, 1 SFP, green shaded region has 0 SFPs, and blank region has 1 FP, 1 SFP. Top panel in both plots shows $\Delta_{db}$-line (purple) and $\DeltaMP$-curve (orange) as a function of $\Delta_{da}$; the orange shaded region is where $\Delta_{db} < \DeltaMP$. }
\label{fig:disp}
\end{figure*}

We introduce the detuning between the linear and nonlinear modes, $\Delta_{ba} = \omega_b-\omega_a$. Then clearly $\Delta_{db} = \Delta_{da}-\Delta_{ba}$, so that $\Delta_{da}$ and $\Delta_{db}$ are related by a constant. The region with multiple fixed points  For this system to exhibit a region with multiple fixed points, we require $\Delta_{db} < \DeltaMP$, as discussed earlier. However, unlike the case where $\Delta_{da}$ is fixed, here as $\Delta_{da}$ varies so does the value of $\DeltaMP$ varies (as shown previously in Fig.~\ref{fig:minG}~(a)). The critical points where $\Delta_{db} = \DeltaMP$ satisfy:
\begin{align}
\Delta_{da}-\Delta_{ba} = -\frac{\sqrt{3}}{2} \left(\gamma+g^2\kappa|\chi_a|^2\right) + g^2\Delta_{da}|\chi_a|^2
\label{cubicDeltaDA}
\end{align}
which is a cubic polynomial in $\Delta_{da}$ for $g \neq 0$, due to the $\Delta_{da}$-dependence of $\chi_a$. A graphical approach allows a number of important conclusions to be made, without the need for explicitly solving Eq.~(\ref{cubicDeltaDA}). In Fig.~\ref{fig:disp}~(a), we plot both sides of Eq.~(\ref{cubicDeltaDA}) as a function of $\Delta_{da}$. The left hand side is always a straight line of unit slope with a $y$-axis translation determined by $\Delta_{ba}$; we label this the $\Delta_{db}$-line (purple). The right hand side is the $\DeltaMP$-curve. For $g=0$, this `curve' is simply a constant since $\DeltaMP = -\frac{\sqrt{3}}{2}\gamma$ (dashed black), the bare critical detuning for the single Kerr oscillator. The intersection, shown by the black square, defines $\Delta_{da}^{\rm cr}$ such that for $\Delta_{da} < \Delta_{da}^{\rm cr}$, we have $\Delta_{db} < \DeltaMP$, and a region with multiple fixed points exists.

As $g$ increases, the $\DeltaMP$-curve acquires some general features: a maximum/minimum at $(\frac{\sqrt{3}}{2}\pm 1)\kappa$, and an asymptotic approach to the bare value $-\frac{\sqrt{3}}{2}\gamma$ for $\Delta_{da} \gg \kappa$, as the effect of the linear mode decreases off-resonance. We find that for $g < g_c$, the modification is weak enough that the $\Delta_{db}$-line only intersects the $\DeltaMP$-curve once, as shown by the black cross in the left panel of Fig.~\ref{fig:disp}~(a); in this case the cubic equation in Eq.~(\ref{cubicDeltaDA}) has only one real root. The critical coupling is achieved when the $\DeltaMP$-curve has maximal derivative that at least equals the (unit) slope of the $\Delta_{db}$-line; imposing this constraint, we find the value of $g_c$ to be:
\begin{align}
\frac{g_c}{\kappa} = \frac{1-2\sqrt{3}\sin\frac{\pi}{9} + 4 \sin^2 \frac{\pi}{9} }{ \sqrt{1-4\sin^2\frac{\pi}{9}} } \approx 0.3881
\end{align}
For $g > g_c$, $\DeltaMP$ is modified strongly enough such that the $\Delta_{db}$-line may intersect the $\DeltaMP$-curve three times, as shown in the right panel of Fig.~\ref{fig:disp}~(a); thus Eq.~(\ref{cubicDeltaDA}) can admit three real roots. However, this is possible only for certain $\Delta_{ba}$. In Fig.~\ref{fig:disp}~(b), we plot $\Delta_{db}$-lines for three different values of $\Delta_{ba}$; recall that positive (negative) $\Delta_{ba}$ translates the $\Delta_{db}$-line down (up). Clearly, for any given set of system parameters, their exists a maximum $\Delta_{ba}^+$ and minimum $\Delta_{ba}^-$, such that beyond these values the $\Delta_{db}$-line only intersects the $\DeltaMP$-curve once, and again only one real root exists. The values $\Delta_{db}^{\pm}$ may be determined graphically, or by finding the zeros of the discriminant of Eq.~(\ref{cubicDeltaDA}) for a given set of a system parameters. Ultimately this range is also determined by the size of the coupling: larger $g$ allows for a larger range of $\Delta_{ba}$ values.

If both $g>g_c$ and $\Delta_{db}^- < \Delta_{db} < \Delta_{db}^+$, the cubic equation admits three real roots $\Delta_{da}^{\rm C,L,U}$, labelled such that $\Delta_{da}^{\rm C} < \Delta_{da}^{\rm L} < \Delta_{da}^{\rm U}$. From Fig.~\ref{fig:disp}~(b), it is clear that there are now \textit{two} regions with multiple fixed points, since $\Delta_{db}<\DeltaMP$ for both $\Delta_{da} < \Delta_{da}^{\rm C}$ \textit{and} $\Delta_{da}^{\rm L} < \Delta_{da} < \Delta_{da}^{\rm U}$. The second, bounded region emerges entirely due to the linear mode's modification of $\DeltaMP$, and hence disappears if the coupling $g$ is too weak. 

Determining the stability of the fixed points analytically is in general a difficult task, again due to the concurrent variation of $\Delta_{db}$ and $\Delta_{da}$. Numerically, however, analyzing stability via Eqs.~(\ref{linEqs}) is still rather straightforward; we now proceed with this analysis. To make connection with a concrete physical system we study a regime analyzed recently in Ref.~\cite{bishop_response_2010} for a nonlinear transmon coupled to a cavity; unlike the present work, the nonlinear element there is modeled as a two-level system. The nonlinear mode is strongly red-detuned relative to the linear mode, $\Delta_{ba} = -1000\kappa$. We also take $g = 200\kappa$, $\Lambda = 200\kappa$, and $\gamma = 0.01\kappa$, with $\kappa/(2\pi) = 1~$MHz; the resulting regime where $|\Delta_{ba}| \gg \frac{g^2}{|\Delta_{ba}|} \gg g \gg \kappa \gg \gamma$ is the strong-dispersive, bad-cavity regime of circuit QED. Fig.~\ref{fig:disp}~(c) shows the phase diagram computed by finding the eigenvalues of the system stability matrix given by Eqs.~(\ref{linEqs}). It clearly indicates a phase with 3 FPs, 2 SFPs (orange hatched region) existing for $\Delta_{da}^{\rm L} < \Delta_{da} < \Delta_{da}^{\rm U}$. The top panel plots the $\Delta_{db}$ line (purple) and $\DeltaMP$ curve (orange); in the orange shaded region $\Delta_{db} < \DeltaMP$, and hence multiple fixed points exist. This region precisely overlaps with the orange hatched region of the phase diagram, as it must. A second region with multiple fixed points exists for $\Delta_{da} < \Delta_{da}^{\rm C}$; here $\Delta_{da}^{\rm C} \approx -25.96\frac{g^2}{|\Delta_{ba}|}$, and is hence outside the plotted region. A very similar phase diagram was uncovered in Ref.~\cite{bishop_response_2010}; hence it appears that the emergence of a bounded region with multiple fixed points is not due to the specific model of the nonlinearity, but rather a consequence of how the properties of the nonlinear mode are affected by coupling to the linear mode.

Note that Fig.~\ref{fig:disp}~(a) does not show the limit cycle phase, where the system has no stable fixed points. This has to do with the particular parameter choice; in the plotted region, $\Delta_{db}$ is \textit{positive}, as shown by the purple line in the top panel. In particular, for $\Delta_{da} = 0$ this clearly puts the system above $\DeltaLC = -\frac{\sqrt{3}}{2}(\gamma+\kappa)$, which is negative. Hence no limit cycle phase can emerge at $\Delta_{da} = 0$. For general $\Delta_{da}$, the determination is more complicated since the critical detuning and minimal coupling are $\Delta_{da}$-dependent; the numerical stability diagram in Fig.~\ref{fig:disp}~(c) takes these into subtleties into account, finding no unstable phase in the plotted region.

Interestingly, a modification in operating conditions allows the emergence of the unstable phase. The strong-dispersive regime may also be reached by detuning the nonlinear mode \textit{above} the linear mode in frequency ~\cite{fitzpatrick_observation_2017}. We thus choose $\Delta_{ba} = 1000\kappa > 0$ instead; all other parameters are unchanged. The resulting phase diagram is plotted in Fig.~\ref{fig:disp}~(d). Regions with multiple fixed points still exists for $\Delta_{da}^{\rm L} < \Delta_{da} < \Delta_{da}^{\rm U}$ and $\Delta_{da} < \Delta_{da}^{\rm C}$; both $\Delta_{da}^{\rm U} \approx 25.96 \frac{g^2}{|\Delta_{ba}|}$ and $\Delta_{da}^{\rm C} \approx -0.98 \frac{g^2}{|\Delta_{ba}|}$ are outside the plotting range. The qualitative features of this region remain the same as in Fig.~\ref{fig:disp}~(c).

Now, however, the stability analysis uncovers a phase with no stable fixed points, in shaded green. At $\Delta_{da} = 0$, the nonlinear mode detuning $\Delta_{db} = \Delta_{da}-\Delta_{ba}$ is now negative (purple line in the top panel) and below $\DeltaLC$, allowing such a phase to emerge. In the region with multiple fixed points, in addition to the 3 FP, 2 SFP phase we also note the emergence of the checkered purple region where only one of the three fixed points is stable. Hence even for otherwise similar dispersive regimes, the sign of $\Delta_{ba}$ can lead to dramatically different phase diagrams. Ultimately, for $|\Delta_{da}| \lesssim |\Delta_{ba}|$, the sign of $\Delta_{ba}$ determines whether the nonlinear mode is driven effectively red- or blue-detuned ($\Delta_{db}<0$ or $>0$ respectively), with red-detuned driving generally leading to unstable phases. 

Finally, we emphasize that while the results in Ref.~\cite{bishop_response_2010} use a dispersive regime model of a nonlinear transmon coupled to a cavity, and make a bad-cavity approximation, the results presented here are not limited to such regimes. In particular, analytic results for the region with multiple fixed points and the dependence of unstable phases on $\Delta_{ba}$ hold in general.

\section{Positive $P$ Representation and Stochastic Differential Equations}
\label{app:corrFuncs}

The positive-$P$ representation is defined as a projection of the density matrix $\rho$ in a \textit{non-diagonal} basis of coherent states~\cite{carmichael_statistical_2002}:
\begin{align}
\rho = \int d\vec{\alpha}~P(\vec{\alpha})\zeta(\vec{\alpha}) \equiv \int d^2\vec{\alpha}~P(\vec{\alpha}) \frac{ \ket{\alpha}\bra{ \alpha^{\dagger *}} }{e^{\alpha\alpha^{\dagger}}} \otimes \frac{ \ket{\beta}\bra{\beta^{\dagger *}} }{e^{\beta\beta^{\dagger}}}
\end{align}
where $\vec{\alpha} = \{\alpha,\alpha^{\dagger},\beta,\beta^{\dagger} \}$, as defined in the main text. The dynamical equation for the positive-$P$ representation is obtained by applying the Liouvillian $\mathcal{L}$ describing the Master equation on this representation of $\rho$; in particular, all we need is the action on the operator $\zeta$, which requires the results:
\begin{align}
\hat{a}\zeta_{\alpha} &= \alpha\zeta_{\alpha}~,~\zeta_{\alpha} \hat{a} = \left(\partial_{\alpha^{\dagger}} + \alpha\right)\zeta_{\alpha} \nonumber \\
\zeta \hat{a}^{\dagger} &= \zeta \alpha^{\dagger}~,~\hat{a}^{\dagger}\zeta_{\alpha} = \left(\partial_{\alpha} + \alpha^{\dagger} \right) \zeta_{\alpha}
\end{align}
Analogous expressions exist for the nonlinear mode operators, with the replacement $\hat{a} \to \hat{b}$. With these results, it is straightforward to convert the Master equation for $\rho$ in the main text term by term to its phase-space form. A standard integration-by-parts procedure then yields the Fokker-Planck equation for the positive-$P$ representation:
\begin{align}
\partial_t P(\vec{\alpha},t) = \left( -\partial_i A_i + \partial_i\partial_j \frac{1}{2} D_{ij} \right)P(\vec{\alpha},t)
\label{fpe}
\end{align}
$A_i$ are the components of the drift vector $\vec{A}$ and $D_{ij}$ are the components of the diffusion matrix $\bm{D}$.  In the above, $\partial_i = \frac{\partial}{\partial \alpha_i}$ where $\alpha_i \in [\alpha,\alpha^{\dagger},\beta,\beta^{\dagger}]$, and repeated indices are summed over. The drift vector $\vec{A}$ describing classical dynamics is given by:
\begin{align}
\vec{A} = 
\begin{pmatrix}
i\Delta_{da} \alpha -i\eta -ig\beta -\alpha \kappa/2 \\
-i\Delta_{da} \alpha^{\dagger} + i\eta + ig\beta^{\dagger} -\alpha^{\dagger} \kappa/2 \\
i\Delta_{db} \beta -ig\alpha + i \Lambda\beta^{\dagger}\beta^2 - \beta\gamma/2 \\
-i\Delta_{db} \beta^{\dagger} +ig\alpha^{\dagger} - i \Lambda \beta(\beta^{\dagger})^2 - \beta^{\dagger}\gamma/2 
\end{pmatrix}
\label{vecA}
\end{align}
while the diffusion matrix takes the form:
\begin{align}
\bm{D} = 
\Lambda
\begin{pmatrix}
0 & 0 & 0 & 0 \\
0 & 0 & 0 & 0 \\
0 & 0 & i\beta^2 & 0 \\
0 & 0 & 0 & -i(\beta^{\dagger})^2 
\end{pmatrix}
\label{matD}
\end{align}

The above Fokker-Planck equation can be mapped to the set of SDEs below:
\begin{align}
d\vec{\alpha} = \vec{A}~dt+ \sqrt{\Lambda}~\bm{B}~\vec{dW}
\label{sdesApp}
\end{align}
In the stochastic term, each component of $\vec{dW}$ is a Wiener increment sampled from \textit{independent} normal distributions $\mathcal{N}(0,1)$. We also define the noise matrix $\bm{B}$, introduced in the main text, via $\bm{D} = \Lambda \bm{B}^{T}\bm{B}$, extracting the explicit dependence on the nonlinearity. It takes the form $\bm{B} = \sqrt{i}~{\rm diag}(0,0,\beta,i\beta^{\dagger})$.

We now scale Eqs.~(\ref{sdesApp}) as discussed in the main text; $\Lambda \to \Lambda/c$ and $(\vec{\alpha},\eta) \to \sqrt{c}(\vec{\alpha},\eta)$. Under this operation, it is clear that $d\vec{\alpha} \to \sqrt{c}~d\vec{\alpha}$, $\vec{A} \to \sqrt{c} \vec{A}$, and $\bm{B} \to \sqrt{c}\bm{B}$. Thus, the SDEs after scaling take the form:
\begin{align}
\sqrt{c}~d\vec{\alpha} &= \sqrt{c}\vec{A}~dt + \sqrt{\frac{\Lambda}{c}}~\sqrt{c}\bm{B}~\vec{dW} \nonumber \\
\implies d\vec{\alpha} &= \vec{A}~dt + \frac{1}{\sqrt{c}} \sqrt{\Lambda}~\bm{B}~\vec{dW} 
\end{align}
Hence we see that while the scaling procedure leaves the deterministic term, defined by the drift vector $\vec{A}$ unchanged, it does scale the stochastic terms that determine quantum effects. In particular, for decreasing $\Lambda$ ($c>1$), the magnitude of the stochastic terms is effectively suppressed.

\section{Linearized spectrum in stable regime}
\label{app:linSpec}

In the regime where classical static fixed points $\dot{\alpha} = \dot{\beta} = 0$ are stable, it is possible to consider a regime of small fluctuations around these classical fixed points. A linearization in these small fluctuations yields a linear model, for which the spectrum of fluctuations can be determined exactly. To do so, we must first determine the linearized Master equation that governs the dynamics of small fluctuations around these classical fixed points. The approach we employ here is to displace the Master equation by the classical solution $(\alphass,\betass)$. We displace the quantum operators by fluctuation operators $(\delta \hat{a},\delta\hat{b})$ as follows:
\begin{align}
\hat{a} = \alphass + \delta \hat{a}~,~\hat{b} = \betass + \delta \hat{b} 
\label{dispModes}
\end{align}
such that $\avg{\delta \hat{a} } = \avg{\delta \hat{b} } = 0$. It is now possible to `linearize' the system Master equation in these fluctuation operators, obtaining a Master equation for the displaced density matrix $\delta \hat{\rho}$. To `linearize' here has a specific meaning: it requires keeping up to \textit{quadratic} terms in the fluctuation operators, which yield linear contributions in the equations of motion. Terms that are linear in the fluctuation operators recover exactly the classical equations of motion, and identically cancel since $(\alphass,\betass)$ satisfy these equations. Making the substitution, we find the effective Master equation:
\begin{align}
\delta\hat{\dot{\rho}} = -i\left[\delta\hat{\mathcal{H}},\delta\hat{\rho} \right] + \kappa \mathcal{D}[\delta \hat{a}] \delta\hat{\rho} + \gamma \mathcal{D}[\delta \hat{b}] \delta\hat{\rho}
\end{align}
where $\delta\hat{\mathcal{H}}$ is the effective quadratic Hamiltonian:
\begin{align}
\delta\hat{\mathcal{H}} = \delta\hat{\mathcal{H}}_a + \delta\hat{\mathcal{H}}_b + \delta\hat{\mathcal{H}}_g
\end{align}
with individual terms defined as:
\begin{align}
\delta\hat{\mathcal{H}}_a &= -\Delta_{da}\delta \hat{a}^{\dagger}\delta \hat{a} \nonumber \\
\delta\hat{\mathcal{H}}_b &= -\Delta_{db}\delta \hat{b}^{\dagger}\delta \hat{b} -\frac{\Lambda}{2}\!\left( 4\betasss\delta \hat{b}^{\dagger}\delta \hat{b} + (\betass^*)^2 \delta \hat{b} \delta \hat{b} + \betass^2 \delta \hat{b}^{\dagger} \delta \hat{b}^{\dagger} \right) \nonumber \\
\delta\hat{\mathcal{H}}_g &= g\left(\delta \hat{a}^{\dagger}\delta \hat{b} + \delta \hat{a} \delta  \hat{b}^{\dagger}\right)
\end{align}
The linear terms corresponding to the linear mode Hamiltonian and the coupling Hamiltonian are unchanged, as expected, while the driving term contributing to the classical fixed points is removed by the displacement operation. The only other change comes to the Kerr term, which takes a linearized form, yielding squeezing terms.

The above Master equation can be mapped to a linearized Fokker-Planck equation for the positive-$P$ distribution \textit{of the fluctuations}, which takes the form:
\begin{align}
\partial_t P(\delta\vec{\alpha},t) = \left( -\partial_i \delta A_i + \partial_i \partial_j \frac{1}{2}\delta D_{ij}\right)P(\delta \vec{\alpha} ,t),
\label{linFPE}
\end{align}
where $\delta \vec{\alpha} \equiv (\delta \alpha, \delta \alpha^{\dagger}, \delta \beta, \delta \beta^{\dagger} )$ and $\partial_i$ is defined as before. The new drift vector $\delta\vec{A}$ and diffusion matrix $\bm{\delta D}$ can be derived exactly as described in the previous section, and these are now dependent on the steady state solution. The Fokker-Planck equation is now linear in the fluctuation variables $\delta \vec{\alpha}$. For convenience, we write the drift vector components $\delta A_i$ in terms of an equivalent drift matrix $\bm{\delta A}$, such that $\delta\vec{A} = \bm{\delta A}\cdot \delta \vec{\alpha}$. The drift matrix is defined as:
\begin{align}
\bm{\delta A} = 
\begin{pmatrix}
\mathbf{C}_a & \mathbf{C}_g \\
\mathbf{C}_g & \mathbf{C}_b
\end{pmatrix}
\end{align}
where we have employed block matrix notation; in the above and for what follows, boldface symbols represent 2-by-2 matrices and italics boldface symbols represent 4-by-4 matrices. The 2-by-2 matrices $\mathbf{C}_a$, $\mathbf{C}_b$, and $\mathbf{C}_g$ take the respective forms:
\begin{align}
\mathbf{C}_a &= 
\begin{pmatrix}
i\Delta_{da} - \frac{\kappa}{2}  & 0 \\
0 & -i\Delta_{da} -\frac{\kappa}{2}
\end{pmatrix} \nonumber \\
\mathbf{C}_b &= 
\begin{pmatrix}
i\Delta_{db} - \frac{\gamma}{2} + i2\Lambda\betasss & i\Lambda\betass^2  \\
-i\Lambda(\betass^*)^2 & -i\Delta_{db} - \frac{\gamma}{2} - i2\Lambda\betasss 
\end{pmatrix} \nonumber \\
\mathbf{C}_g &= 
\begin{pmatrix}
-ig & 0 \\
0 & ig 
\end{pmatrix} 
\end{align}
Finally, the diffusion matrix is given by:
\begin{align}
\bm{\delta D} =
\Lambda 
\begin{pmatrix}
\mathbf{0} & \mathbf{0} \\
\mathbf{0} & \mathbf{Z}
\end{pmatrix},~ \mathbf{Z} = 
\begin{pmatrix}
i\betass^2 & 0 \\
0 & -i(\betass^*)^2
\end{pmatrix}
\end{align}
where we have also written $\bm{\delta D}$ in block matrix form, with $\mathbf{0}$ being the 2-by-2 matrix of zeroes. Note that $\bm{\delta D}$ is just the full diffusion matrix $\bm{D}$ evaluated at the stable fixed point. 

For a linear Fokker-Planck equation of the form of Eq.~(\ref{linFPE}), the steady state spectrum of fluctuations around a stable fixed point may be analytically determined using the expression~\cite{chaturvedi_stochastic_1977, drummond_quantum_1980, carmichael_statistical_2002}:
\begin{align}
\bm{\bar{T}}[\omega] = \left( \bm{\delta A} -i\omega \bm{I} \right)^{-1} \bm{\delta D}  \left( \bm{\delta A}^T + i\omega \bm{I} \right)^{-1}
\label{linSpec}
\end{align} 
where the matrix elements $\bar{T}_{ij}[\omega] = \mathcal{F}\left\{\avg{ \delta \alpha_i(\tau)\delta \alpha_j(0) }\right\}$ define the spectrum via the Wiener-Khinchin theorem ($\mathcal{F}\left\{\cdot\right\}$ being the Fourier transform), for $\delta \alpha_i \in \left\{\delta \alpha, \delta \alpha^{\dagger}, \delta \beta, \delta \beta^{\dagger} \right\}$ for $i = 1,\ldots 4$. Therefore, the linear mode fluctuation spectrum is given by ${\bar{T}}_{21}[\omega]$, while the nonlinear mode fluctuation spectrum is given by ${\bar{T}}_{43}[\omega]$. The sparsity of the diffusion matrix $\bm{\delta D}$ allows the spectra to be computed relatively simply. In particular, the matrix of spectra, $\bm{\bar{T}}[\omega]$ takes a block matrix form:
\begin{align}
\bm{\bar{T}}[\omega] = 
\begin{pmatrix}
\mathbf{t}_{11} & \mathbf{t}_{12} \\
\mathbf{t}_{21} & \mathbf{t}_{22} 
\end{pmatrix}
\end{align}
where the $\mathbf{t}_{ij}$ are 2-by-2 matrices. We are interested only in the matrices lying on the diagonals of $\bm{\bar{T}}[\omega]$, namely $\mathbf{t}_{11}$ and $\mathbf{t}_{22}$, since these will contain the matrix elements ${\bar{T}}_{21}[\omega]$ and ${\bar{T}}_{43}[\omega]$ respectively. By virtue of the sparse form that $\bm{\delta D}$ takes, together with useful expressions for matrix inversion in block form, we find from Eq.~(\ref{linSpec}) that $\mathbf{t}_{22}$ can be efficiently expressed as:
\begin{align}
\mathbf{t}_{22} = \mathbf{G}_{-}^{-1}[\omega]~\mathbf{Z}~\mathbf{G}_{+}^{-1}[\omega]
\label{t22}
\end{align}
where the matrices $\mathbf{G}_{\pm}[\omega]$ are defined as:
\begin{widetext}
\begin{align}
\mathbf{G}_{+}[\omega] &= \left(\mathbf{C}_b^{T} + i\omega \mathbf{I}\right) - \mathbf{C}_g \left( \mathbf{C}_a^T + i\omega\mathbf{I} \right)^{-1}\mathbf{C}_g   = \begin{pmatrix}
i\omega + i\widetilde{\Delta}_{db}[-\omega] -\frac{\widetilde{\gamma}[-\omega]}{2} + i2\Lambda\betasss  & - i\Lambda(\betass^*)^2  \\
 i\Lambda\betass^2 &  i\omega - i\widetilde{\Delta}_{db}[-\omega] -\frac{\widetilde{\gamma}[-\omega] }{2} - i2\Lambda\betasss 
\end{pmatrix}  \nonumber \\
\mathbf{G}_{-}[\omega] &= \left(\mathbf{C}_b - i\omega \mathbf{I}\right) - \mathbf{C}_g \left( \mathbf{C}_a - i\omega\mathbf{I} \right)^{-1}\mathbf{C}_g  ~~= \begin{pmatrix}
-i\omega + i\widetilde{\Delta}_{db}[\omega] -\frac{\widetilde{\gamma}[\omega]}{2} + i2\Lambda\betasss  &  i\Lambda\betass^2  \\
- i\Lambda(\betass^*)^2 & - i\omega - i\widetilde{\Delta}_{db}[\omega] -\frac{\widetilde{\gamma}[\omega] }{2} - i2\Lambda\betasss 
\end{pmatrix}
\end{align}
\end{widetext}
The expressions $\widetilde{\Delta}_{db}[\pm\omega]$, $\widetilde{\gamma}[\pm\omega]$ are exactly as defined in Eq.~(\ref{sParams}), but with $s \to \pm i\omega$. Now, the bottom left-matrix element of $\mathbf{t}_{22}$ yields the nonlinear mode spectrum ${\bar{T}}_{43}[\omega]$; by multiplying out Eq.~(\ref{t22}), this is found to be:
\begin{align}
{\bar{T}}_{43}[\omega] &= \frac{ \Lambda^2 |\betass|^4 }{|\mathcal{X}[\omega]|^2} \left(\gamma + \frac{g^2 \kappa}{(\omega+\Delta_{da})^2+\frac{\kappa^2}{4} } \right) 
\end{align}
where $\mathcal{X}[\omega] \equiv {\rm det}~\mathbf{G}_{-}[\omega] = {\rm det}~\mathbf{G}_{+}[-\omega]$ takes the form:
\begin{align}
\mathcal{X}[\omega] = \Bigg[ &\left(i\omega - i\widetilde{\Delta}_{db}[\omega]-2i\Lambda \betasss + \frac{\widetilde{\gamma}[\omega]}{2} \right) \times \nonumber \\
	    &\left(i\omega + i\widetilde{\Delta}_{db}[\omega]+2i\Lambda\betasss + \frac{\widetilde{\gamma}[\omega]}{2} \right) -\Lambda^2|\bar{\beta}|^4 \Bigg] 
\end{align}
Similarly, we find that $\mathbf{t}_{11}$ can be easily related to $\mathbf{t}_{22}$:
\begin{align}
\mathbf{t}_{11} =  \left( \mathbf{C}_a - i\omega\mathbf{I} \right)^{-1}\mathbf{C}_g \left(\mathbf{t}_{22} \right)\mathbf{C}_g \left( \mathbf{C}_a + i\omega\mathbf{I} \right)^{-1} 
\label{t11}
\end{align}
Again, the bottom-left matrix element can be easily found, yielding the linear mode fluctuation spectrum $\bar{T}_{21}[\omega]$:
\begin{align}
{\bar{T}}_{21}[\omega] &= \left[ \frac{g^2}{ (\omega+\Delta_{da})^2 + \frac{\kappa^2}{4} } \right] {\bar{T}}_{43}[\omega]
\label{T21}
\end{align}
Note that if $g \to 0$, ${\bar{T}}_{43}[\omega]$ reduces to the zero temperature spectrum of a single Kerr oscillator~\cite{drummond_quantum_1980}, and if $\Lambda \to 0$, the entire spectrum vanishes, as expected. In general, the spectrum has a complicated multi-peak structure depending on the detunings and driving strength. Lastly, we note that the same expressions may be obtained a little more transparently within a linearized Heisenberg-Langevin approach, where both modes experience input vacuum noise from the zero temperature baths they are coupled to.

Having obtained the steady state spectrum of \textit{fluctuations} around the classical fixed point, we may finally compute the full spectrum of the linear mode, which is plotted in Fig. 4 of the main text. Under the displacement operation [defined in Eq.~(\ref{dispModes})], $S_a[\omega]$ takes the form:
\begin{align}
S_a[\omega] &= \int_{-\infty}^{\infty} d\tau~e^{-i\omega \tau}\avg{\hat{a}^{\dagger}(\tau)\hat{a}(0) } \nonumber \\
            &= \int_{-\infty}^{\infty} d\tau~e^{-i\omega \tau}|\bar{\alpha}|^2 + \int_{-\infty}^{\infty} d\tau~e^{-i\omega \tau}\avg{\delta\hat{a}^{\dagger}(\tau)\delta\hat{a}(0)} \nonumber \\
            &= |\bar{\alpha}|^2\delta[\omega] + \bar{T}_{21}[\omega]
\end{align}
where we have used $\avg{\delta\hat{a}} = \avg{\delta\hat{a}^{\dagger}} = 0$ in the steady state. $S_a[\omega]$ in this linearized case has an interesting feature: note that the spectrum of fluctuations $\bar{T}_{21}[\omega]$, Eq.~(\ref{T21}), depends on $\Lambda$ and $\betasss$ only via the product form $\Lambda\betasss$. Thus the aforementioned scaling procedure $\Lambda \to \Lambda/c$, $(\bar{\alpha},\bar{\beta}) \to \sqrt{c}(\bar{\alpha},\bar{\beta})$ leaves $\bar{T}_{21}[\omega]$ unchanged. However, this procedure \textit{does} scale the classical occupations, $(|\bar{\alpha}|^2,\betasss) \to c(|\bar{\alpha}|^2,\betasss)$. Hence for decreasing $\Lambda$ ($c>1$), the magnitude of quantum fluctuations remains unchanged (in the linearized case), but the mode occupation, is larger; thus for weaker $\Lambda$ fluctuations are smaller relative to the classical occupation, making linearization a better approximation. Conversely, as $\Lambda$ increases fluctuations become more important and linearization ultimately breaks down. We emphasize that the same feature is present in the linearized spectrum of fluctuations of the single coherently driven Kerr oscillator~\cite{drummond_quantum_1980}.

\section{Phase diffusion in quantum limit cycle regime}
\label{app:analyticDiff}

In this section, we briefly discuss dynamics in the limit cycle region in the weak noise regime, closely following Ref.~\cite{louca_stable_2015}. In the absence of noise, the governing SDEs [Eqs.~(4) of the main text, and equivalently Eqs.~(\ref{sdesApp}) in the Appendix] become ODEs (ordinary differential equations), describing classical dynamics. Unlike the previous Appendix section, in the limit cycle regime the classical fixed points $\dot{\bar{\alpha}} = \dot{\bar{\beta}} = 0$ are no longer stable, and linearizing around these points is no longer valid. To proceed, we note that a new set of stable solutions emerges in this regime instead, namely limit cycles that follow a fixed periodic orbit (dimension 1) embedded in an $n$-dimensional phase space spanned by the complex variables $\{\alpha,\alpha^{\dagger},\beta,\beta^{\dagger}\}$ (here, $n = 8$). We parametrize the periodic orbit defining the deterministic limit cycle as the vector curve $\vec{\mathcal{P}}(t)$, such that,
\begin{align}
\frac{d}{dt}\vec{\mathcal{P}}(t) = \vec{A}[\vec{\mathcal{P}}(t)]
\label{detLC}
\end{align}
which is just Eq.~(\ref{sdesApp}) without the stochastic term. Here $\vec{\mathcal{P}}(t+\frac{2\pi}{\Delta\omega}) = \vec{\mathcal{P}}(t)$, where $\Delta\omega$ is the frequency spacing mentioned in the main text. 

Once the stochastic terms in Eqs.~(\ref{sdesApp}) are included, one may ask what happens in a weak noise limit, where the sample paths that are solutions to Eqs.~(\ref{sdesApp}) closely follow the deterministic limit cycle trajectory $\mathcal{P}(t)$. Noise can perturb the deterministic limit cycle by pushing sample paths off the fixed orbit, i.e. in the $(n$-$1)$-dimensional hyperplane perpendicular to the limit cycle trajectory. However, noise can also cause perturbations parallel to the limit cycle; sample paths perturbed in this way would still follow the deterministic trajectory, but would correspond to a different time parameter.

Given this intuition, it proves useful to write sample paths of Eqs.~(\ref{sdesApp}) in terms of a part parallel to the deterministic limit cycle trajectory, and a part perpendicular to the trajectory:
\begin{align}
\vec{\alpha}(t) = \vec{\alpha}_{\parallel}(\tau(t)) + \vec{\alpha}_{\perp}(\tau(t)) = \vec{\mathcal{P}}(\tau(t)) + \vec{\alpha}_{\perp}(\tau(t))
\label{sampleDecomp}
\end{align}
Here, $\tau(t)$ is an effective time parameter; in the absence of noise, $\tau \to t$ and $\vec{\alpha}_{\perp} \to 0$, which returns us to the deterministic case. In the presence of noise, the above decomposition connects a sample path at any time $t$ to a point on the deterministic limit cycle, which is determined by $\tau(t)$. The choice of $\tau(t)$ further defines a tangent vector $\hat{T}(\tau(t))$ and a normal hyperplane $\Pi(\tau(t)) \perp \hat{T}(\tau(t))$, which together comprise the co-moving Frenet frame of the deterministic limit cycle trajectory. We make the decomposition unique by requiring that the deviation $\vec{\alpha}_{\perp}(\tau(t))$ of the sample paths from the deterministic limit cycle lie in this normal hyperplane $\Pi(\tau(t))$ at any time $t$. 

The dynamics parallel to the deterministic limit cycle are entirely determined by the dynamics of $\tau(t)$. On the other hand, deviations perpendicular to the trajectory evolve in two ways: first, as $\tau(t)$ advances, the plane $\Pi(\tau(t))$ in which they lie is moving \textit{along} the trajectory. Secondly, the deviations themselves can move within the plane $\Pi(\tau(t))$; these dynamics are \textit{normal} to the deterministic limit cycle. To make this connection clear, we define a deviation coordinate $\vec{z}(t)$ on a \textit{fixed} plane normal to a given (initial) point on the deterministic limit cycle trajectory; in particular $\vec{z}(t) \in \Pi(\tau(t=0))$. Then, with $t$, $\vec{z}(t)$ remains in $\Pi(\tau(t=0))$, but can move on this plane. The movement of the plane itself happens \textit{along} the limit cycle trajectory, and is governed by a transformation matrix $\bm{U}(\tau(t))$. Such a matrix can be shown to exist under quite general conditions~\cite{louca_stable_2015}. Thus, for the sample path dynamics perpendicular to the deterministic limit cycle, we can make the ans\"atz:
\begin{align}
\vec{\alpha}_{\perp}(\tau(t)) = \bm{U}(\tau(t))\vec{z}(t),~\vec{z}(t) \in \Pi(\tau(t=0))
\label{ansatzAperp}
\end{align}

Our goal is now to determine more quantitatively how the introduction of noise affects the dynamics of $\tau(t)$ and $\vec{\alpha}_{\perp}(\tau(t))$. In what follows, for clarity we suppress the $t$-dependence of $\tau$. Using the decomposition of sample paths written in Eq.~(\ref{sampleDecomp}), the dynamics are formally given by:
\begin{align}
\frac{d}{dt}\vec{\alpha} = \frac{d\tau}{dt} \frac{d}{d\tau} \vec{\mathcal{P}}(\tau) + \frac{d}{dt} \vec{\alpha}_{\perp}(\tau) 
\end{align}
Using Eq.~(\ref{detLC}) and the ans\"atz for Eq.~(\ref{ansatzAperp}), we may rewrite the above as:
\begin{align}
\frac{d}{dt}\vec{\alpha} = \frac{d\tau}{dt} \vec{A}[\vec{\mathcal{P}}(\tau)] + \frac{d\tau}{dt} \frac{d\bm{U}(\tau)}{d\tau} \vec{z}(t) + \bm{U}(\tau) \frac{d}{dt} \vec{z}(t)  
\label{evolZ}
\end{align}
To now introduce noise, we consider the linearized dynamics as governed by Eq.~(\ref{sdesApp}), parallel and perpendicular to the limit cycle trajectory; this yields:
\begin{align}
\frac{d}{dt} \vec{\alpha} &= \vec{A}[ \vec{\mathcal{P}}(\tau) + \vec{\alpha}_{\perp}(\tau) ] + \sqrt{\Lambda}\bm{B}[ \vec{\mathcal{P}}(\tau) + \vec{\alpha}_{\perp}(\tau) ] \frac{\vec{dW}}{dt} \nonumber \\
&\approx \vec{A} [ \vec{\mathcal{P}}(\tau) ] + \bm{J}_A[\vec{\mathcal{P}}(\tau)] \vec{\alpha}_{\perp}(\tau) + \sqrt{\Lambda}\bm{B}[\vec{\mathcal{P}}(\tau)] \frac{\vec{dW}}{dt}
\end{align}
For brevity, we suppress the $t$ dependence of $\tau$. In the second line we perform the linearization, keeping only terms to first order in the noise; since the deviation $\vec{\alpha}_{\perp}$ is already first order in noise, we neglect any terms $O(||\vec{\alpha}_{\perp}||^2,||\vec{\alpha}_{\perp} \cdot \vec{dW}||)$. $\bm{J}_A[\vec{\mathcal{P}}(\tau)]$ is the Jacobian matrix of $\vec{A}$, evaluated on the deterministic limit cycle trajectory. The diffusion matrix $\bm{B}$ is also evaluated on this trajectory. In the following, for ease of notation we define $f^{\mathcal{P}} \equiv f[\vec{\mathcal{P}}(\tau)]$ for generic $f$. We may now equate the above linearized expression with the formal evolution equation, Eq.~(\ref{evolZ}):
\begin{align}
\frac{d}{dt}\vec{\alpha} &= \frac{d\tau}{dt} \vec{A}^{\mathcal{P}} + \frac{d\tau}{dt} \frac{d\bm{U}(\tau)}{d\tau} \vec{z}(t) + \bm{U}(\tau) \frac{d}{dt} \vec{z}(t) \nonumber \\
&= \vec{A}^{\mathcal{P}} + \bm{J}_A^{\mathcal{P}} \bm{U}(\tau)\vec{z}(t) + \sqrt{\Lambda}\bm{B}^{\mathcal{P}}  \frac{\vec{dW}}{dt}
\label{linSDE}
\end{align}
where we have used Eq.~(\ref{ansatzAperp}) on the right hand side. Decomposing the sample paths into parallel and perpendicular components relative to the deterministic limit cycle allows us to separate the dynamics in the above equation. To this end,  we now define $\mathbb{P}_T(\tau)$ as the projection onto the tangent vector $\hat{T}(\tau)$, which is just the dot product with $\hat{T}(\tau)$, and $\mathbb{P}_{\Pi}(\tau)$ as the projection onto the normal hyperplane $\Pi(\tau)$. Then, projecting the dynamics onto the tangent vector at $\tau(t)$ and rearranging, we eventually find:
\begin{align}
\frac{d\tau}{dt} = 1 +  \frac{\langle  \hat{T}(\tau) , \bm{J}_A^{\mathcal{P}} \bm{U}(\tau) \vec{z}(t) -\frac{d\bm{U}(\tau)}{d\tau} \vec{z}(t) + \sqrt{\Lambda}\bm{B}^{\mathcal{P}} \frac{\vec{dW}}{dt} \rangle }{\langle \hat{T}(\tau), \vec{A}^{\mathcal{P}} + \frac{d\bm{U}(\tau)}{d\tau} \vec{z}(t) \rangle}
 \end{align}
Here, we use the fact that $\frac{d}{dt}\vec{z}(t) \perp \hat{T}(\tau)$~\cite{louca_stable_2015}. The above is a complicated SDE for $\tau(t)$; however, it can be simplified by considering the equation only to leading order in the noise. In this case we drop the influence of noise via the normal perturbations $\vec{z}(t)$, as well as the implicit $t$-dependence of $\tau$ on the right hand side. We thus obtain an SDE describing the lowest order, direct effect of noise:
\begin{align}
d\tau = dt + \sqrt{\Lambda}~\frac{ \langle  \hat{T}(t) , \bm{B}[\vec{\mathcal{P}}(t)] \vec{dW} \rangle }{ \langle \hat{T}(t) , \vec{A}[\vec{\mathcal{P}}(t)] \rangle }
\end{align}
The lowest order effect of noise is then to perturb the time parameter, modulating it with a stochastic drift in addition to a deterministic evolution. The stochastic term is the projection of the noise vector onto the tangent vector at every time $t$. Note that the noise term is time dependent, and more precisely periodic with the limit cycle period $\frac{2\pi}{\Delta\omega}$. In situations where it is accurate to replace the stochastic term by its time average, the explicit dependence on $\sqrt{\Lambda}$ indicates clearly that the effective phase diffusion \textit{rate}, determined via the correlator $\avg{\tau(t)\tau(0)}$, scales as $\Lambda$ (thus the phase diffusion \textit{time} scales as $1/\Lambda$)~\cite{rayanov_frequency_2015, navarrete-benlloch_general_2017}

\begin{figure*}
\includegraphics[scale=0.35]{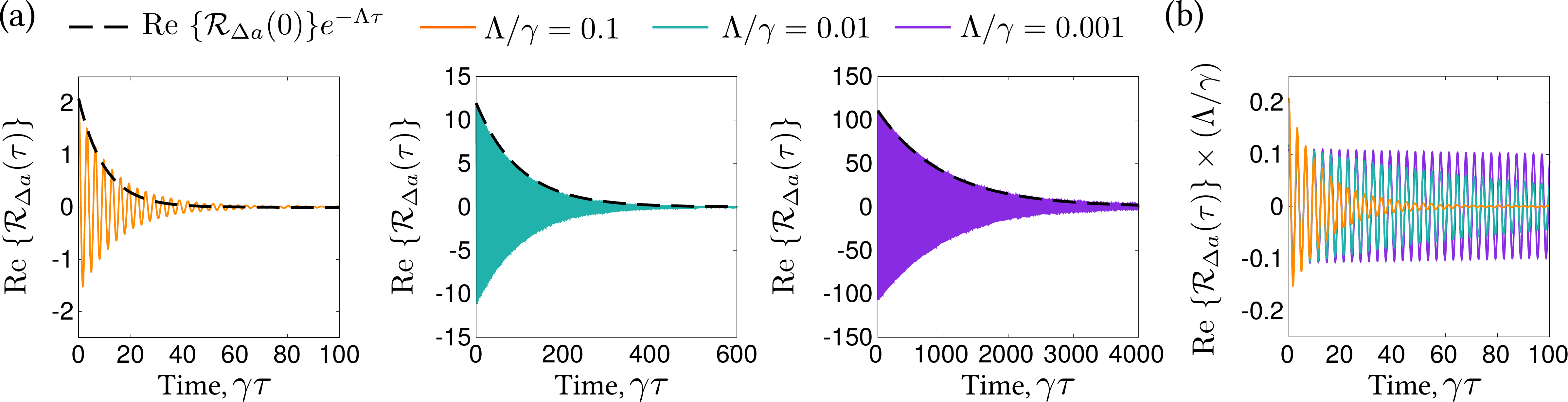}
\caption{Plot of real part of reduced linear mode autocorrelation function $\mathcal{R}_{\Delta a}(\tau)$, for $\Lambda \in [0.1,0.01,0.001]\gamma$, from left to right (orange, green, and purple respectively). In dashed black, we show an exponential envelope decaying with rate $\Lambda$ for each plot. Note the different time axis range for each plot. (b) Real part of linear mode autocorrelation function for the same $\Lambda$ values as (a), but now plotted over the same time range. Also, the correlation function is scaled by $\Lambda/\gamma$ to normalize the $y$-axis.}
\label{fig:corrPlot}
\end{figure*}

For completeness, we indicate that the dynamical equation for perturbations normal to the limit cycle can be determined by projecting Eq.~(\ref{linSDE}) onto the plane $\Pi(\tau)$; this yields~\cite{louca_stable_2015}:
\begin{align}
\frac{d}{dt}\vec{z}(t) = \bm{U}^{-1}(\tau)\mathbb{P}_{\Pi}(\tau) \left[  \bm{J}_A^{\mathcal{P}}\bm{U}(\tau)\vec{z}(t) + \sqrt{\Lambda}\bm{B}^{\mathcal{P}} \frac{\vec{dW}}{dt} \right]
\end{align}
where we have now used the fact that $\mathbb{P}_{\Pi}(\tau)\vec{A}^{\mathcal{P}}(\tau) = 0$ since $\vec{A}^{\mathcal{P}}(\tau) \propto \hat{T}(\tau)$. The first term describes linearized dynamics as governed by the Jacobian; for a stable limit cycle, one expects these dynamics to decay in the long time limit, so that in the absence of noise, $\vec{z} \to 0$. However, the noise term can act to perturb $\vec{z}(t)$, competing against the local stabilizing dynamics~\cite{louca_stable_2015}. 

\section{Numerical simulations of correlation functions}

For computations of the linear mode spectrum, we are interested in the steady state correlation function of the linear mode,
\begin{align}
\mathcal{R}_a(\tau) = \lim_{t\to \infty}\avg{\hat{a}^{\dagger}(t+\tau)\hat{a}(t)} \equiv \avg{\hat{a}^{\dagger}(\tau)\hat{a}(0)}
\end{align}
In the long time limit. the steady state correlation function $\mathcal{R}_a(\tau)$ becomes a function of the single time coordinate $\tau$ only, as indicated above. Knowing $\mathcal{R}_a(\tau)$ for $\tau > 0$ is sufficient in the steady state to find its value for negative $\tau$ as well, as we now describe~\cite{gardiner_quantum_2004}. Note that one may translate the correlation function by $-\tau$ without changing its value, by virtue of the correlation function having a steady state value. This leads us to:
\begin{align}
\mathcal{R}_a(\tau) = \avg{\hat{a}^{\dagger}(\tau)\hat{a}(0)} = \avg{\hat{a}^{\dagger}(0)\hat{a}(-\tau)} = \avg{\hat{a}^{\dagger}(-\tau)\hat{a}(0)}^* 
\end{align}
The above implies $\mathcal{R}_a(-\tau) = \mathcal{R}_a^*(\tau)$. For stochastic simulations based on the positive-P representation, the above correlation function can be estimated by computing an average over stochastic trajectories instead:
\begin{align}
\mathcal{R}_a(\tau) = \lim_{t\to \infty} \frac{1}{N_s} \sum_{i=1}^{N_s} \alpha_i^{\dagger}(t+\tau)\alpha_i(t)
\end{align} 
where the index $i$ now labels specific stochastic realizations of the variables $\alpha^{\dagger}(t)$, $\alpha(t)$. The initial time $t=t_{\rm ss}$ is chosen to be long enough such that initial transients have decayed away and a steady state is reached; this time depends on the dynamical regime of operation. The number of trajectories simulated is taken to be at least $N_s = 10^5$, as mentioned in the main text.

Once the correlation functions are obtained, the power spectral density can be computed using the Wiener-Khinchin theorem:
\begin{align}
S_a[\omega] = \int_{-\infty}^{\infty} d\tau~e^{-i\omega \tau} \mathcal{R}_a(\tau)
\end{align}
Using the above relation between the correlation function for positive and negative times, we can write:
\begin{align}
S_a[\omega] &= \int_{-\infty}^{0} d\tau~e^{-i\omega \tau} \mathcal{R}_a(\tau) + \int_{0}^{\infty} d\tau~e^{-i\omega \tau} \mathcal{R}_a(\tau) \nonumber \\
\implies S_a[\omega] &= 2~\text{Re}~\left\{\int_{0}^{\infty} d\tau~e^{-i\omega \tau} \mathcal{R}_a(\tau)  \right\}
\end{align}
Hence the steady state power spectrum of the linear mode is purely real and can be easily computed using the Wiener-Khinchin theorem. 

For the quantum limit cycle regime studied in Fig.~4 of the main text, we consider the \textit{reduced} steady state correlation function $\mathcal{R}_{\Delta a}(\tau)$, defined as:
\begin{align}
\mathcal{R}_{\Delta a}(\tau) &= \lim_{t\to\infty}\avg{ [\hat{a}^{\dagger}(t+\tau)-\avg{\hat{a}^{\dagger}(t+\tau)} ] [\hat{a}(t)-\avg{\hat{a}(t)} ] } \nonumber \\
&\equiv \mathcal{R}_a(\tau) - \avg{\hat{a}^{\dagger}(\tau)}\avg{\hat{a}(0)}
\end{align}
This definition subtracts away the nonzero steady state amplitude expectation from the correlation function (equivalently, it removes the zero frequency peak visible in the spectra in Fig.~4 of the main text). We plot $\text{Re}~\{\mathcal{R}_{\Delta a}(\tau)\}$ in Fig.~\ref{fig:corrPlot}~(a) for three values of the nonlinearity strength. Also shown is a fit to an exponential decay with decay rate $\Lambda$. Note that different time axes in each plot. In Fig.~\ref{fig:corrPlot}~(b), the three functions are plotted on the same time axes, indicating the difference in correlation decay. The overlap of oscillations indicates the same dominant frequency component in each case.
\vspace{10mm}

\onecolumngrid
\begin{center}
\rule{0.5\textwidth}{1pt}
\end{center}

%\bibliography{Bibliography/NLO-PRLNotes.bib}

\begin{thebibliography}{54}%
\makeatletter
\providecommand \@ifxundefined [1]{%
 \@ifx{#1\undefined}
}%
\providecommand \@ifnum [1]{%
 \ifnum #1\expandafter \@firstoftwo
 \else \expandafter \@secondoftwo
 \fi
}%
\providecommand \@ifx [1]{%
 \ifx #1\expandafter \@firstoftwo
 \else \expandafter \@secondoftwo
 \fi
}%
\providecommand \natexlab [1]{#1}%
\providecommand \enquote  [1]{``#1''}%
\providecommand \bibnamefont  [1]{#1}%
\providecommand \bibfnamefont [1]{#1}%
\providecommand \citenamefont [1]{#1}%
\providecommand \href@noop [0]{\@secondoftwo}%
\providecommand \href [0]{\begingroup \@sanitize@url \@href}%
\providecommand \@href[1]{\@@startlink{#1}\@@href}%
\providecommand \@@href[1]{\endgroup#1\@@endlink}%
\providecommand \@sanitize@url [0]{\catcode `\\12\catcode `\$12\catcode
  `\&12\catcode `\#12\catcode `\^12\catcode `\_12\catcode `\%12\relax}%
\providecommand \@@startlink[1]{}%
\providecommand \@@endlink[0]{}%
\providecommand \url  [0]{\begingroup\@sanitize@url \@url }%
\providecommand \@url [1]{\endgroup\@href {#1}{\urlprefix }}%
\providecommand \urlprefix  [0]{URL }%
\providecommand \Eprint [0]{\href }%
\providecommand \doibase [0]{http://dx.doi.org/}%
\providecommand \selectlanguage [0]{\@gobble}%
\providecommand \bibinfo  [0]{\@secondoftwo}%
\providecommand \bibfield  [0]{\@secondoftwo}%
\providecommand \translation [1]{[#1]}%
\providecommand \BibitemOpen [0]{}%
\providecommand \bibitemStop [0]{}%
\providecommand \bibitemNoStop [0]{.\EOS\space}%
\providecommand \EOS [0]{\spacefactor3000\relax}%
\providecommand \BibitemShut  [1]{\csname bibitem#1\endcsname}%
\let\auto@bib@innerbib\@empty
%</preamble>
\bibitem [{\citenamefont {Clarke}\ and\ \citenamefont
  {Wilhelm}(2008)}]{clarke_superconducting_2008}%
  \BibitemOpen
  \bibfield  {author} {\bibinfo {author} {\bibfnamefont {J.}~\bibnamefont
  {Clarke}}\ and\ \bibinfo {author} {\bibfnamefont {F.~K.}\ \bibnamefont
  {Wilhelm}},\ }\href {\doibase 10.1038/nature07128} {\bibfield  {journal}
  {\bibinfo  {journal} {Nature}\ }\textbf {\bibinfo {volume} {453}},\ \bibinfo
  {pages} {1031} (\bibinfo {year} {2008})}\BibitemShut {NoStop}%
\bibitem [{\citenamefont {Bertet}\ \emph {et~al.}(2011)\citenamefont {Bertet},
  \citenamefont {Ong}, \citenamefont {Boissonneault}, \citenamefont {Bolduc},
  \citenamefont {Mallet}, \citenamefont {Doherty}, \citenamefont {Blais},
  \citenamefont {Vion},\ and\ \citenamefont {Esteve}}]{bertet_circuit_2011}%
  \BibitemOpen
  \bibfield  {author} {\bibinfo {author} {\bibfnamefont {P.}~\bibnamefont
  {Bertet}}, \bibinfo {author} {\bibfnamefont {F.~R.}\ \bibnamefont {Ong}},
  \bibinfo {author} {\bibfnamefont {M.}~\bibnamefont {Boissonneault}}, \bibinfo
  {author} {\bibfnamefont {A.}~\bibnamefont {Bolduc}}, \bibinfo {author}
  {\bibfnamefont {F.}~\bibnamefont {Mallet}}, \bibinfo {author} {\bibfnamefont
  {A.~C.}\ \bibnamefont {Doherty}}, \bibinfo {author} {\bibfnamefont
  {A.}~\bibnamefont {Blais}}, \bibinfo {author} {\bibfnamefont
  {D.}~\bibnamefont {Vion}}, \ and\ \bibinfo {author} {\bibfnamefont
  {D.}~\bibnamefont {Esteve}},\ }\href {http://arxiv.org/abs/1111.0501}
  {\bibfield  {journal} {\bibinfo  {journal} {arXiv:1111.0501 [quant-ph]}\ }
  (\bibinfo {year} {2011})},\ \bibinfo {note} {arXiv: 1111.0501}\BibitemShut
  {NoStop}%
\bibitem [{\citenamefont {Devoret}\ and\ \citenamefont
  {Schoelkopf}(2013)}]{devoret_superconducting_2013}%
  \BibitemOpen
  \bibfield  {author} {\bibinfo {author} {\bibfnamefont {M.~H.}\ \bibnamefont
  {Devoret}}\ and\ \bibinfo {author} {\bibfnamefont {R.~J.}\ \bibnamefont
  {Schoelkopf}},\ }\href {\doibase 10.1126/science.1231930} {\bibfield
  {journal} {\bibinfo  {journal} {Science}\ }\textbf {\bibinfo {volume}
  {339}},\ \bibinfo {pages} {1169} (\bibinfo {year} {2013})}\BibitemShut
  {NoStop}%
\bibitem [{\citenamefont {Blais}\ \emph {et~al.}(2004)\citenamefont {Blais},
  \citenamefont {Huang}, \citenamefont {Wallraff}, \citenamefont {Girvin},\
  and\ \citenamefont {Schoelkopf}}]{blais_cavity_2004}%
  \BibitemOpen
  \bibfield  {author} {\bibinfo {author} {\bibfnamefont {A.}~\bibnamefont
  {Blais}}, \bibinfo {author} {\bibfnamefont {R.-S.}\ \bibnamefont {Huang}},
  \bibinfo {author} {\bibfnamefont {A.}~\bibnamefont {Wallraff}}, \bibinfo
  {author} {\bibfnamefont {S.~M.}\ \bibnamefont {Girvin}}, \ and\ \bibinfo
  {author} {\bibfnamefont {R.~J.}\ \bibnamefont {Schoelkopf}},\ }\href
  {\doibase 10.1103/PhysRevA.69.062320} {\bibfield  {journal} {\bibinfo
  {journal} {Phys. Rev. A}\ }\textbf {\bibinfo {volume} {69}},\ \bibinfo
  {pages} {062320} (\bibinfo {year} {2004})}\BibitemShut {NoStop}%
\bibitem [{\citenamefont {Wallraff}\ \emph {et~al.}(2004)\citenamefont
  {Wallraff}, \citenamefont {Schuster}, \citenamefont {Blais}, \citenamefont
  {Frunzio}, \citenamefont {Huang}, \citenamefont {Majer}, \citenamefont
  {Kumar}, \citenamefont {Girvin},\ and\ \citenamefont
  {Schoelkopf}}]{wallraff_strong_2004}%
  \BibitemOpen
  \bibfield  {author} {\bibinfo {author} {\bibfnamefont {A.}~\bibnamefont
  {Wallraff}}, \bibinfo {author} {\bibfnamefont {D.~I.}\ \bibnamefont
  {Schuster}}, \bibinfo {author} {\bibfnamefont {A.}~\bibnamefont {Blais}},
  \bibinfo {author} {\bibfnamefont {L.}~\bibnamefont {Frunzio}}, \bibinfo
  {author} {\bibfnamefont {R.-S.}\ \bibnamefont {Huang}}, \bibinfo {author}
  {\bibfnamefont {J.}~\bibnamefont {Majer}}, \bibinfo {author} {\bibfnamefont
  {S.}~\bibnamefont {Kumar}}, \bibinfo {author} {\bibfnamefont {S.~M.}\
  \bibnamefont {Girvin}}, \ and\ \bibinfo {author} {\bibfnamefont {R.~J.}\
  \bibnamefont {Schoelkopf}},\ }\href {\doibase 10.1038/nature02851} {\bibfield
   {journal} {\bibinfo  {journal} {Nature}\ }\textbf {\bibinfo {volume}
  {431}},\ \bibinfo {pages} {162} (\bibinfo {year} {2004})}\BibitemShut
  {NoStop}%
\bibitem [{\citenamefont {Majer}\ \emph {et~al.}(2007)\citenamefont {Majer},
  \citenamefont {Chow}, \citenamefont {Gambetta}, \citenamefont {Koch},
  \citenamefont {Johnson}, \citenamefont {Schreier}, \citenamefont {Frunzio},
  \citenamefont {Schuster}, \citenamefont {Houck}, \citenamefont {Wallraff},
  \citenamefont {Blais}, \citenamefont {Devoret}, \citenamefont {Girvin},\ and\
  \citenamefont {Schoelkopf}}]{majer_coupling_2007}%
  \BibitemOpen
  \bibfield  {author} {\bibinfo {author} {\bibfnamefont {J.}~\bibnamefont
  {Majer}}, \bibinfo {author} {\bibfnamefont {J.~M.}\ \bibnamefont {Chow}},
  \bibinfo {author} {\bibfnamefont {J.~M.}\ \bibnamefont {Gambetta}}, \bibinfo
  {author} {\bibfnamefont {J.}~\bibnamefont {Koch}}, \bibinfo {author}
  {\bibfnamefont {B.~R.}\ \bibnamefont {Johnson}}, \bibinfo {author}
  {\bibfnamefont {J.~A.}\ \bibnamefont {Schreier}}, \bibinfo {author}
  {\bibfnamefont {L.}~\bibnamefont {Frunzio}}, \bibinfo {author} {\bibfnamefont
  {D.~I.}\ \bibnamefont {Schuster}}, \bibinfo {author} {\bibfnamefont {A.~A.}\
  \bibnamefont {Houck}}, \bibinfo {author} {\bibfnamefont {A.}~\bibnamefont
  {Wallraff}}, \bibinfo {author} {\bibfnamefont {A.}~\bibnamefont {Blais}},
  \bibinfo {author} {\bibfnamefont {M.~H.}\ \bibnamefont {Devoret}}, \bibinfo
  {author} {\bibfnamefont {S.~M.}\ \bibnamefont {Girvin}}, \ and\ \bibinfo
  {author} {\bibfnamefont {R.~J.}\ \bibnamefont {Schoelkopf}},\ }\href
  {\doibase 10.1038/nature06184} {\bibfield  {journal} {\bibinfo  {journal}
  {Nature}\ }\textbf {\bibinfo {volume} {449}},\ \bibinfo {pages} {443}
  (\bibinfo {year} {2007})}\BibitemShut {NoStop}%
\bibitem [{\citenamefont {DiCarlo}\ \emph {et~al.}(2009)\citenamefont
  {DiCarlo}, \citenamefont {Chow}, \citenamefont {Gambetta}, \citenamefont
  {Bishop}, \citenamefont {Johnson}, \citenamefont {Schuster}, \citenamefont
  {Majer}, \citenamefont {Blais}, \citenamefont {Frunzio}, \citenamefont
  {Girvin},\ and\ \citenamefont {Schoelkopf}}]{dicarlo_demonstration_2009}%
  \BibitemOpen
  \bibfield  {author} {\bibinfo {author} {\bibfnamefont {L.}~\bibnamefont
  {DiCarlo}}, \bibinfo {author} {\bibfnamefont {J.~M.}\ \bibnamefont {Chow}},
  \bibinfo {author} {\bibfnamefont {J.~M.}\ \bibnamefont {Gambetta}}, \bibinfo
  {author} {\bibfnamefont {L.~S.}\ \bibnamefont {Bishop}}, \bibinfo {author}
  {\bibfnamefont {B.~R.}\ \bibnamefont {Johnson}}, \bibinfo {author}
  {\bibfnamefont {D.~I.}\ \bibnamefont {Schuster}}, \bibinfo {author}
  {\bibfnamefont {J.}~\bibnamefont {Majer}}, \bibinfo {author} {\bibfnamefont
  {A.}~\bibnamefont {Blais}}, \bibinfo {author} {\bibfnamefont
  {L.}~\bibnamefont {Frunzio}}, \bibinfo {author} {\bibfnamefont {S.~M.}\
  \bibnamefont {Girvin}}, \ and\ \bibinfo {author} {\bibfnamefont {R.~J.}\
  \bibnamefont {Schoelkopf}},\ }\href {\doibase 10.1038/nature08121} {\bibfield
   {journal} {\bibinfo  {journal} {Nature}\ }\textbf {\bibinfo {volume}
  {460}},\ \bibinfo {pages} {240} (\bibinfo {year} {2009})}\BibitemShut
  {NoStop}%
\bibitem [{\citenamefont {Schuster}\ \emph {et~al.}(2005)\citenamefont
  {Schuster}, \citenamefont {Wallraff}, \citenamefont {Blais}, \citenamefont
  {Frunzio}, \citenamefont {Huang}, \citenamefont {Majer}, \citenamefont
  {Girvin},\ and\ \citenamefont {Schoelkopf}}]{schuster_ac_2005}%
  \BibitemOpen
  \bibfield  {author} {\bibinfo {author} {\bibfnamefont {D.~I.}\ \bibnamefont
  {Schuster}}, \bibinfo {author} {\bibfnamefont {A.}~\bibnamefont {Wallraff}},
  \bibinfo {author} {\bibfnamefont {A.}~\bibnamefont {Blais}}, \bibinfo
  {author} {\bibfnamefont {L.}~\bibnamefont {Frunzio}}, \bibinfo {author}
  {\bibfnamefont {R.-S.}\ \bibnamefont {Huang}}, \bibinfo {author}
  {\bibfnamefont {J.}~\bibnamefont {Majer}}, \bibinfo {author} {\bibfnamefont
  {S.~M.}\ \bibnamefont {Girvin}}, \ and\ \bibinfo {author} {\bibfnamefont
  {R.~J.}\ \bibnamefont {Schoelkopf}},\ }\href {\doibase
  10.1103/PhysRevLett.94.123602} {\bibfield  {journal} {\bibinfo  {journal}
  {Phys. Rev. Lett.}\ }\textbf {\bibinfo {volume} {94}},\ \bibinfo {pages}
  {123602} (\bibinfo {year} {2005})}\BibitemShut {NoStop}%
\bibitem [{\citenamefont {Gambetta}\ \emph {et~al.}(2006)\citenamefont
  {Gambetta}, \citenamefont {Blais}, \citenamefont {Schuster}, \citenamefont
  {Wallraff}, \citenamefont {Frunzio}, \citenamefont {Majer}, \citenamefont
  {Devoret}, \citenamefont {Girvin},\ and\ \citenamefont
  {Schoelkopf}}]{gambetta_qubit-photon_2006}%
  \BibitemOpen
  \bibfield  {author} {\bibinfo {author} {\bibfnamefont {J.}~\bibnamefont
  {Gambetta}}, \bibinfo {author} {\bibfnamefont {A.}~\bibnamefont {Blais}},
  \bibinfo {author} {\bibfnamefont {D.~I.}\ \bibnamefont {Schuster}}, \bibinfo
  {author} {\bibfnamefont {A.}~\bibnamefont {Wallraff}}, \bibinfo {author}
  {\bibfnamefont {L.}~\bibnamefont {Frunzio}}, \bibinfo {author} {\bibfnamefont
  {J.}~\bibnamefont {Majer}}, \bibinfo {author} {\bibfnamefont {M.~H.}\
  \bibnamefont {Devoret}}, \bibinfo {author} {\bibfnamefont {S.~M.}\
  \bibnamefont {Girvin}}, \ and\ \bibinfo {author} {\bibfnamefont {R.~J.}\
  \bibnamefont {Schoelkopf}},\ }\href {\doibase 10.1103/PhysRevA.74.042318}
  {\bibfield  {journal} {\bibinfo  {journal} {Phys. Rev. A}\ }\textbf {\bibinfo
  {volume} {74}},\ \bibinfo {pages} {042318} (\bibinfo {year}
  {2006})}\BibitemShut {NoStop}%
\bibitem [{\citenamefont {Bishop}\ \emph {et~al.}(2010)\citenamefont {Bishop},
  \citenamefont {Ginossar},\ and\ \citenamefont
  {Girvin}}]{bishop_response_2010}%
  \BibitemOpen
  \bibfield  {author} {\bibinfo {author} {\bibfnamefont {L.~S.}\ \bibnamefont
  {Bishop}}, \bibinfo {author} {\bibfnamefont {E.}~\bibnamefont {Ginossar}}, \
  and\ \bibinfo {author} {\bibfnamefont {S.~M.}\ \bibnamefont {Girvin}},\
  }\href {\doibase 10.1103/PhysRevLett.105.100505} {\bibfield  {journal}
  {\bibinfo  {journal} {Phys. Rev. Lett.}\ }\textbf {\bibinfo {volume} {105}},\
  \bibinfo {pages} {100505} (\bibinfo {year} {2010})}\BibitemShut {NoStop}%
\bibitem [{\citenamefont {Sundaresan}\ \emph {et~al.}(2015)\citenamefont
  {Sundaresan}, \citenamefont {Liu}, \citenamefont {Sadri}, \citenamefont
  {Sz{\H o}cs}, \citenamefont {Underwood}, \citenamefont {Malekakhlagh},
  \citenamefont {T{\"u}reci},\ and\ \citenamefont
  {Houck}}]{sundaresan_beyond_2015}%
  \BibitemOpen
  \bibfield  {author} {\bibinfo {author} {\bibfnamefont {N.~M.}\ \bibnamefont
  {Sundaresan}}, \bibinfo {author} {\bibfnamefont {Y.}~\bibnamefont {Liu}},
  \bibinfo {author} {\bibfnamefont {D.}~\bibnamefont {Sadri}}, \bibinfo
  {author} {\bibfnamefont {L.~J.}\ \bibnamefont {Sz{\H o}cs}}, \bibinfo
  {author} {\bibfnamefont {D.~L.}\ \bibnamefont {Underwood}}, \bibinfo {author}
  {\bibfnamefont {M.}~\bibnamefont {Malekakhlagh}}, \bibinfo {author}
  {\bibfnamefont {H.~E.}\ \bibnamefont {T{\"u}reci}}, \ and\ \bibinfo {author}
  {\bibfnamefont {A.~A.}\ \bibnamefont {Houck}},\ }\href {\doibase
  10.1103/PhysRevX.5.021035} {\bibfield  {journal} {\bibinfo  {journal} {Phys.
  Rev. X}\ }\textbf {\bibinfo {volume} {5}},\ \bibinfo {pages} {021035}
  (\bibinfo {year} {2015})}\BibitemShut {NoStop}%
\bibitem [{\citenamefont {Reed}\ \emph {et~al.}(2010)\citenamefont {Reed},
  \citenamefont {DiCarlo}, \citenamefont {Johnson}, \citenamefont {Sun},
  \citenamefont {Schuster}, \citenamefont {Frunzio},\ and\ \citenamefont
  {Schoelkopf}}]{reed_high-fidelity_2010}%
  \BibitemOpen
  \bibfield  {author} {\bibinfo {author} {\bibfnamefont {M.~D.}\ \bibnamefont
  {Reed}}, \bibinfo {author} {\bibfnamefont {L.}~\bibnamefont {DiCarlo}},
  \bibinfo {author} {\bibfnamefont {B.~R.}\ \bibnamefont {Johnson}}, \bibinfo
  {author} {\bibfnamefont {L.}~\bibnamefont {Sun}}, \bibinfo {author}
  {\bibfnamefont {D.~I.}\ \bibnamefont {Schuster}}, \bibinfo {author}
  {\bibfnamefont {L.}~\bibnamefont {Frunzio}}, \ and\ \bibinfo {author}
  {\bibfnamefont {R.~J.}\ \bibnamefont {Schoelkopf}},\ }\href {\doibase
  10.1103/PhysRevLett.105.173601} {\bibfield  {journal} {\bibinfo  {journal}
  {Phys. Rev. Lett.}\ }\textbf {\bibinfo {volume} {105}},\ \bibinfo {pages}
  {173601} (\bibinfo {year} {2010})}\BibitemShut {NoStop}%
\bibitem [{\citenamefont {Vijay}\ \emph {et~al.}(2009)\citenamefont {Vijay},
  \citenamefont {Devoret},\ and\ \citenamefont {Siddiqi}}]{vijay_invited_2009}%
  \BibitemOpen
  \bibfield  {author} {\bibinfo {author} {\bibfnamefont {R.}~\bibnamefont
  {Vijay}}, \bibinfo {author} {\bibfnamefont {M.~H.}\ \bibnamefont {Devoret}},
  \ and\ \bibinfo {author} {\bibfnamefont {I.}~\bibnamefont {Siddiqi}},\ }\href
  {\doibase 10.1063/1.3224703} {\bibfield  {journal} {\bibinfo  {journal}
  {Review of Scientific Instruments}\ }\textbf {\bibinfo {volume} {80}},\
  \bibinfo {pages} {111101} (\bibinfo {year} {2009})}\BibitemShut {NoStop}%
\bibitem [{\citenamefont {Mallet}\ \emph {et~al.}(2009)\citenamefont {Mallet},
  \citenamefont {Ong}, \citenamefont {Palacios-Laloy}, \citenamefont {Nguyen},
  \citenamefont {Bertet}, \citenamefont {Vion},\ and\ \citenamefont
  {Esteve}}]{mallet_single-shot_2009}%
  \BibitemOpen
  \bibfield  {author} {\bibinfo {author} {\bibfnamefont {F.}~\bibnamefont
  {Mallet}}, \bibinfo {author} {\bibfnamefont {F.~R.}\ \bibnamefont {Ong}},
  \bibinfo {author} {\bibfnamefont {A.}~\bibnamefont {Palacios-Laloy}},
  \bibinfo {author} {\bibfnamefont {F.}~\bibnamefont {Nguyen}}, \bibinfo
  {author} {\bibfnamefont {P.}~\bibnamefont {Bertet}}, \bibinfo {author}
  {\bibfnamefont {D.}~\bibnamefont {Vion}}, \ and\ \bibinfo {author}
  {\bibfnamefont {D.}~\bibnamefont {Esteve}},\ }\href {\doibase
  10.1038/nphys1400} {\bibfield  {journal} {\bibinfo  {journal} {Nat Phys}\
  }\textbf {\bibinfo {volume} {5}},\ \bibinfo {pages} {791} (\bibinfo {year}
  {2009})}\BibitemShut {NoStop}%
\bibitem [{\citenamefont {Yurke}\ \emph {et~al.}(1989)\citenamefont {Yurke},
  \citenamefont {Corruccini}, \citenamefont {Kaminsky}, \citenamefont {Rupp},
  \citenamefont {Smith}, \citenamefont {Silver}, \citenamefont {Simon},\ and\
  \citenamefont {Whittaker}}]{yurke_observation_1989}%
  \BibitemOpen
  \bibfield  {author} {\bibinfo {author} {\bibfnamefont {B.}~\bibnamefont
  {Yurke}}, \bibinfo {author} {\bibfnamefont {L.~R.}\ \bibnamefont
  {Corruccini}}, \bibinfo {author} {\bibfnamefont {P.~G.}\ \bibnamefont
  {Kaminsky}}, \bibinfo {author} {\bibfnamefont {L.~W.}\ \bibnamefont {Rupp}},
  \bibinfo {author} {\bibfnamefont {A.~D.}\ \bibnamefont {Smith}}, \bibinfo
  {author} {\bibfnamefont {A.~H.}\ \bibnamefont {Silver}}, \bibinfo {author}
  {\bibfnamefont {R.~W.}\ \bibnamefont {Simon}}, \ and\ \bibinfo {author}
  {\bibfnamefont {E.~A.}\ \bibnamefont {Whittaker}},\ }\href {\doibase
  10.1103/PhysRevA.39.2519} {\bibfield  {journal} {\bibinfo  {journal} {Phys.
  Rev. A}\ }\textbf {\bibinfo {volume} {39}},\ \bibinfo {pages} {2519}
  (\bibinfo {year} {1989})}\BibitemShut {NoStop}%
\bibitem [{\citenamefont {Castellanos-Beltran}\ and\ \citenamefont
  {Lehnert}(2007)}]{castellanos-beltran_widely_2007}%
  \BibitemOpen
  \bibfield  {author} {\bibinfo {author} {\bibfnamefont {M.~A.}\ \bibnamefont
  {Castellanos-Beltran}}\ and\ \bibinfo {author} {\bibfnamefont {K.~W.}\
  \bibnamefont {Lehnert}},\ }\href {\doibase 10.1063/1.2773988} {\bibfield
  {journal} {\bibinfo  {journal} {Appl. Phys. Lett.}\ }\textbf {\bibinfo
  {volume} {91}},\ \bibinfo {pages} {083509} (\bibinfo {year}
  {2007})}\BibitemShut {NoStop}%
\bibitem [{\citenamefont {Abdo}\ \emph {et~al.}(2013)\citenamefont {Abdo},
  \citenamefont {Sliwa}, \citenamefont {Frunzio},\ and\ \citenamefont
  {Devoret}}]{abdo_directional_2013}%
  \BibitemOpen
  \bibfield  {author} {\bibinfo {author} {\bibfnamefont {B.}~\bibnamefont
  {Abdo}}, \bibinfo {author} {\bibfnamefont {K.}~\bibnamefont {Sliwa}},
  \bibinfo {author} {\bibfnamefont {L.}~\bibnamefont {Frunzio}}, \ and\
  \bibinfo {author} {\bibfnamefont {M.}~\bibnamefont {Devoret}},\ }\href
  {\doibase 10.1103/PhysRevX.3.031001} {\bibfield  {journal} {\bibinfo
  {journal} {Phys. Rev. X}\ }\textbf {\bibinfo {volume} {3}},\ \bibinfo {pages}
  {031001} (\bibinfo {year} {2013})}\BibitemShut {NoStop}%
\bibitem [{\citenamefont {Eichler}\ \emph {et~al.}(2014)\citenamefont
  {Eichler}, \citenamefont {Salathe}, \citenamefont {Mlynek}, \citenamefont
  {Schmidt},\ and\ \citenamefont {Wallraff}}]{eichler_quantum-limited_2014}%
  \BibitemOpen
  \bibfield  {author} {\bibinfo {author} {\bibfnamefont {C.}~\bibnamefont
  {Eichler}}, \bibinfo {author} {\bibfnamefont {Y.}~\bibnamefont {Salathe}},
  \bibinfo {author} {\bibfnamefont {J.}~\bibnamefont {Mlynek}}, \bibinfo
  {author} {\bibfnamefont {S.}~\bibnamefont {Schmidt}}, \ and\ \bibinfo
  {author} {\bibfnamefont {A.}~\bibnamefont {Wallraff}},\ }\href {\doibase
  10.1103/PhysRevLett.113.110502} {\bibfield  {journal} {\bibinfo  {journal}
  {Phys. Rev. Lett.}\ }\textbf {\bibinfo {volume} {113}},\ \bibinfo {pages}
  {110502} (\bibinfo {year} {2014})}\BibitemShut {NoStop}%
\bibitem [{\citenamefont {Macklin}\ \emph {et~al.}(2015)\citenamefont
  {Macklin}, \citenamefont {O{\textquoteright}Brien}, \citenamefont {Hover},
  \citenamefont {Schwartz}, \citenamefont {Bolkhovsky}, \citenamefont {Zhang},
  \citenamefont {Oliver},\ and\ \citenamefont
  {Siddiqi}}]{macklin_nearquantum-limited_2015}%
  \BibitemOpen
  \bibfield  {author} {\bibinfo {author} {\bibfnamefont {C.}~\bibnamefont
  {Macklin}}, \bibinfo {author} {\bibfnamefont {K.}~\bibnamefont
  {O{\textquoteright}Brien}}, \bibinfo {author} {\bibfnamefont
  {D.}~\bibnamefont {Hover}}, \bibinfo {author} {\bibfnamefont {M.~E.}\
  \bibnamefont {Schwartz}}, \bibinfo {author} {\bibfnamefont {V.}~\bibnamefont
  {Bolkhovsky}}, \bibinfo {author} {\bibfnamefont {X.}~\bibnamefont {Zhang}},
  \bibinfo {author} {\bibfnamefont {W.~D.}\ \bibnamefont {Oliver}}, \ and\
  \bibinfo {author} {\bibfnamefont {I.}~\bibnamefont {Siddiqi}},\ }\href
  {\doibase 10.1126/science.aaa8525} {\bibfield  {journal} {\bibinfo  {journal}
  {Science}\ }\textbf {\bibinfo {volume} {350}},\ \bibinfo {pages} {307}
  (\bibinfo {year} {2015})}\BibitemShut {NoStop}%
\bibitem [{\citenamefont {Fink}\ \emph {et~al.}(2017)\citenamefont {Fink},
  \citenamefont {Dombi}, \citenamefont {Vukics}, \citenamefont {Wallraff},\
  and\ \citenamefont {Domokos}}]{fink_observation_2017}%
  \BibitemOpen
  \bibfield  {author} {\bibinfo {author} {\bibfnamefont {J.~M.}\ \bibnamefont
  {Fink}}, \bibinfo {author} {\bibfnamefont {A.}~\bibnamefont {Dombi}},
  \bibinfo {author} {\bibfnamefont {A.}~\bibnamefont {Vukics}}, \bibinfo
  {author} {\bibfnamefont {A.}~\bibnamefont {Wallraff}}, \ and\ \bibinfo
  {author} {\bibfnamefont {P.}~\bibnamefont {Domokos}},\ }\href {\doibase
  10.1103/PhysRevX.7.011012} {\bibfield  {journal} {\bibinfo  {journal} {Phys.
  Rev. X}\ }\textbf {\bibinfo {volume} {7}},\ \bibinfo {pages} {011012}
  (\bibinfo {year} {2017})}\BibitemShut {NoStop}%
\bibitem [{\citenamefont {Fitzpatrick}\ \emph {et~al.}(2017)\citenamefont
  {Fitzpatrick}, \citenamefont {Sundaresan}, \citenamefont {Li}, \citenamefont
  {Koch},\ and\ \citenamefont {Houck}}]{fitzpatrick_observation_2017}%
  \BibitemOpen
  \bibfield  {author} {\bibinfo {author} {\bibfnamefont {M.}~\bibnamefont
  {Fitzpatrick}}, \bibinfo {author} {\bibfnamefont {N.~M.}\ \bibnamefont
  {Sundaresan}}, \bibinfo {author} {\bibfnamefont {A.~C.~Y.}\ \bibnamefont
  {Li}}, \bibinfo {author} {\bibfnamefont {J.}~\bibnamefont {Koch}}, \ and\
  \bibinfo {author} {\bibfnamefont {A.~A.}\ \bibnamefont {Houck}},\ }\href
  {\doibase 10.1103/PhysRevX.7.011016} {\bibfield  {journal} {\bibinfo
  {journal} {Phys. Rev. X}\ }\textbf {\bibinfo {volume} {7}},\ \bibinfo {pages}
  {011016} (\bibinfo {year} {2017})}\BibitemShut {NoStop}%
\bibitem [{\citenamefont {Boissonneault}\ \emph {et~al.}(2010)\citenamefont
  {Boissonneault}, \citenamefont {Gambetta},\ and\ \citenamefont
  {Blais}}]{boissonneault_improved_2010}%
  \BibitemOpen
  \bibfield  {author} {\bibinfo {author} {\bibfnamefont {M.}~\bibnamefont
  {Boissonneault}}, \bibinfo {author} {\bibfnamefont {J.~M.}\ \bibnamefont
  {Gambetta}}, \ and\ \bibinfo {author} {\bibfnamefont {A.}~\bibnamefont
  {Blais}},\ }\href {\doibase 10.1103/PhysRevLett.105.100504} {\bibfield
  {journal} {\bibinfo  {journal} {Phys. Rev. Lett.}\ }\textbf {\bibinfo
  {volume} {105}},\ \bibinfo {pages} {100504} (\bibinfo {year}
  {2010})}\BibitemShut {NoStop}%
\bibitem [{\citenamefont {Dykman}\ and\ \citenamefont
  {Krivoglaz}(1979)}]{dykman_theory_1979}%
  \BibitemOpen
  \bibfield  {author} {\bibinfo {author} {\bibfnamefont {M.~I.}\ \bibnamefont
  {Dykman}}\ and\ \bibinfo {author} {\bibfnamefont {M.~A.}\ \bibnamefont
  {Krivoglaz}},\ }\href {http://web.pa.msu.edu/people/dykman/pub06/DK79.pdf}
  {\bibfield  {journal} {\bibinfo  {journal} {Soviet Journal of Experimental
  and Theoretical Physics}\ }\textbf {\bibinfo {volume} {50}},\ \bibinfo
  {pages} {30} (\bibinfo {year} {1979})}\BibitemShut {NoStop}%
\bibitem [{\citenamefont {Siddiqi}\ \emph {et~al.}(2004)\citenamefont
  {Siddiqi}, \citenamefont {Vijay}, \citenamefont {Pierre}, \citenamefont
  {Wilson}, \citenamefont {Metcalfe}, \citenamefont {Rigetti}, \citenamefont
  {Frunzio},\ and\ \citenamefont {Devoret}}]{siddiqi_RF-driven_2004}%
  \BibitemOpen
  \bibfield  {author} {\bibinfo {author} {\bibfnamefont {I.}~\bibnamefont
  {Siddiqi}}, \bibinfo {author} {\bibfnamefont {R.}~\bibnamefont {Vijay}},
  \bibinfo {author} {\bibfnamefont {F.}~\bibnamefont {Pierre}}, \bibinfo
  {author} {\bibfnamefont {C.~M.}\ \bibnamefont {Wilson}}, \bibinfo {author}
  {\bibfnamefont {M.}~\bibnamefont {Metcalfe}}, \bibinfo {author}
  {\bibfnamefont {C.}~\bibnamefont {Rigetti}}, \bibinfo {author} {\bibfnamefont
  {L.}~\bibnamefont {Frunzio}}, \ and\ \bibinfo {author} {\bibfnamefont
  {M.~H.}\ \bibnamefont {Devoret}},\ }\href {\doibase
  10.1103/PhysRevLett.93.207002} {\bibfield  {journal} {\bibinfo  {journal}
  {Phys. Rev. Lett.}\ }\textbf {\bibinfo {volume} {93}},\ \bibinfo {pages}
  {207002} (\bibinfo {year} {2004})}\BibitemShut {NoStop}%
\bibitem [{\citenamefont {Siddiqi}\ \emph {et~al.}(2005)\citenamefont
  {Siddiqi}, \citenamefont {Vijay}, \citenamefont {Pierre}, \citenamefont
  {Wilson}, \citenamefont {Frunzio}, \citenamefont {Metcalfe}, \citenamefont
  {Rigetti}, \citenamefont {Schoelkopf}, \citenamefont {Devoret}, \citenamefont
  {Vion},\ and\ \citenamefont {Esteve}}]{siddiqi_direct_2005}%
  \BibitemOpen
  \bibfield  {author} {\bibinfo {author} {\bibfnamefont {I.}~\bibnamefont
  {Siddiqi}}, \bibinfo {author} {\bibfnamefont {R.}~\bibnamefont {Vijay}},
  \bibinfo {author} {\bibfnamefont {F.}~\bibnamefont {Pierre}}, \bibinfo
  {author} {\bibfnamefont {C.~M.}\ \bibnamefont {Wilson}}, \bibinfo {author}
  {\bibfnamefont {L.}~\bibnamefont {Frunzio}}, \bibinfo {author} {\bibfnamefont
  {M.}~\bibnamefont {Metcalfe}}, \bibinfo {author} {\bibfnamefont
  {C.}~\bibnamefont {Rigetti}}, \bibinfo {author} {\bibfnamefont {R.~J.}\
  \bibnamefont {Schoelkopf}}, \bibinfo {author} {\bibfnamefont {M.~H.}\
  \bibnamefont {Devoret}}, \bibinfo {author} {\bibfnamefont {D.}~\bibnamefont
  {Vion}}, \ and\ \bibinfo {author} {\bibfnamefont {D.}~\bibnamefont
  {Esteve}},\ }\href {\doibase 10.1103/PhysRevLett.94.027005} {\bibfield
  {journal} {\bibinfo  {journal} {Phys. Rev. Lett.}\ }\textbf {\bibinfo
  {volume} {94}},\ \bibinfo {pages} {027005} (\bibinfo {year}
  {2005})}\BibitemShut {NoStop}%
\bibitem [{\citenamefont {Dykman}(2012)}]{dykman_fluctuating_2012}%
  \BibitemOpen
  \bibfield  {author} {\bibinfo {author} {\bibfnamefont {M.}~\bibnamefont
  {Dykman}},\ }\href@noop {} {\emph {\bibinfo {title}
  {Fluctuating {Nonlinear} {Oscillators}: {From} {Nanomechanics} to {Quantum}
  {Superconducting} {Circuits}}}}\ (\bibinfo  {publisher} {OUP Oxford},\
  \bibinfo {year} {2012})\BibitemShut {NoStop}%
\bibitem [{\citenamefont {Del{\textquoteright}Haye}\ \emph
  {et~al.}(2007)\citenamefont {Del{\textquoteright}Haye}, \citenamefont
  {Schliesser}, \citenamefont {Arcizet}, \citenamefont {Wilken}, \citenamefont
  {Holzwarth},\ and\ \citenamefont {Kippenberg}}]{delhaye_optical_2007}%
  \BibitemOpen
  \bibfield  {author} {\bibinfo {author} {\bibfnamefont {P.}~\bibnamefont
  {Del{\textquoteright}Haye}}, \bibinfo {author} {\bibfnamefont
  {A.}~\bibnamefont {Schliesser}}, \bibinfo {author} {\bibfnamefont
  {O.}~\bibnamefont {Arcizet}}, \bibinfo {author} {\bibfnamefont
  {T.}~\bibnamefont {Wilken}}, \bibinfo {author} {\bibfnamefont
  {R.}~\bibnamefont {Holzwarth}}, \ and\ \bibinfo {author} {\bibfnamefont
  {T.~J.}\ \bibnamefont {Kippenberg}},\ }\href {\doibase 10.1038/nature06401}
  {\bibfield  {journal} {\bibinfo  {journal} {Nature}\ }\textbf {\bibinfo
  {volume} {450}},\ \bibinfo {pages} {1214} (\bibinfo {year}
  {2007})}\BibitemShut {NoStop}%
\bibitem [{\citenamefont {Del{\textquoteright}Haye}\ \emph
  {et~al.}(2008)\citenamefont {Del{\textquoteright}Haye}, \citenamefont
  {Arcizet}, \citenamefont {Schliesser}, \citenamefont {Holzwarth},\ and\
  \citenamefont {Kippenberg}}]{delhaye_full_2008}%
  \BibitemOpen
  \bibfield  {author} {\bibinfo {author} {\bibfnamefont {P.}~\bibnamefont
  {Del{\textquoteright}Haye}}, \bibinfo {author} {\bibfnamefont
  {O.}~\bibnamefont {Arcizet}}, \bibinfo {author} {\bibfnamefont
  {A.}~\bibnamefont {Schliesser}}, \bibinfo {author} {\bibfnamefont
  {R.}~\bibnamefont {Holzwarth}}, \ and\ \bibinfo {author} {\bibfnamefont
  {T.~J.}\ \bibnamefont {Kippenberg}},\ }\href {\doibase
  10.1103/PhysRevLett.101.053903} {\bibfield  {journal} {\bibinfo  {journal}
  {Phys. Rev. Lett.}\ }\textbf {\bibinfo {volume} {101}},\ \bibinfo {pages}
  {053903} (\bibinfo {year} {2008})}\BibitemShut {NoStop}%
\bibitem [{\citenamefont {Levy}\ \emph {et~al.}(2010)\citenamefont {Levy},
  \citenamefont {Gondarenko}, \citenamefont {Foster}, \citenamefont
  {Turner-Foster}, \citenamefont {Gaeta},\ and\ \citenamefont
  {Lipson}}]{levy_cmos-compatible_2010}%
  \BibitemOpen
  \bibfield  {author} {\bibinfo {author} {\bibfnamefont {J.~S.}\ \bibnamefont
  {Levy}}, \bibinfo {author} {\bibfnamefont {A.}~\bibnamefont {Gondarenko}},
  \bibinfo {author} {\bibfnamefont {M.~A.}\ \bibnamefont {Foster}}, \bibinfo
  {author} {\bibfnamefont {A.~C.}\ \bibnamefont {Turner-Foster}}, \bibinfo
  {author} {\bibfnamefont {A.~L.}\ \bibnamefont {Gaeta}}, \ and\ \bibinfo
  {author} {\bibfnamefont {M.}~\bibnamefont {Lipson}},\ }\href {\doibase
  10.1038/nphoton.2009.259} {\bibfield  {journal} {\bibinfo  {journal} {Nat
  Photon}\ }\textbf {\bibinfo {volume} {4}},\ \bibinfo {pages} {37} (\bibinfo
  {year} {2010})}\BibitemShut {NoStop}%
\bibitem [{\citenamefont {Herr}\ \emph {et~al.}(2014)\citenamefont {Herr},
  \citenamefont {Brasch}, \citenamefont {Jost}, \citenamefont {Wang},
  \citenamefont {Kondratiev}, \citenamefont {Gorodetsky},\ and\ \citenamefont
  {Kippenberg}}]{herr_temporal_2014}%
  \BibitemOpen
  \bibfield  {author} {\bibinfo {author} {\bibfnamefont {T.}~\bibnamefont
  {Herr}}, \bibinfo {author} {\bibfnamefont {V.}~\bibnamefont {Brasch}},
  \bibinfo {author} {\bibfnamefont {J.~D.}\ \bibnamefont {Jost}}, \bibinfo
  {author} {\bibfnamefont {C.~Y.}\ \bibnamefont {Wang}}, \bibinfo {author}
  {\bibfnamefont {N.~M.}\ \bibnamefont {Kondratiev}}, \bibinfo {author}
  {\bibfnamefont {M.~L.}\ \bibnamefont {Gorodetsky}}, \ and\ \bibinfo {author}
  {\bibfnamefont {T.~J.}\ \bibnamefont {Kippenberg}},\ }\href {\doibase
  10.1038/nphoton.2013.343} {\bibfield  {journal} {\bibinfo  {journal} {Nat
  Photon}\ }\textbf {\bibinfo {volume} {8}},\ \bibinfo {pages} {145} (\bibinfo
  {year} {2014})}\BibitemShut {NoStop}%
\bibitem [{\citenamefont {Erickson}\ \emph {et~al.}(2014)\citenamefont
  {Erickson}, \citenamefont {Vissers}, \citenamefont {Sandberg}, \citenamefont
  {Jefferts},\ and\ \citenamefont {Pappas}}]{erickson_frequency_2014}%
  \BibitemOpen
  \bibfield  {author} {\bibinfo {author} {\bibfnamefont {R.~P.}\ \bibnamefont
  {Erickson}}, \bibinfo {author} {\bibfnamefont {M.~R.}\ \bibnamefont
  {Vissers}}, \bibinfo {author} {\bibfnamefont {M.}~\bibnamefont {Sandberg}},
  \bibinfo {author} {\bibfnamefont {S.~R.}\ \bibnamefont {Jefferts}}, \ and\
  \bibinfo {author} {\bibfnamefont {D.~P.}\ \bibnamefont {Pappas}},\ }\href
  {\doibase 10.1103/PhysRevLett.113.187002} {\bibfield  {journal} {\bibinfo
  {journal} {Phys. Rev. Lett.}\ }\textbf {\bibinfo {volume} {113}},\ \bibinfo
  {pages} {187002} (\bibinfo {year} {2014})}\BibitemShut {NoStop}%
\bibitem [{\citenamefont {Haus}(2000)}]{haus_mode-locking_2000}%
  \BibitemOpen
  \bibfield  {author} {\bibinfo {author} {\bibfnamefont {H.~A.}\ \bibnamefont
  {Haus}},\ }\href {\doibase 10.1109/2944.902165} {\bibfield  {journal}
  {\bibinfo  {journal} {IEEE Journal of Selected Topics in Quantum
  Electronics}\ }\textbf {\bibinfo {volume} {6}},\ \bibinfo {pages} {1173}
  (\bibinfo {year} {2000})}\BibitemShut {NoStop}%
\bibitem [{\citenamefont {Faist}\ \emph {et~al.}(2016)\citenamefont {Faist},
  \citenamefont {Villares}, \citenamefont {Scalari}, \citenamefont {R{\"o}sch},
  \citenamefont {Bonzon}, \citenamefont {Hugi},\ and\ \citenamefont
  {Beck}}]{faist_quantum_2016}%
  \BibitemOpen
  \bibfield  {author} {\bibinfo {author} {\bibfnamefont {J.}~\bibnamefont
  {Faist}}, \bibinfo {author} {\bibfnamefont {G.}~\bibnamefont {Villares}},
  \bibinfo {author} {\bibfnamefont {G.}~\bibnamefont {Scalari}}, \bibinfo
  {author} {\bibfnamefont {M.}~\bibnamefont {R{\"o}sch}}, \bibinfo {author}
  {\bibfnamefont {C.}~\bibnamefont {Bonzon}}, \bibinfo {author} {\bibfnamefont
  {A.}~\bibnamefont {Hugi}}, \ and\ \bibinfo {author} {\bibfnamefont
  {M.}~\bibnamefont {Beck}},\ }\href {\doibase 10.1515/nanoph-2016-0015}
  {\bibfield  {journal} {\bibinfo  {journal} {Nanophotonics}\ }\textbf
  {\bibinfo {volume} {5}} (\bibinfo {year} {2016}),\
  10.1515/nanoph-2016-0015}\BibitemShut {NoStop}%
\bibitem [{\citenamefont {Ganesan}\ \emph
  {et~al.}(2017{\natexlab{a}})\citenamefont {Ganesan}, \citenamefont {Do},\
  and\ \citenamefont {Seshia}}]{ganesan_frequency_2017}%
  \BibitemOpen
  \bibfield  {author} {\bibinfo {author} {\bibfnamefont {A.}~\bibnamefont
  {Ganesan}}, \bibinfo {author} {\bibfnamefont {C.}~\bibnamefont {Do}}, \ and\
  \bibinfo {author} {\bibfnamefont {A.}~\bibnamefont {Seshia}},\ }\href
  {\doibase 10.1063/1.4985266} {\bibfield  {journal} {\bibinfo  {journal}
  {Appl. Phys. Lett.}\ }\textbf {\bibinfo {volume} {111}},\ \bibinfo {pages}
  {064101} (\bibinfo {year} {2017}{\natexlab{a}})}\BibitemShut {NoStop}%
\bibitem [{\citenamefont {Ganesan}\ \emph
  {et~al.}(2017{\natexlab{b}})\citenamefont {Ganesan}, \citenamefont {Do},\
  and\ \citenamefont {Seshia}}]{ganesan_phononic_2017}%
  \BibitemOpen
  \bibfield  {author} {\bibinfo {author} {\bibfnamefont {A.}~\bibnamefont
  {Ganesan}}, \bibinfo {author} {\bibfnamefont {C.}~\bibnamefont {Do}}, \ and\
  \bibinfo {author} {\bibfnamefont {A.}~\bibnamefont {Seshia}},\ }\href
  {\doibase 10.1103/PhysRevLett.118.033903} {\bibfield  {journal} {\bibinfo
  {journal} {Phys. Rev. Lett.}\ }\textbf {\bibinfo {volume} {118}},\ \bibinfo
  {pages} {033903} (\bibinfo {year} {2017}{\natexlab{b}})}\BibitemShut
  {NoStop}%
\bibitem [{\citenamefont {Chembo}\ and\ \citenamefont
  {Yu}(2010)}]{chembo_modal_2010}%
  \BibitemOpen
  \bibfield  {author} {\bibinfo {author} {\bibfnamefont {Y.~K.}\ \bibnamefont
  {Chembo}}\ and\ \bibinfo {author} {\bibfnamefont {N.}~\bibnamefont {Yu}},\
  }\href {\doibase 10.1103/PhysRevA.82.033801} {\bibfield  {journal} {\bibinfo
  {journal} {Phys. Rev. A}\ }\textbf {\bibinfo {volume} {82}},\ \bibinfo
  {pages} {033801} (\bibinfo {year} {2010})}\BibitemShut {NoStop}%
\bibitem [{\citenamefont {Mansuripur}\ \emph {et~al.}(2016)\citenamefont
  {Mansuripur}, \citenamefont {Vernet}, \citenamefont {Chevalier},
  \citenamefont {Aoust}, \citenamefont {Schwarz}, \citenamefont {Xie},
  \citenamefont {Caneau}, \citenamefont {Lascola}, \citenamefont {Zah},
  \citenamefont {Caffey}, \citenamefont {Day}, \citenamefont {Missaggia},
  \citenamefont {Connors}, \citenamefont {Wang}, \citenamefont {Belyanin},\
  and\ \citenamefont {Capasso}}]{mansuripur_single-mode_2016}%
  \BibitemOpen
  \bibfield  {author} {\bibinfo {author} {\bibfnamefont {T.~S.}\ \bibnamefont
  {Mansuripur}}, \bibinfo {author} {\bibfnamefont {C.}~\bibnamefont {Vernet}},
  \bibinfo {author} {\bibfnamefont {P.}~\bibnamefont {Chevalier}}, \bibinfo
  {author} {\bibfnamefont {G.}~\bibnamefont {Aoust}}, \bibinfo {author}
  {\bibfnamefont {B.}~\bibnamefont {Schwarz}}, \bibinfo {author} {\bibfnamefont
  {F.}~\bibnamefont {Xie}}, \bibinfo {author} {\bibfnamefont {C.}~\bibnamefont
  {Caneau}}, \bibinfo {author} {\bibfnamefont {K.}~\bibnamefont {Lascola}},
  \bibinfo {author} {\bibfnamefont {C.-e.}\ \bibnamefont {Zah}}, \bibinfo
  {author} {\bibfnamefont {D.~P.}\ \bibnamefont {Caffey}}, \bibinfo {author}
  {\bibfnamefont {T.}~\bibnamefont {Day}}, \bibinfo {author} {\bibfnamefont
  {L.~J.}\ \bibnamefont {Missaggia}}, \bibinfo {author} {\bibfnamefont {M.~K.}\
  \bibnamefont {Connors}}, \bibinfo {author} {\bibfnamefont {C.~A.}\
  \bibnamefont {Wang}}, \bibinfo {author} {\bibfnamefont {A.}~\bibnamefont
  {Belyanin}}, \ and\ \bibinfo {author} {\bibfnamefont {F.}~\bibnamefont
  {Capasso}},\ }\href {\doibase 10.1103/PhysRevA.94.063807} {\bibfield
  {journal} {\bibinfo  {journal} {Phys. Rev. A}\ }\textbf {\bibinfo {volume}
  {94}},\ \bibinfo {pages} {063807} (\bibinfo {year} {2016})}\BibitemShut
  {NoStop}%
\bibitem [{\citenamefont {Rayanov}\ \emph {et~al.}(2015)\citenamefont
  {Rayanov}, \citenamefont {Altshuler}, \citenamefont {Rubo},\ and\
  \citenamefont {Flach}}]{rayanov_frequency_2015}%
  \BibitemOpen
  \bibfield  {author} {\bibinfo {author} {\bibfnamefont {K.}~\bibnamefont
  {Rayanov}}, \bibinfo {author} {\bibfnamefont {B.~L.}\ \bibnamefont
  {Altshuler}}, \bibinfo {author} {\bibfnamefont {Y.~G.}\ \bibnamefont {Rubo}},
  \ and\ \bibinfo {author} {\bibfnamefont {S.}~\bibnamefont {Flach}},\ }\href
  {\doibase 10.1103/PhysRevLett.114.193901} {\bibfield  {journal} {\bibinfo
  {journal} {Phys. Rev. Lett.}\ }\textbf {\bibinfo {volume} {114}},\ \bibinfo
  {pages} {193901} (\bibinfo {year} {2015})}\BibitemShut {NoStop}%
\bibitem [{\citenamefont {D.Sc}(1926)}]{d.sc_lxxxviii._1926}%
  \BibitemOpen
  \bibfield  {author} {\bibinfo {author} {\bibfnamefont {B.~v. d. P.~J.}\
  \bibnamefont {D.Sc}},\ }\href {\doibase 10.1080/14786442608564127} {\bibfield
   {journal} {\bibinfo  {journal} {The London, Edinburgh, and Dublin
  Philosophical Magazine and Journal of Science}\ }\textbf {\bibinfo {volume}
  {2}},\ \bibinfo {pages} {978} (\bibinfo {year} {1926})}\BibitemShut {NoStop}%
\bibitem [{\citenamefont {Strogatz}(1994)}]{strogatz_nonlinear_1994}%
  \BibitemOpen
  \bibfield  {author} {\bibinfo {author} {\bibfnamefont {S.~H.}\ \bibnamefont
  {Strogatz}},\ }\href@noop {} {\emph {\bibinfo {title}
  {Nonlinear {Dynamics} and {Chaos}: {With} {Applications} to {Physics},
  {Biology}, {Chemistry}, and {Engineering}}}}\ (\bibinfo  {publisher} {Avalon
  Publishing},\ \bibinfo {year} {1994})\ \bibinfo {note} {google-Books-ID:
  FIYHiBLWCJMC}\BibitemShut {NoStop}%
\bibitem [{\citenamefont {Louca}(2015)}]{louca_stable_2015}%
  \BibitemOpen
  \bibfield  {author} {\bibinfo {author} {\bibfnamefont {S.}~\bibnamefont
  {Louca}},\ }\href {http://arxiv.org/abs/1506.00756} {\bibfield  {journal}
  {\bibinfo  {journal} {arXiv:1506.00756 [math]}\ } (\bibinfo {year} {2015})},\
  \bibinfo {note} {arXiv: 1506.00756}\BibitemShut {NoStop}%
\bibitem [{\citenamefont {L{\"o}rch}\ \emph {et~al.}(2014)\citenamefont
  {L{\"o}rch}, \citenamefont {Qian}, \citenamefont {Clerk}, \citenamefont
  {Marquardt},\ and\ \citenamefont {Hammerer}}]{lorch_laser_2014}%
  \BibitemOpen
  \bibfield  {author} {\bibinfo {author} {\bibfnamefont {N.}~\bibnamefont
  {L{\"o}rch}}, \bibinfo {author} {\bibfnamefont {J.}~\bibnamefont {Qian}},
  \bibinfo {author} {\bibfnamefont {A.}~\bibnamefont {Clerk}}, \bibinfo
  {author} {\bibfnamefont {F.}~\bibnamefont {Marquardt}}, \ and\ \bibinfo
  {author} {\bibfnamefont {K.}~\bibnamefont {Hammerer}},\ }\href {\doibase
  10.1103/PhysRevX.4.011015} {\bibfield  {journal} {\bibinfo  {journal} {Phys.
  Rev. X}\ }\textbf {\bibinfo {volume} {4}},\ \bibinfo {pages} {011015}
  (\bibinfo {year} {2014})}\BibitemShut {NoStop}%
\bibitem [{\citenamefont {Navarrete-Benlloch}\ \emph
  {et~al.}(2017)\citenamefont {Navarrete-Benlloch}, \citenamefont {Weiss},
  \citenamefont {Walter},\ and\ \citenamefont
  {de~Valc{\'a}rcel}}]{navarrete-benlloch_general_2017}%
  \BibitemOpen
  \bibfield  {author} {\bibinfo {author} {\bibfnamefont {C.}~\bibnamefont
  {Navarrete-Benlloch}}, \bibinfo {author} {\bibfnamefont {T.}~\bibnamefont
  {Weiss}}, \bibinfo {author} {\bibfnamefont {S.}~\bibnamefont {Walter}}, \
  and\ \bibinfo {author} {\bibfnamefont {G.~J.}\ \bibnamefont
  {de~Valc{\'a}rcel}},\ }\href {\doibase 10.1103/PhysRevLett.119.133601}
  {\bibfield  {journal} {\bibinfo  {journal} {Phys. Rev. Lett.}\ }\textbf
  {\bibinfo {volume} {119}},\ \bibinfo {pages} {133601} (\bibinfo {year}
  {2017})}\BibitemShut {NoStop}%
\bibitem [{\citenamefont {Bourassa}\ \emph {et~al.}(2012)\citenamefont
  {Bourassa}, \citenamefont {Beaudoin}, \citenamefont {Gambetta},\ and\
  \citenamefont {Blais}}]{bourassa_josephson-junction-embedded_2012}%
  \BibitemOpen
  \bibfield  {author} {\bibinfo {author} {\bibfnamefont {J.}~\bibnamefont
  {Bourassa}}, \bibinfo {author} {\bibfnamefont {F.}~\bibnamefont {Beaudoin}},
  \bibinfo {author} {\bibfnamefont {J.~M.}\ \bibnamefont {Gambetta}}, \ and\
  \bibinfo {author} {\bibfnamefont {A.}~\bibnamefont {Blais}},\ }\href
  {\doibase 10.1103/PhysRevA.86.013814} {\bibfield  {journal} {\bibinfo
  {journal} {Phys. Rev. A}\ }\textbf {\bibinfo {volume} {86}},\ \bibinfo
  {pages} {013814} (\bibinfo {year} {2012})}\BibitemShut {NoStop}%
\bibitem [{\citenamefont {Khan}(2017)}]{khan_supplementary_2017}%
  \BibitemOpen
  \href {} {\bibinfo {title} {See Supplementary Material [url]}, which includes Ref.~\cite{girvin_circuit_2014}}{\ }\BibitemShut {NoStop}%
\bibitem [{\citenamefont {Drummond}\ and\ \citenamefont
  {Walls}(1980)}]{drummond_quantum_1980}%
  \BibitemOpen
  \bibfield  {author} {\bibinfo {author} {\bibfnamefont {P.~D.}\ \bibnamefont
  {Drummond}}\ and\ \bibinfo {author} {\bibfnamefont {D.~F.}\ \bibnamefont
  {Walls}},\ }\href {\doibase 10.1088/0305-4470/13/2/034} {\bibfield  {journal}
  {\bibinfo  {journal} {J. Phys. A: Math. Gen.}\ }\textbf {\bibinfo {volume}
  {13}},\ \bibinfo {pages} {725} (\bibinfo {year} {1980})}\BibitemShut
  {NoStop}%
\bibitem [{\citenamefont {da~Silva}\ \emph {et~al.}(2010)\citenamefont
  {da~Silva}, \citenamefont {Bozyigit}, \citenamefont {Wallraff},\ and\
  \citenamefont {Blais}}]{da_silva_schemes_2010}%
  \BibitemOpen
  \bibfield  {author} {\bibinfo {author} {\bibfnamefont {M.~P.}\ \bibnamefont
  {da~Silva}}, \bibinfo {author} {\bibfnamefont {D.}~\bibnamefont {Bozyigit}},
  \bibinfo {author} {\bibfnamefont {A.}~\bibnamefont {Wallraff}}, \ and\
  \bibinfo {author} {\bibfnamefont {A.}~\bibnamefont {Blais}},\ }\href
  {\doibase 10.1103/PhysRevA.82.043804} {\bibfield  {journal} {\bibinfo
  {journal} {Phys. Rev. A}\ }\textbf {\bibinfo {volume} {82}},\ \bibinfo
  {pages} {043804} (\bibinfo {year} {2010})}\BibitemShut {NoStop}%
\bibitem [{\citenamefont {Carmichael}(2002)}]{carmichael_statistical_2002}%
  \BibitemOpen
  \bibfield  {author} {\bibinfo {author} {\bibfnamefont {H.~J.}\ \bibnamefont
  {Carmichael}},\ }\href {http://www.springer.com/us/book/9783540548829} {\emph
  {\bibinfo {title} {Statistical {Methods} in {Quantum} {Optics} 1 - {Master}
  {\textbar} {Howard} {J}. {Carmichael} {\textbar} {Springer}}}}\ (\bibinfo
  {year} {2002})\BibitemShut {NoStop}%
\bibitem [{\citenamefont {Gilchrist}\ \emph {et~al.}(1997)\citenamefont
  {Gilchrist}, \citenamefont {Gardiner},\ and\ \citenamefont
  {Drummond}}]{gilchrist_positive_1997}%
  \BibitemOpen
  \bibfield  {author} {\bibinfo {author} {\bibfnamefont {A.}~\bibnamefont
  {Gilchrist}}, \bibinfo {author} {\bibfnamefont {C.~W.}\ \bibnamefont
  {Gardiner}}, \ and\ \bibinfo {author} {\bibfnamefont {P.~D.}\ \bibnamefont
  {Drummond}},\ }\href {\doibase 10.1103/PhysRevA.55.3014} {\bibfield
  {journal} {\bibinfo  {journal} {Phys. Rev. A}\ }\textbf {\bibinfo {volume}
  {55}},\ \bibinfo {pages} {3014} (\bibinfo {year} {1997})}\BibitemShut
  {NoStop}%
\bibitem [{\citenamefont {Gardiner}(2004)}]{gardiner_quantum_2004}%
  \BibitemOpen
  \bibfield  {author} {\bibinfo {author} {\bibfnamefont {C.}~\bibnamefont
  {Gardiner}},\ }\href {http://www.springer.com/us/book/9783540223016} {\emph
  {\bibinfo {title} {Quantum {Noise} - {A} {Handbook} of {Markovian} and
  {Non}-{Markovian} {\textbar} {Crispin} {Gardiner} {\textbar} {Springer}}}}\
  (\bibinfo {year} {2004})\BibitemShut {NoStop}%
\bibitem [{\citenamefont {Johansson}\ \emph {et~al.}(2013)\citenamefont
  {Johansson}, \citenamefont {Nation},\ and\ \citenamefont
  {Nori}}]{johansson_qutip_2013}%
  \BibitemOpen
  \bibfield  {author} {\bibinfo {author} {\bibfnamefont {J.~R.}\ \bibnamefont
  {Johansson}}, \bibinfo {author} {\bibfnamefont {P.~D.}\ \bibnamefont
  {Nation}}, \ and\ \bibinfo {author} {\bibfnamefont {F.}~\bibnamefont
  {Nori}},\ }\href {\doibase 10.1016/j.cpc.2012.11.019} {\bibfield  {journal}
  {\bibinfo  {journal} {Computer Physics Communications}\ }\textbf {\bibinfo
  {volume} {184}},\ \bibinfo {pages} {1234} (\bibinfo {year}
  {2013})}\BibitemShut {NoStop}%
\bibitem [{\citenamefont {Kamal}\ \emph {et~al.}(2009)\citenamefont {Kamal},
  \citenamefont {Marblestone},\ and\ \citenamefont
  {Devoret}}]{kamal_signal--pump_2009}%
  \BibitemOpen
  \bibfield  {author} {\bibinfo {author} {\bibfnamefont {A.}~\bibnamefont
  {Kamal}}, \bibinfo {author} {\bibfnamefont {A.}~\bibnamefont {Marblestone}},
  \ and\ \bibinfo {author} {\bibfnamefont {M.}~\bibnamefont {Devoret}},\ }\href
  {\doibase 10.1103/PhysRevB.79.184301} {\bibfield  {journal} {\bibinfo
  {journal} {Phys. Rev. B}\ }\textbf {\bibinfo {volume} {79}},\ \bibinfo
  {pages} {184301} (\bibinfo {year} {2009})}\BibitemShut {NoStop}%
\bibitem [{\citenamefont {Casteels}\ and\ \citenamefont
  {Ciuti}(2017)}]{casteels_quantum_2017}%
  \BibitemOpen
  \bibfield  {author} {\bibinfo {author} {\bibfnamefont {W.}~\bibnamefont
  {Casteels}}\ and\ \bibinfo {author} {\bibfnamefont {C.}~\bibnamefont
  {Ciuti}},\ }\href {\doibase 10.1103/PhysRevA.95.013812} {\bibfield  {journal}
  {\bibinfo  {journal} {Phys. Rev. A}\ }\textbf {\bibinfo {volume} {95}},\
  \bibinfo {pages} {013812} (\bibinfo {year} {2017})}\BibitemShut {NoStop}%
\bibitem [{\citenamefont {Chaturvedi}\ \emph {et~al.}(1977)\citenamefont
  {Chaturvedi}, \citenamefont {Gardiner}, \citenamefont {Matheson},\ and\
  \citenamefont {Walls}}]{chaturvedi_stochastic_1977}%
  \BibitemOpen
  \bibfield  {author} {\bibinfo {author} {\bibfnamefont {S.}~\bibnamefont
  {Chaturvedi}}, \bibinfo {author} {\bibfnamefont {C.~W.}\ \bibnamefont
  {Gardiner}}, \bibinfo {author} {\bibfnamefont {I.~S.}\ \bibnamefont
  {Matheson}}, \ and\ \bibinfo {author} {\bibfnamefont {D.~F.}\ \bibnamefont
  {Walls}},\ }\href {\doibase 10.1007/BF01014350} {\bibfield  {journal}
  {\bibinfo  {journal} {J Stat Phys}\ }\textbf {\bibinfo {volume} {17}},\
  \bibinfo {pages} {469} (\bibinfo {year} {1977})}\BibitemShut {NoStop}%
\bibitem [{\citenamefont {Eichler}\ and\ \citenamefont
  {Wallraff}(2014)}]{eichler_controlling_2014}%
  \BibitemOpen
  \bibfield  {author} {\bibinfo {author} {\bibfnamefont {C.}~\bibnamefont
  {Eichler}}\ and\ \bibinfo {author} {\bibfnamefont {A.}~\bibnamefont
  {Wallraff}},\ }\href {\doibase 10.1140/epjqt2} {\bibfield  {journal}
  {\bibinfo  {journal} {EPJ Quantum Technology}\ }\textbf {\bibinfo {volume}
  {1}},\ \bibinfo {pages} {2} (\bibinfo {year} {2014})}\BibitemShut {NoStop}%
\bibitem [{\citenamefont {Zhou}\ \emph {et~al.}(2014)\citenamefont {Zhou},
  \citenamefont {Schmitt}, \citenamefont {Bertet}, \citenamefont {Vion},
  \citenamefont {Wustmann}, \citenamefont {Shumeiko},\ and\ \citenamefont
  {Esteve}}]{zhou_high-gain_2014}%
  \BibitemOpen
  \bibfield  {author} {\bibinfo {author} {\bibfnamefont {X.}~\bibnamefont
  {Zhou}}, \bibinfo {author} {\bibfnamefont {V.}~\bibnamefont {Schmitt}},
  \bibinfo {author} {\bibfnamefont {P.}~\bibnamefont {Bertet}}, \bibinfo
  {author} {\bibfnamefont {D.}~\bibnamefont {Vion}}, \bibinfo {author}
  {\bibfnamefont {W.}~\bibnamefont {Wustmann}}, \bibinfo {author}
  {\bibfnamefont {V.}~\bibnamefont {Shumeiko}}, \ and\ \bibinfo {author}
  {\bibfnamefont {D.}~\bibnamefont {Esteve}},\ }\href {\doibase
  10.1103/PhysRevB.89.214517} {\bibfield  {journal} {\bibinfo  {journal} {Phys.
  Rev. B}\ }\textbf {\bibinfo {volume} {89}},\ \bibinfo {pages} {214517}
  (\bibinfo {year} {2014})}\BibitemShut {NoStop}%
\bibitem [{\citenamefont {Girvin}(2014)}]{girvin_circuit_2014}%
  \BibitemOpen
  \bibfield  {author} {\bibinfo {author} {\bibfnamefont {S.~M.}\ \bibnamefont
  {Girvin}},\ }in\ \href
  {http://www.oxfordscholarship.com/view/10.1093/acprof:oso/9780199681181.001.0001/acprof-9780199681181-chapter-3}
  {\emph {\bibinfo {booktitle} {Quantum {Machines}: {Measurement} and {Control}
  of {Engineered} {Quantum} {Systems}}}}\ (\bibinfo  {publisher} {Oxford
  University Press},\ \bibinfo {year} {2014})\ pp.\ \bibinfo {pages}
  {113--256},\ \bibinfo {note} {dOI:
  10.1093/acprof:oso/9780199681181.003.0003}\BibitemShut {NoStop}%
\end{thebibliography}

%\addbibresource{Bibliography/NLO-PRLNotes.bib}
%\printbibliography

\end{document}